\documentclass[aps,prl,twocolumn,hyperref,superscriptaddress]{revtex4-2}

\usepackage{hyperref}
\hypersetup{colorlinks,linkcolor=blue,citecolor=blue,anchorcolor=blue}
\usepackage{soul}
\usepackage{xcolor}

\usepackage{amssymb}
\usepackage{multirow}
\usepackage{bm}
\usepackage{graphicx}
\usepackage{epstopdf}
\usepackage{subfigure}
\usepackage{natbib}
\usepackage{epsfig}
\usepackage{amsfonts}
\usepackage{mathrsfs}
\usepackage[toc,page,title,titletoc,header]{appendix}
\usepackage{dsfont,amsthm,amsbsy}
\usepackage{amsmath}
\usepackage{mathtools}
\usepackage{hyperref}
\usepackage{ulem}

\usepackage{etoolbox}
\patchcmd{\section}
  {\centering}
  {\raggedright}
  {}
  {}

\usepackage{bold-extra}
\newcommand{\inlinesubsection}[1]{
    \addcontentsline{toc}{subsection}{#1}
    \noindent{\bfseries #1}
}

\definecolor{lightred}{rgb}{0.88, 0.75, 0.78} 

\definecolor{lightblue}{rgb}{0.68, 0.85, 0.9}

\begin{document}

\title{Self-Supervised Generative Models for Crystal Structures}

\author{Fangze~Liu}
\email{fangzel@stanford.edu}
\affiliation{Department of Physics, Stanford University, Stanford, CA 94305, USA}
\affiliation{Stanford Institute for Materials and Energy Sciences, SLAC National Accelerator Laboratory, Menlo Park, CA 94025, USA}

\author{Zhantao~Chen}
\email{zhantao@stanford.edu}
\affiliation{Stanford Institute for Materials and Energy Sciences, SLAC National Accelerator Laboratory, Menlo Park, CA 94025, USA}
\affiliation{Linac Coherent Light Source, SLAC National Accelerator Laboratory, Menlo Park, CA 94025, USA}

\author{Tianyi~Liu}
\affiliation{Department of Chemistry, Stanford University, Stanford, CA 94305, USA}
\affiliation{Stanford Institute for Materials and Energy Sciences, SLAC National Accelerator Laboratory, Menlo Park, CA 94025, USA}

\author{Ruyi~Song}
\affiliation{Stanford Institute for Materials and Energy Sciences, SLAC National Accelerator Laboratory, Menlo Park, CA 94025, USA}

\author{Yu~Lin}
\affiliation{Stanford Institute for Materials and Energy Sciences, SLAC National Accelerator Laboratory, Menlo Park, CA 94025, USA}

\author{Joshua~J.~Turner}
\email{joshuat@slac.stanford.edu}
\affiliation{Stanford Institute for Materials and Energy Sciences, SLAC National Accelerator Laboratory, Menlo Park, CA 94025, USA}
\affiliation{Linac Coherent Light Source, SLAC National Accelerator Laboratory, Menlo Park, CA 94025, USA}

\author{Chunjing~Jia}
\email{chunjing@phys.ufl.edu}
\affiliation{Stanford Institute for Materials and Energy Sciences, SLAC National Accelerator Laboratory, Menlo Park, CA 94025, USA}
\affiliation{Department of Physics, University of Florida, Gainesville, FL 32611, USA}

\begin{abstract}
    Drawing inspiration from the achievements of natural language processing, we adopt self-supervised learning and utilize an equivariant graph neural network to develop a unified platform designed for training generative models capable of generating crystal structures, as well as efficiently adapting to downstream tasks in material property prediction. To mitigate the challenge of incorporating large-scale assessment on the reliability of generated structures into the training process, we utilize the generative adversarial network (GAN) with its discriminator being a cost-effective evaluator for the generated structures, resulting in notable improvements in model performance. 
    We demonstrate the utility of our model in finding the optimal crystal structure under predefined conditions. Without reliance on properties acquired experimentally or numerically, our model further displays its capability to comprehend the mechanism of crystal structure formation through its ability to grouping chemically similar elements. Therefore, this paper extends an invitation to explore deeper into the scientific understanding of material structures through generative models, offering a fresh perspective on broadening the scope and efficacy of machine learning in material science.
\end{abstract}

\maketitle

\section{Introduction}

Structure and property predictions of crystalline materials have been a longstanding and central focus in condensed matter physics and material sciences. Recent advancements have demonstrated the efficacy of machine learning techniques in predicting various materials properties, including electronic topology, thermodynamic properties, mechanical moduli, etc.~\cite{xie2018crystal, schmidt2019recent, chen2019graph, tawfik2020predicting, andrejevic2022machine, kong2022density, choudhary2022recent, moosavi2022data, liu2023machine}. Furthermore, there has been a growing focus towards applying generative models inspired from computer vision, such as generative adversarial networks (GANs)~\cite{kim2020generative, long2021constrained, zhao2021high}, diffusion models~\cite{xie2021crystal, lyngby2022data, yang2023scalable}, and variational autoencoders (VAEs)~\cite{xie2021crystal, zhu2023wycryst} for crystal structure generation. GANs, which involve adversarial training of two neural networks, have been applied to a limited range of materials, like specific compositional families~\cite{kim2020generative} and two-dimensional materials~\cite{long2021constrained, zhao2021high}. VAEs, which encode material representation into a compressed latent space and generate materials by decoding sampled latent codes in the space, is a favored architecture that can be effectively combined with other training methods like GAN~\cite{kim2020generative} and diffusion processes~\cite{xie2021crystal}, though they demand additional efforts for effective latent space interpretation and utilization. Diffusion models, known as a physics-induced models and celebrated in computer vision~\cite{yang2023diffusion}, have seen successful adaptation for crystal generation~\cite{xie2021crystal}, a trend underscored by a recent surge of studies~\cite{yang2023scalable, zeni2023mattergen}. Instead of using generative models, the combination of simple element substitutions and density function theory (DFT)~\cite{merchant2023scaling, szymanski2023autonomous} hints at the benefits of incorporating DFT into the generative learning frameworks for enhancing crystal generation.

\begin{figure*}[htb!]
\includegraphics[width=0.8\linewidth]{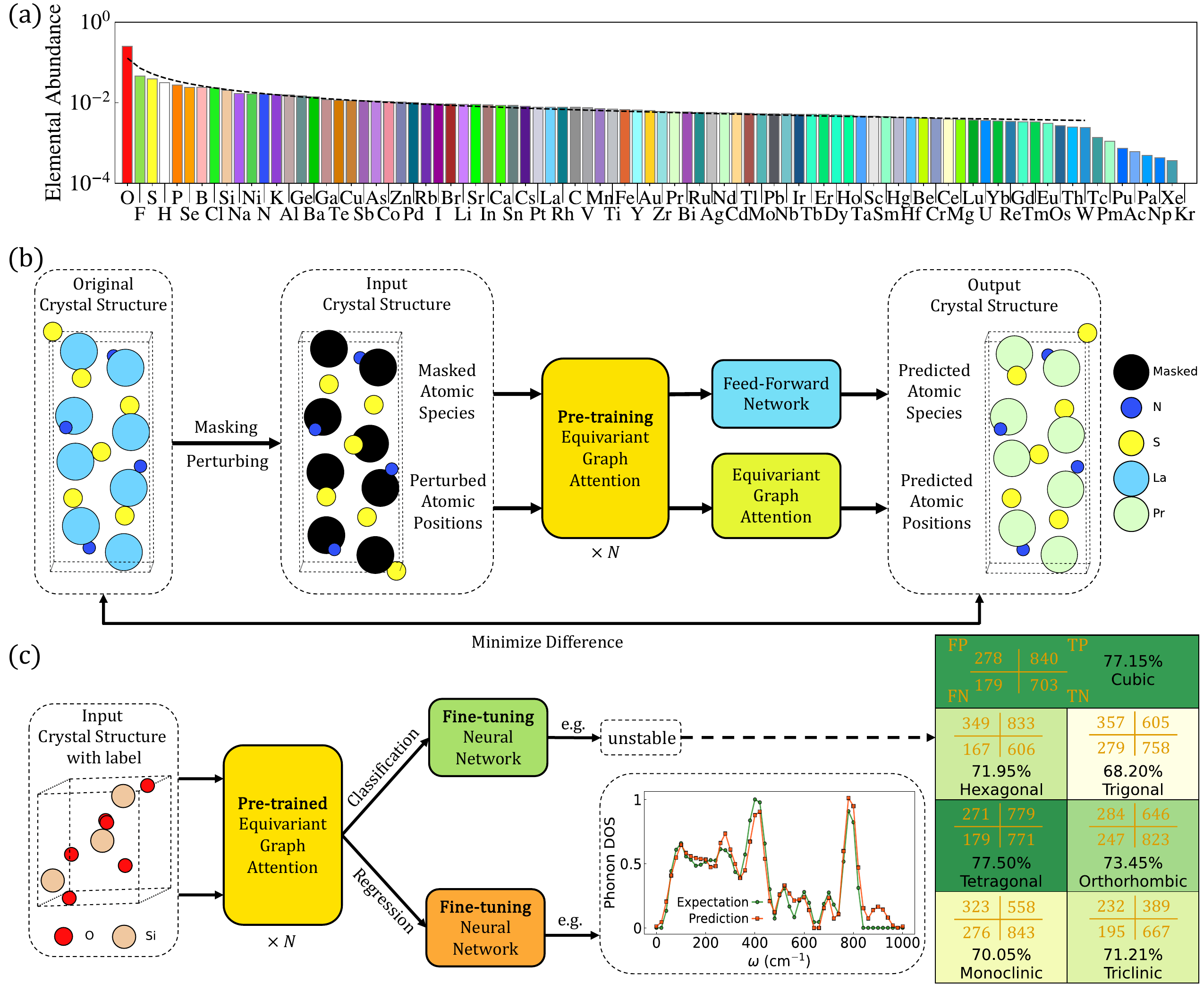}
\caption{\label{self_supervise}
\textbf{Self-supervised learning in crystal structure generation.} 
(a) Training samples show a power law elemental abundance distribution, $p(X=x) \propto 1/x^\alpha$, where $X$ is the rank of an element and $\alpha=0.8163$ with excluding the the eight least frequently occurring elements. This resulting distribution (dashed line) resembles the natural language token distribution, typically with an exponent close to $1$~\cite{chan2022data}, supporting the utilization of self-supervised learning in crystal generation.
(b) The self-supervised pre-training employs an equivariant graph attention transformer architecture, EquiformerV2~\cite{liao2023equiformerv2}, processing contaminated structures produced by masking parts of the atoms and perturbing atomic positions in the original stable crystal structures, and being trained on tasks to predict the masked atoms and restore atomic positions. Primitive lattice vectors are predefined to compute the edge distance embedding, serving as input for Equiformer. 
The pre-trained model is well-equipped for various supervised down-stream tasks through the fine-tuning procedure as illustrated in (c). During fine-tuning, the pre-trained model is concatenated with a randomly initialized feed-forward neural network and all model parameters undergo fine-tuning. 
To showcase the model's classification ability, an example of predicting crystal stability, using a labeled training set based solely on cubic system, is illustrated. The accompanying diagram displays classification results (True Positives, True Negatives, False Negatives, and False Positives in a clockwise order), with accuracy indicated by percentage numbers and reflected by colors. 
An example of the regression tasks, predicting phonon Density of States (DOS), is also depicted. The training set is sourced from the Materials Project. More details are provided in Supplementary Materials. 
}
\end{figure*}

One fundamental question is focused on how to understand the connection between atomic structures and the properties of materials. Machine learning techniques have proven powerful in predicting the latter, given the former, especially as the dataset of stable crystal structures is currently much larger than the dataset of stable structures with properties obtained either numerically or experimentally. However, predicting atomic structures presents considerably greater difficulties, including the vast structure design space that exists for materials discovery and the lack of suitable evaluation metrics for generated materials. 
This situation is similar to the challenges faced in natural language processing (NLP), as the volume of unlabeled textual data is much larger than that of labeled question-answer data. Another level of similarity is observed in the distribution of atomic species in crystal structures compared to the distribution of vocabulary in natural language data~\cite{piantadosi2014zipf} (see Fig.~\ref{self_supervise}a). As studies have found that the emergent learning ability of large transformer-based language models~\cite{vaswani2017attention} is attributed to the skewed and long-tailed distributional properties of the training data~\cite{chan2022data}, this suggests that the established strategies adopted in training large language models could potentially be successful in training machine learning models for materials science. Comparing to the costly in-context learning~\cite{brown2020language}, self-supervised learning is a more approachable training method~\cite{devlin2018bert, galassi2020attention, niu2021review}: a neural network is pre-trained by large volumes of unlabeled and augmented data, like randomly masking tokens or shuffling the order of sentences~\cite{lewis2019bart}. The pre-trained model can capture the nuanced patterns and structures in training data, and further be fine-tuned on task-specific labeled data and reduce the reliance on expensive labeled datasets.

Taking inspiration from NLP, we present a novel machine learning framework for material generation and properties prediction, empowered by an efficient pre-training strategy without the need of human knowledge as a prerequisite for labeling. Analogous to a typical training procedure for NLP models, where fill-in-the-blank and sentence-arranging exercises are utilized, we design a self-supervised training procedure for models based on the state-of-the-art transformer-based equivariant graph neural network, EquiformerV2~\cite{liao2023equiformerv2}. During pre-training, the model takes contaminated crystal structures as inputs, in which atomic species and positions are randomly masked and perturbed, and learns to reconstruct the complete and noiseless structures. In tests with crystal structures previously unseen by the model, the pre-trained model has shown a preliminary ability to identify local optimal solutions from incomplete structures. Hence, when provided with a basic design for the structure of novel materials, our model can produce a completion plan, generating the most likely stable crystal structures under given conditions. In addition, we also demonstrate the adaptability of our pre-trained model for down-stream regression and classification tasks through supervised fine-tuning.

The challenge, however, lies in the lack of a critic to evaluate the generated structures efficiently. We consider using the actor-critic learning framework to guide generative training, similar to the application of reinforcement learning from human feedback (RLHF) in NLP~\cite{christiano2017deep}. While numerical methods like DFT calculations are potential options for the critic's role, leveraging GANs offers a significantly more computationally efficient solution for early-stage training. More importantly, our goal extends beyond mere generative tasks. We aim for models to reveal the intrinsic information embedded within material structures. For example, our model serves as a conditional probabilistic model for investigating the likelihood of various compositions being stable under given crystallographic conditions, so it offers a probabilistic insight into the nature of crystals and delivers richer information compared to other unconditional probabilistic models derived through data mining~\cite{hautier2011data, glawe2016optimal}. Consequently, we incorporate the GAN architecture into our training pipeline to fully exploit existing data without relying on external information, such as DFT-calculated stability labels, and thereby circumvent the limitations imposed by human preconceptions.

In this paper, we explore the application of self-supervised learning to crystal structure generation and demonstrate that the incorporation of a discriminative model can enhance the reliability of the generated crystal structures with minimal additional effort. This enhancement is evidenced by a comprehensive comparison between the outcomes of self-supervised learning and its combination with GAN. Our methodology is unique as it relies solely on unlabeled crystal structure data, offering a data- and computation-efficient training strategy for early-stage training. Furthermore, this approach paves the way for a first-principles understanding of the intrinsic information hidden in existing material structures, thereby serving as an invitation for further exploration into the analysis of material structures through generative models.

\begin{figure*}[htb!]
\includegraphics[width=0.6\linewidth]{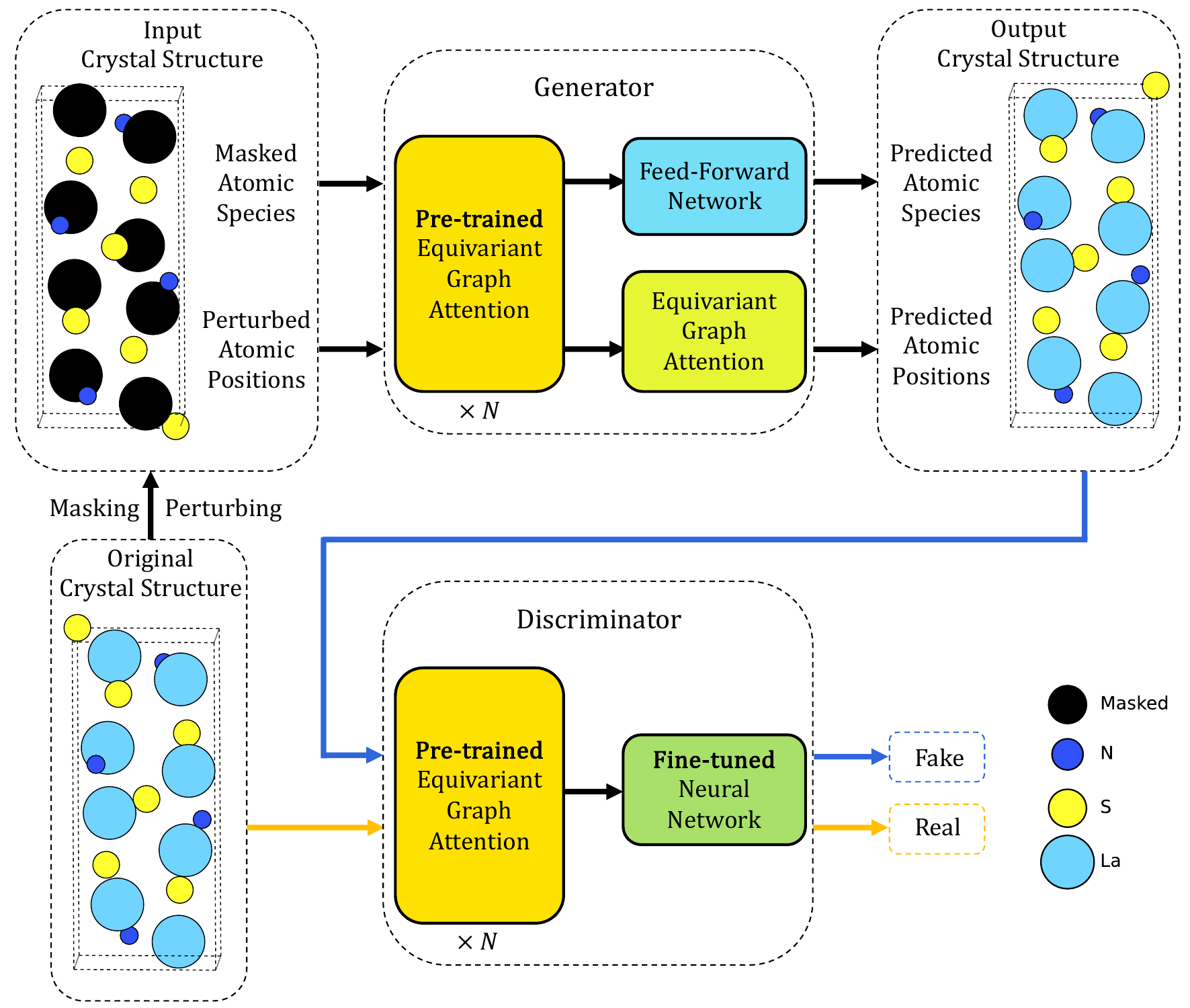}
\caption{\label{gan_illustration}
\textbf{Generative adversarial network for crystal structure generation.} We further adopt a GAN framework as a strategic enhancement to the pre-trained generative model. The discriminator, initialized with the fine-tuned stability-predicting model, is to distinguish between original stable and generated crystal structures, while the generator, initialized with the pre-trained model, aims to produce crystal structures indistinguishable from the real ones to the discriminator. All model parameters undergo fine-tuning.}
\end{figure*}

\begin{figure*}[htb!]
\includegraphics[width=1\linewidth]{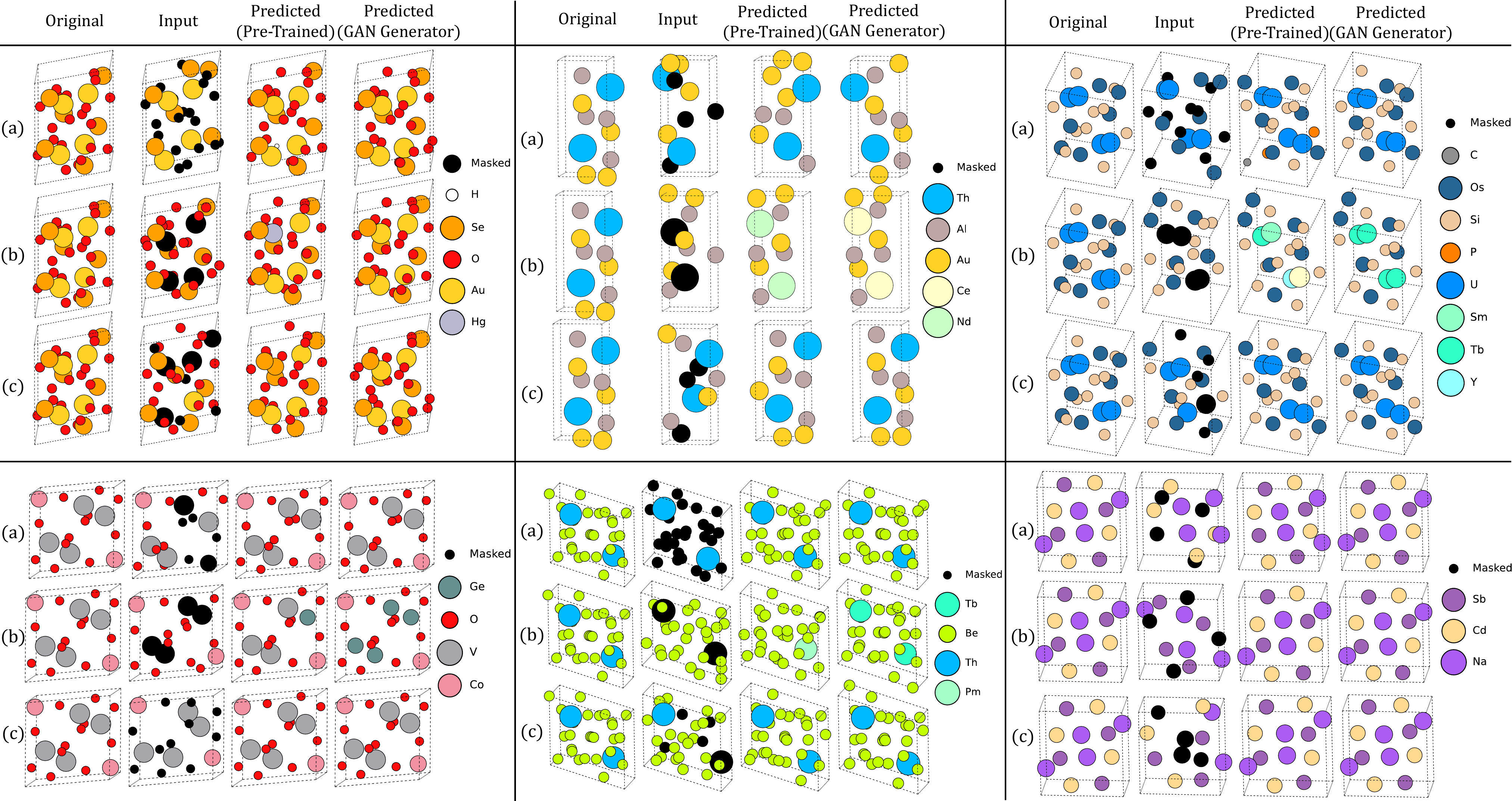}
\caption{\label{generators_results}
\textbf{Generative/reconstructive capability illustration.} 
Showcasing the capability of the pre-trained generative model and the GAN generator to find the optimal crystal structures from inputs with species masked in three ways: (a) frequent occurring species, (b) infrequent occurring species, and (c) $30\%$ random atoms, along with all positions perturbed by noise following a normal distribution with a standard deviation of $\tilde{\sigma}_{\text{noise}}\times\min(\text{edges})$ ($\tilde{\sigma}_{\text{noise}}=0.2$). Note that the pre-trained model is trained with $\tilde{\sigma}_{\text{noise}}=0.1$, and the GAN generator with $\tilde{\sigma}_{\text{noise}}=0.2$. 
More examples are provided in Supplementary Materials.
}
\end{figure*}

\section{Results}

\inlinesubsection{Pre-training for crystal generation.} To obtain a generative model, we integrate the concepts of masked training and machine learning denoising into equivariant graph neural networks. Given a stable crystal structure as a valid training sample, we prepare an imperfect input structure by masking a portion of the atoms, \textit{i.e.}, setting their atomic numbers to zero, and perturbing the equilibrium atomic positions. In particular, the incomplete input structures are generated through the following operations: (1) masking all atoms of a randomly selected species or randomly masking $15\%$ of all atoms in the unit cell, regardless of species, where the choice between these masking strategies is made randomly; (2) adding random displacements to the positions of all atoms, including the masked ones. The primitive lattice vectors and periodicity of the input structures are predefined to compute the edge distance embedding, serving as input for the generative model. As illustrated in Fig.~\ref{self_supervise}b, the model is designed to reconstruct the complete structures from the given inputs by predicting the atomic species and positions in separate task-specific layers, and the discrepancies between the reconstructed structures and the pristine structures are used to update model weights.

Specifically, the model incorporates the main structure of the EquiformerV2 \cite{liao2023equiformerv2}, an equivariant graph neural network with the attention mechanism, and two auxiliary, shallow neural networks that further map the equivariant features to the desired atomic species and position information. The equivariant backbone model efficiently represents crystal structures with all symmetries preserved and provides informative embedding for the subsequent layers. To train the model, we adopt a hybrid loss function that combines the negative log-likelihood for atomic species predictions and the mean squared error for position predictions. Further details about the model and its training process are provided in the Methods section and Supplementary Materials. \\

\inlinesubsection{Fine-tuning for downstream tasks.} The pre-trained backbone model can serve as a versatile pre-trained network block adaptable for various downstream property-prediction tasks by connecting with additional shallow layers. For each specific supervised task, this pre-trained model is integrated with a feed-forward layer, which is randomly initialized and is adapted to convert the output of the pre-trained model to meet requirements of the new task. During fine-tuning, both the pre-trained model and the feed-forward layer are trained on a task-specific dataset. This approach allows the integrated model to be rapidly and efficiently tailored for a variety of tasks, offering significant advantages over training a new model from scratch.  

To illustrate the capability of performing classification as an exemplary downstream task, we showcase the model performance on predicting the stability of a material structure in the inset table of Fig.~\ref{self_supervise}c. Here, we use the stability labels from the Materials Project database determined by the convex hull. Notably, the fine-tuning network is trained on structures with cubic lattices only and reaches a commendable accuracy of $77.15\%$. It further generalizes reasonably well to diverse lattice types, yielding close accuracy for tetragonal lattices and around $70\%$ accuracies for other lattice types that the model has not been trained with. Such fine-tuning ideas can also be extended to regression tasks, such as predicting the phonon density of states (DOS), as shown in the lower part of Fig.~\ref{self_supervise}c. A diverse set of other regression tasks including predictions of Fermi energy, bulk moduli, and shear moduli are further discussed in Supplementary Materials. It is worth mentioning that, for all the demonstrated downstream tasks, the material structures--specifically, the atomic species and positions--are the only inputs. We intentionally exclude any additional atomic properties, such as covalent radius and electronegativity, to explore the feasibility of predicting physical properties based solely on first principles. The remarkable prediction performances on most examined tasks indicate the promising potential of our approach.\\

\begin{figure*}[htb!]
\includegraphics[width=0.8\linewidth]{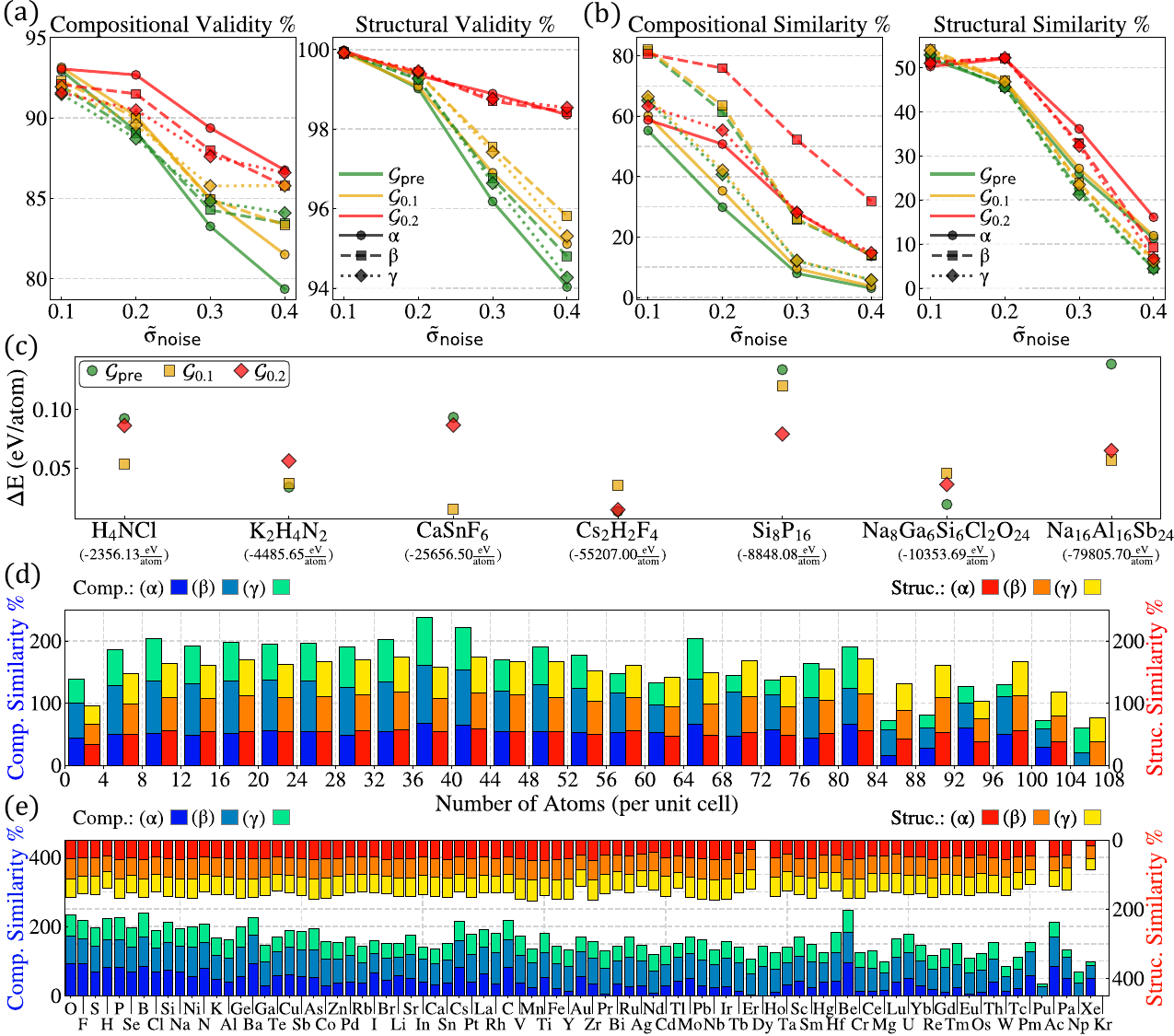}
\caption{\label{metrics}
\textbf{Performance evaluation.} 
This figure presents the performance of generative models, covering: (a) validity, (b) similarity and (c) DFT calculations. 
For panel (c), crystal structures, randomly selected from the test set and perturbed by positional noise with $\tilde{\sigma}_{\text{noise}}$, are analyzed for energy difference ($\Delta E$) compared to originals. The total energy per atom of originals are shown in brackets. 
Panels (d) and (e) evaluate the impacts of the number of atoms per unit cell and the atomic species being masked on compositional similarity (blue-series histograms, left axis) and structural similarity (red-series histograms, right axis), using model $\mathcal{G}_{0.2}$ and testing on samples with $\tilde{\sigma}_{\text{noise}}=0.2$. Note that the right axis in panel (e) is inverted. More detailed analyses are provided in Supplementary Materials.
Positional noise for the test samples follows a Gaussian distribution with a standard deviation $\sigma_{\text{noise}}=\tilde{\sigma}_{\text{noise}}\times \min{(\text{edges})}$. 
Masking strategies include: ($\alpha$) masking all atoms of a specific species within each crystal structure, repeated for each species; ($\beta$) randomly selecting and masking $15\%$ of atoms, repeated five times per structure; and ($\gamma$) a similar approach with $30\%$ of atoms. Evaluation is performed on a test set augmented from $2{,}000$ samples to approximately $6{,}500$ for mask type ($\alpha$) and $10{,}000$ for mask type ($\beta$) and ($\gamma$). 
Model notion: $\mathcal{G}_{\text{pre}}$ represents the pre-trained generative model, trained with samples perturbed at $\tilde{\sigma}_{\text{noise}}=0.1$; $\mathcal{G}_{0.1}$ and $\mathcal{G}_{0.2}$ denote GAN generators trained with data at noise levels of $\tilde{\sigma}_{\text{noise}}=0.1$ and $\tilde{\sigma}_{\text{noise}}=0.2$, respectively.
}
\end{figure*}

\begin{figure}[htb!]
\includegraphics[width=1\linewidth]{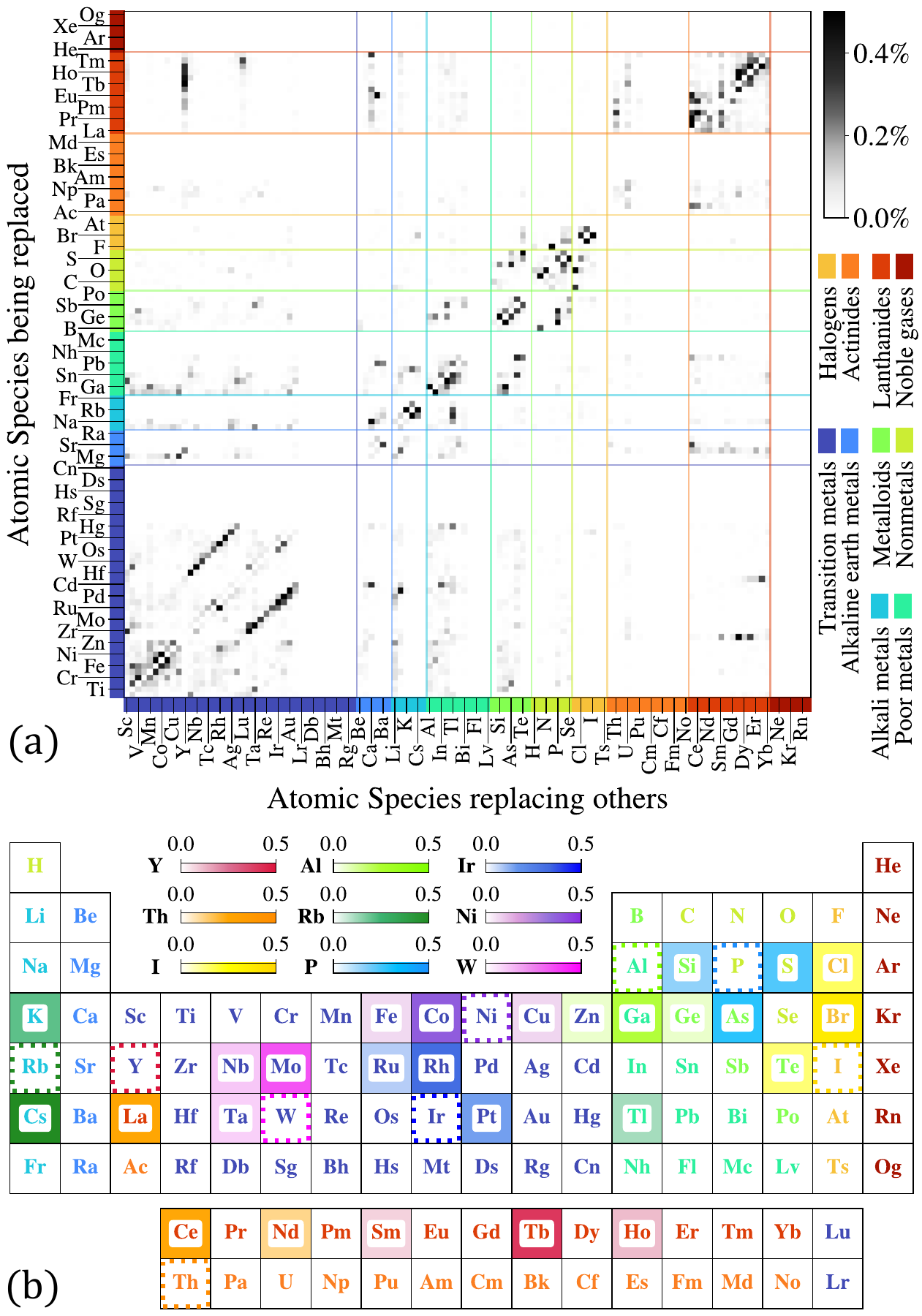}
\caption{\label{novelty}
\textbf{Species replacement correlation.} (a) The correlation between the species being replaced and those replacing them in the valid, novel crystal structures generated by model $\mathcal{G}_{\text{0.2}}$. The color bar represents average correlations across all test data used for similarity calculations in Fig.~\ref{metrics}. Species are organized based on their chemical families.  
Panel (b) visualizes the correlation in (a) through a periodic table: selected species (enclosed with dashed lines) and their top three replacements are displayed, with the replacement ratio indicated by the block color. Different color maps are assigned to various species being replaced. The color of each species symbol reflects their chemical families, as identified in panel (a). 
Additional details are available in the Supplementary Materials.
}
\end{figure}

\inlinesubsection{GAN for more reliable structure generation.}\\ 
Through self-supervised learning on a dataset consisting of over $20{,}000$ diverse, stable crystal structures from the Materials Project \cite{jain2013commentary}, the pre-trained model displays proficiency in accurately predicting atom types and positions in the majority of cases, with a few examples displayed in Fig.~\ref{generators_results} and additional examples available in the Supplementary Materials. 
However, the model with the architecture presented in Fig.~\ref{self_supervise}b encounters limitations. For example, the model faces challenges in accurately reconstructing species that occur infrequently within our dataset. It tends to replace the masked species with alternatives possessing similar chemical properties, as depicted in Fig.~\ref{generators_results}(b)), thereby leading to a lack of compositional validity in some outputs.

To mitigate the aforementioned challenges in generating physically valid structures, it becomes essential to incorporate a mechanism for evaluating the generated structures. Utilizing human pre-knowledge, such as DFT, to train discriminator models is a common practice and has lead to successful outcomes in some recent studies~\cite{merchant2023scaling, szymanski2023autonomous}. However, it can become exceedingly demanding in terms of computational resources, as DFT calculations for each crystal structure can take several minutes to even hours, depending on the number and species of atoms and their types in the unit cell. Considering the need to evaluate as many generated crystals as training samples in each training epoch, a DFT-based evaluation mechanism is thus impractical or at least inefficient, highlighting the pressing need for developing more computationally efficient evaluation methods.

Given these constraints, generative adversarial networks (GANs) emerge as a more practical solution. Following the GAN framework illustrated in Fig.~\ref{gan_illustration}, we combine our pre-trained generative model (Fig.~\ref{self_supervise}b) with a discriminative model that has been initialized using fine-tuned parameters for stability predictions (Fig.~\ref{self_supervise}c), motivated by the similarities between stability classification and the discriminator's role. 
Specifically, the generator is tasked with reconstructing material structures from incomplete and perturbed input structures. The output structures are further passed into the discriminative model along with the pristine structures. The goal of the generator is to produce seemingly realistic materials that can pass the test by the discriminator, while the discriminator tries to tell the difference of generated (fake) structures from the real ones. 

To provide an intuitive understanding of our models' performance, we provide representative examples of the original material structures, contaminated inputs, and the corresponding outputs generated by different models in Fig.~\ref{generators_results}. The generative model, trained under the GAN framework, demonstrates visibly improved performance. It is particularly evident when atoms of a less common species are masked: the GAN generator outperforms the pre-trained model in reasonably repositioning atoms and providing more reliable predictions for the masked species. While evaluating the enhancement offered by GANs over pre-trained model purely from a visual perspective is not persuasive, we quantitatively assess their differences based on three metrics: validity, similarity and novelty.\\

\inlinesubsection{Validity.} We conduct an analysis of the validity of generated crystal structures using two fundamental criteria commonly employed in the field: structural validity and compositional validity~\cite{xie2021crystal}. \textit{Structural validity} assesses whether the minimum Euclidean distance between atoms is greater than a threshold of $0.5$ \AA, ensuring appropriate atomic spacing. \textit{Compositional validity} is determined by verifying if the chemical composition maintains a neutral charge, which is calculated by SMACT~\cite{davies2019smact}. 

As shown in Fig.~\ref{metrics}a, when the noise level in the testing inputs matches that of the training inputs, the GAN generator, though not outperforming the pre-trained model in constructing structurally valid crystals, demonstrates enhanced robustness in generating compositionally valid structures. Despite all generative models being trained on inputs with only $15\%$ of atoms masked, the GAN approach displays remarkable extrapolation capability, especially in scenarios where $30\%$ atoms are randomly masked (mask type $\gamma$).  

Interestingly, the GAN generator trained with a higher noise level ($\mathcal{G}_{0.2}$) performs similarly to other models on the test set with a lower noise level ($\tilde{\sigma}_{\text{noise}}=0.1$), however, when test samples include higher noise levels, GAN generators outperform the pre-trained model ($\mathcal{G}_{\text{pre}}$) in creating compositionally and structurally valid crystals, specially the structural validity score of $\mathcal{G}_{0.2}$ is distinctly higher than that of $\mathcal{G}_{\text{pre}}$. 
$\mathcal{G}_{0.2}$ uses a training set with $\tilde{\sigma}_{\text{noise}}=0.2$ during GAN training, but since it utilizes the model parameters of the pre-trained model $\mathcal{G}_{\text{pre}}$, it has been effectively exposed to training set with different noise levels. This observation suggests that diversifying the training set by including variations in $\tilde{\sigma}_{\text{noise}}$ may enhance the model's versatility. \\

\inlinesubsection{Similarity.} We introduce two metrics designed to evaluate the effectiveness of generative models in accurately reconstructing the crystal structures of ground truth materials, namely compositional similarity and structural similarity. \textit{Compositional similarity} quantifies the probability of generative models in reproducing the same compositions as the originals (see Eqn.~\ref{eq:struc_similarity}). \textit{Structural similarity} reflects their accuracy in replicating the atomic positions in alignment with their original counterparts, scaled by the noise introduced to the inputs (as defined by Eqn.~\ref{eq:comp_similarity}). 

Figure~\ref{metrics}b shows that the distribution of compositional and structural similarities approximates a Gaussian distribution, with the mean values locating at the standard deviation of the noise distribution introduced to the training samples, $\sigma_{\text{noise}}=\tilde{\sigma}_{\text{noise}}\times \min{(\text{edges})}$. As the noise level of the test set increases, the models quickly become incapable of reverting the compositions and structures back to their originals. This inability could be attributed to the generated structures converging to other local optima, distinct from the original structures. 

Figure~\ref{metrics}d-e visualizes the impact of the crystal size and the species to be unmasked on models' performance. It demonstrates that, despite the non-uniform distribution of the size of crystals (illustrated in Supplementary Materials) and species abundance within the datasets (as shown in Fig.~\ref{self_supervise}a), the structural similarity scores for all models remain largely unaffected by variations in the number of atoms and the types of species masked. This observation suggests that the generative models exhibit significant robustness in reconstructing crystal structures, regardless of the difference in size and composition encountered during training. \\

\inlinesubsection{DFT-based stability check.} Another aspect of the performance of our model is the stability of the generated structures. Due to the relatively high computational cost and complexity of DFT simulations (especially for complex systems, \textit{e.g.}, large unit cells containing hundreds of atoms \cite{ratcliff2017challenges, blum2022fhi, gao2019molecular, park2023thickness, song2023structure, pederson2022large} and systems containing polarized electrons \cite{cramer2009density, xue2023extended, wang2023exploration}), we randomly select a small subset of systems studied in this work (with various unit cell sizes) to conceptually show the relative stability of structures generated by different models. 

Here we use the DFT total energy to feature the relative stability. A lower total energy value means a higher structural stability for systems with the same chemical composition. To eliminate the influence of unit cell size, we further divide the total energy values by the number of atoms in the unit cell. Figure~\ref{metrics}c shows that structures produced by GAN generators usually exhibit a non-negligible lower total energy (per atom) $\Delta E$, highlighting the effectiveness of GANs in constructing more stable structures in comparison to those generated by the pre-trained model. Additionally, the generated structure with the lowest total energy for each composition achieves the highest structural similarity score (refer to Supplementary Materials). This observation suggests that structural similarity may act as a proxy for the structural stability suggested by DFT calculations. \\

\inlinesubsection{Novelty.} To gain deeper insights into the information encoded by generative models, as shown in Fig.~\ref{novelty}, we plot the correlation between the species being replaced and those replacing them in the generated crystal structures. In this analysis, we only consider structures that differ in composition from their original counterparts and are compositionally and structurally valid.
Our observations suggest that the model may automatically capture fundamental chemical properties from the self-supervised learning process without the needs of explicit labels, such as electron configurations, ionic radii, oxidation states, and the position of the element in the periodic table. For example, elements within the 4$d$ series are preferred to be interchangeable with those in the 5$d$ series within the same group, due to their similar electronic structures. Yttrium, for instance, is observed to be replaceable by Terbium and Holmium, as they share the same number of outer electrons, comparable atomic radii and oxidation states. Moreover, a common trend is noted where many elements exhibit a preference for replacement with others in close proximity on the periodic table (see Fig.~\ref{novelty}b). Such high level information provided by our pre-trained model can also be further utilized to help find novel stable compounds by narrowing down most promising candidate elements.

We further compare our method with simple elemental substitution method~\cite{glawe2016optimal}, which computes the likelihood of substituting species $A$ with species $B$, denoted as $P(A,B)$, through data mining. This method is particularly useful when the probability of one species being replaced by another remains constant, irrespective of different atomic and positional information within various crystals---although such scenario rarely applies. On the other hand, our generative model is essentially a more sophisticated probabilistic model, since it outputs the probability of specific atomic species at each masked site. Therefore, our model can offer conditional probabilities $P(A,B|\text{condition})$ under diverse conditions, such as varying structures and compositions. We also find that the averaged probabilities over conditions given by our generative models bears a resemblance to those produced by simple substitution (see Supplementary Materials). 

\section{Discussion}

In this study, we propose an innovative machine learning framework that employs equivariant graph neural networks and utilizes the training strategies of reconstructing corrupted inputs (Fig.~\ref{self_supervise}b) and adversarial learning (Fig.~\ref{gan_illustration}) to study the mechanisms underlying crystal structures formation. Our approach is designed to pursue three primary applications: (1) generating or reconstructing crystalline material structures, (2) predicting material properties, and (3) conducting behavioral and structural evaluations for the generative models trained exclusively on crystal structures.

Our model demonstrates a high capability in generating or reconstructing crystal structures: in case where, given any partial information about the original structure, the original remains the most optimal (stable) among all possibilities, our model achieves up to $80\%$ accuracy in reconstructing the optimal composition and possesses a $60\%$ efficiency in denoising the corrupted input towards the optimal structure (as depicted Fig.~\ref{metrics}). We also expect that our generative model has the potential to discover novel materials, a hypothesis that requires further empirical validation. Moreover, our model can also be effectively generalized to various crystal generation scenarios. For example, in situations where the input is a crystal with fixed compositions but a randomized structure, adjusting the training inputs accordingly can facilitate this objective.

Regarding the material property prediction, this model shows its adaptability to a broad range of physical property prediction problems that only take crystal structure as input (as illustrated in Fig.~\ref{self_supervise}c and further elaborated in Supplementary Materials). Our goal is to showcase that data-driven machine learning models are capable of not only accomplishing tasks defined by their training loss functions but also extending their applicability to related downstream tasks with minimal cost. This versatility is rooted in the universal and foundational contents learned by the models, rather than specific, narrowly defined tasks. Analogous to the process of learning language structure before composing poetry, our approach focuses on training the model to learn crystal structures rather than directly instructing it to generate band structures.

In model evaluation, we assess the performance of the generative models on various test sets by examining aspects of validity, similarity, and novelty. We provide a detailed analysis of the model's capability across different input samples, focusing particularly on the influence of the number of atoms and the types of atomic species involved, with more features (such as symmetry groups) awaiting further exploration. One noteworthy observation is that the structural similarity, as defined in Eqn.~\ref{eq:struc_similarity}, remains consistent across structures of varying sizes and compositions (see Fig.~\ref{metrics}). 
Additionally, we endeavor to examine whether crystal structures inherently contain all the information necessary for a comprehensive understanding of their formation, so our approach intentionally minimizes the reliance on pre-established human knowledge and assumptions. In attempting to analyze the generative model's underlying logic in selecting replacement elements, we find that the model may be capable of finding chemically similar species and even reinventing the periodic table of elements (refer to Fig.~\ref{novelty}). Beyond behavioral evaluations, conducting structural evaluations, such as feature attribution, is a promising avenue for future research.

Although our work marks a step forward, it encounters limitations: 
(1) Achieving ``large crystal models'' capable of accurately predicting crystal structures requires a more comprehensive dataset and larger model parameters.
(2) Currently, models have not been trained on dataset that include perturbed lattice vectors, but this limitation can be readily addressed by including a feed-forward network for lattice vector prediction into the generator. 
(3) Our generative model is presently confined to cases where the basic design of the desired crystal structure should be predefined. In other words, unlike other diffusion-based studies that aim for optimal structures with given compositions, our current work practically focuses on finding the optimal composition based on a basic design of the structure. However, our training framework is adaptable for structural optimization by introducing higher noise levels. Employing a diffusion process as the generator within the GAN framework improves the outcomes in the absence of specific structural designs. 
(4) Analogous to enhancements of language models through RLHF, for a thorough examination of the generated structures and a further refinement for practical use, incorporating human insights becomes an inevitable next step. 
(5) To achieve the generation of materials with constrained properties, our model needs to be further fine-tuned using property-specific datasets, though this is primarily a consideration under resource limitations. We expect the ultimate objective as conditioning the generation process by text representations and facilitating training through zero-shot learning.

Despite these challenges, our work opens the door for significant future advancements. Drawing inspiration from natural language processing, we implement an efficient self-supervised training method, offering a comprehensive framework for crystal generation and property prediction. To overcome the lack of an accessible supervisor, we demonstrate that incorporating a GAN framework enables the generator to produce more reliable outcomes. This approach, free from prior knowledge, further permits the investigation of information intrinsically encoded within crystal structures. There, our study serves as an invitation for deeper exploration into understanding materials from first principles using generative machine learning techniques.

\section{Methods} 
\label{sec:methods}

\inlinesubsection{Model setup.} We represent the atomic numbers of species of $N$ atoms in a unit cell of real crystal structures as $Z \in \mathbb{R}^{N\times C}$, where $Z$ is a set of one-hot vectors with length $C=118$ for $N$ atoms in the unit cell. The perturbations applied to the atomic positions in $D=3$ dimensional space are denoted as $P \in \mathbb{R}^{N\times D}$. 
Consequently, our model's output regarding atomic species is described as $\tilde{\mathcal{Z}}\in \mathbb{R}^{N\times C}$, the type-0 output of the equivariant graph neural network with $C$ output channels, and the output for positional perturbations is $\tilde{P} \in \mathbb{R}^{N\times D}$, the type-1 output of the equivariant graph neural network.

For pre-training, we train EquiformerV2~\cite{liao2023equiformerv2} with $12$ Transformer blocks, $8$ attention heads, $118$ output channels, a maximum degree of $6$ and a maximum order of $2$. We use the Adam optimizer with a learning rate of $1\times 10^{-4}$. Batch sizes increases throughout the training process (see details in Supplementary Materials).\\

\inlinesubsection{Loss functions.} 
For the pre-train model, we choose negative log-likelihood loss (NLLLoss) applied after LogSoftmax as the loss function for atomic species, and mean squared error (MSE) for atomic positions:
\begin{equation} \label{eq:loss_pretrain}
\begin{aligned}
    &\mathcal{L}_{\text{pretrain}} = \mathcal{L}_{\text{species}} + w_{\text{position}} \times \mathcal{L}_{\text{position}}, \\
    &\mathcal{L}_{\text{species}} = -\sum_{i=1}^{N} \sum_{j=1}^{C} Z_{ij}  \text{LogSoftmax}\big( \tilde{\mathcal{Z}}_{i} \big)_{j}, \\
    &\mathcal{L}_{\text{position}} = \frac{1}{N} \sum_{i=1}^{N} \sum_{j=1}^{D} (P_{ij} - \tilde{P}_{ij})^2. \\
\end{aligned}
\end{equation}

For fine-tuning regression tasks, we use MSE as the loss function. For fine-tuning classification tasks, the loss function is NLLLoss applied after LogSoftmax. 

For GAN, having found that the discriminator tends to outperform the generator rapidly while using cross entropy as the generator loss, and leads to gradient vanishing in our tasks, we use Wasserstein distance $\mathcal{L}_{\text{W}} = \frac{1}{M}\sum_{m=1}^{M}\sum_{k=1}^{K} \left|y_{k}^{(m)} - \tilde{y}_{k}^{(m)}\right|$ as the training loss. Here, $M$ denotes the number of samples in each batch and $K=2$ for binary classification. 
The Wasserstein loss for the generator $\hat{G}: \mathbb{R}^{N\times (C+D)} \rightarrow \mathbb{R}^{N\times (C+D)}$ is formulated as
\begin{equation}
\begin{aligned}
\mathcal{L}^G_{\text{W}} = \frac{1}{M}\sum_{m=1}^{M}\sum_{k=1}^{2} \left|y_{k}^{(m)}- \text{LogSoftmax}\left(\hat{D}\left(\hat{G}(X^{(m)})\right)\right)_{k}\right|,
\end{aligned}
\end{equation}
with $y^{(m)} = (0,1)$ indicating the label vector for each sample $m$. The input of the generate $X^{(m)}$ consists of masked atomic species $\tilde{\mathcal{Z}}^{(m)} \in \mathbb{R}^{N\times C}$ and perturbed positions $\tilde{P}^{(m)} \in \mathbb{R}^{N\times D}$ augmented based on the $m$-th crystal structure. 

On the other hand, for the Wasserstein loss in the discriminator $\hat{D}: \mathbb{R}^{N\times (C+D)} \rightarrow \mathbb{R}^{2}$, $y^{(m)}$ is set to $(1,0)$ for real structure inputs, \textit{i.e.}, $\tilde{y}^{(m)}=\text{LogSoftmax}\left(\hat{D}\left(X^{(m)}\right)\right)$; $y^{(m)}=(0,1)$ for inputs from the generated structures, \textit{i.e.}, $\tilde{y}^{(m)}=\text{LogSoftmax}\left(\hat{D}\left(\hat{G}(X^{(m)})\right)\right)$. When training the generator, we keep the parameters of the discriminator frozen, and vice versa.

To prevent the generator from producing features specifically tailored to the weakness of the discriminator, rather than learning to produce the realistic crystal structures, we incorporate $\mathcal{L}_{\text{species}}$ and $\mathcal{L}_{\text{position}}$, as specified Eqn.~\ref{eq:loss_pretrain}, into the generator loss:
\begin{equation}
\begin{aligned}
\mathcal{L}^G = w_{\text{W}} \times \mathcal{L}^G_{\text{W}} + \mathcal{L}_{\text{species}} + w_{\text{position}} \times \mathcal{L}_{\text{position}}.
\end{aligned}
\end{equation}
Note that we use $w_{\text{position}}=200$ as for pre-training and GAN. The training of GAN is done with $w_{\text{W}}=0.1$, though both $w_{\text{W}}=1$ and $0.1$ have been tested with no significant difference observed.

Modifications to the training procedure are also made to avoid this issue, such as reducing the learning rate of the discriminator (lr$_{G}=10^{-4}$ and lr$_{D}=10^{-5}$), introducing infinitesimal positional noise to the real crystal structures (the noise follows a normal distribution with standard deviation of $\sigma = 0.001\times\min(\text{edges})$), and adjusting the training schedule to update the generator ten times more frequent than the discriminator. \\

\inlinesubsection{Related metrics.} 
Compositional similarity is defined as the average percentage match between the generated and original material compositions across the test set:
\begin{equation}\label{eq:comp_similarity}
    \text{Comp. Similarity} = \frac{1}{|\text{test set}|} \sum_{m=1}^{|\text{test set}|} \delta(Z^{(m)}-\tilde{Z}^{(m)}),
\end{equation}
Here $\tilde{Z}^{(m)}_{ij}=\delta \left(j-\arg \max_k \left(\text{LogSoftmax} \left(\tilde{\mathcal{Z}}^{(m)}_{i}\right)_k \right) \right)$, where $\tilde{\mathcal{Z}}^{(m)}\in \mathbb{R}^{N\times C}$ denotes the compositional output from the generative model, as shown in Eqn.~\ref{eq:loss_pretrain}. This metric provides insight into the model's ability to accurately restore the composition of materials. However, it is important to note that (1) while high compositional similarity indicates model accuracy, it is contradictory to assessing its capability to innovate novel structures; (2) for computational simplicity, a generated structure is considered compositionally identical to the original structure only if $Z_{i}=\tilde{Z}_{i}$ for all $i$. This definition, however, overlooks the scenario where two atoms of different species ($Z_{i}\neq Z_{j}$) may have their locations exchanged due to the large positional noise. Such an exchange would lead the generative models to swap their species and positions, resulting in $\tilde{Z}_{i} = Z_{j}$, $\tilde{Z}_{j} = Z_{i}$, $\tilde{P}_{i} = P_{j}$, and $\tilde{P}_{j} = P_{i}$. In this case, the generated structure should still be considered equivalent to the original counterpart. With low noise levels, the probability of this occurrence is sufficiently low. However, as our analysis in Fig.~\ref{metrics}b includes high noise levels, this could lead to inaccuracies in assessing compositional similarity.

Structural similarity, on the other hand, aims to evaluate the model's ability in replicating the structural integrity when the model decides to generate compositions identical to the originals. It is defined as:
\begin{widetext}
\begin{equation}\label{eq:struc_similarity}
    \text{Struc. Similarity} = 
    1- \frac{1}{|\text{test set}|} \sum_{m=1}^{|\text{test set}|} \delta(Z^{(m)}-\tilde{Z}^{(m)}) \left( \frac{1}{N_m}\sum_{i=1}^{N_m} \sqrt{\sum_{j=1}^{D} \left(P^{(m)}_{ij}-\tilde{P}^{(m)}_{ij} \right)^2} \right)/S^{(m)}.
\end{equation}
\end{widetext}

Here, the equation considers the Euclidean distance between the positions in the original structure ($P^{(m)}\in \mathbb{R}^{N\times D}$) and generated structure ($\tilde{P}^{(m)}\in \mathbb{R}^{N\times D}$) for the $m$-th test sample, normalized by the average positional deviation from the equilibrium introduced to the input ($S^{(m)}\in \mathbb{R}$). A higher structural similarity score, approaching 1, indicates a closer match to the original material's structural positioning, offering a direct measure of the model performance in restoring material structure. \\

For DFT calculations, we employ the all-electron electronic structure package FHI-AIMS \cite{blum2009ab} to conduct single-point DFT calculations. We specifically exclude systems containing partially-filled \textit{d}- or \textit{f}-orbitals in this part, so the combination of semilocal Perdew-Burke-Ernzerhof (PBE) exchange-correlation (XC) functional \cite{perdew1996generalized} and the ``atomic ZORA'' scalar relativity \cite{blum2009ab, huhn2017one, zhao2021quasi} is considered to be suitable for total energy prediction. Moreover, ``intermediate" numerical settings of the numeric atom-centered-orbital basis are used \cite{havu2009efficient}, and the \textit{k}-grids for different calculations are adjusted according to different unit cell sizes to ensure adequate density of the sampled points in the reciprocal space. \\

All datasets used in this work are sourced from the Materials Project. More details about the model setups and data are available in the Supplementary Materials. Our code will be available at \url{https://github.com/fangzel/CrystalGenerativeModels}.

\section{Acknowledgement}

We thank the insightful discussions with Cheng Peng, Wei Shao, Jay Qu, and Mingkyung Han. This work is supported by the U.S.\@ Department of Energy, Office of Science, Basic Energy Sciences under Award No.\@ DE-SC0022216. Y.L. acknowledges the support by the U.S. Department of Energy, Laboratory Directed Research and Development program at SLAC National Accelerator Laboratory, under contract No. DE-AC02-76SF00515. This research used computational resources of the National Energy Research Scientific Computing Center (NERSC), a U.S.\@ Department of Energy Office of Science User Facility located at Lawrence Berkeley National Laboratory, operated under Contract No.\@ DE-AC02-05CH11231.

\section{Declaration of generative AI and AI-assisted technologies in the writing process}

During the preparation of this work the authors used the large language model ChatGPT by OpenAI to refine the language and enhance the readability of this paper. All content was independently written by the authors before utilizing this tool, and the authors thoroughly reviewed and edited the content after its use, taking full responsibility for the content of the publication.

\section{Author Contributions}

F.L. and Z.C. contributed equally to this research. F.L. conceived and designed the research with input from Z.C.. Z.C. built the pre-train model and collected dataset. F.L. refined, expanded, and trained all models, performed experiments, and analyzed and visualized the results, with support from Z.C. and T.L.. R.S. performed DFT calculations. Z.C., F.L. and J.J.T. wrote the manuscript with feedback and contributions from all authors. C.J.J., J.J.T. and Y.L. supervised this research.

\bibliography{ref}

\begin{thebibliography}{48}%
\makeatletter
\providecommand \@ifxundefined [1]{%
 \@ifx{#1\undefined}
}%
\providecommand \@ifnum [1]{%
 \ifnum #1\expandafter \@firstoftwo
 \else \expandafter \@secondoftwo
 \fi
}%
\providecommand \@ifx [1]{%
 \ifx #1\expandafter \@firstoftwo
 \else \expandafter \@secondoftwo
 \fi
}%
\providecommand \natexlab [1]{#1}%
\providecommand \enquote  [1]{``#1''}%
\providecommand \bibnamefont  [1]{#1}%
\providecommand \bibfnamefont [1]{#1}%
\providecommand \citenamefont [1]{#1}%
\providecommand \href@noop [0]{\@secondoftwo}%
\providecommand \href [0]{\begingroup \@sanitize@url \@href}%
\providecommand \@href[1]{\@@startlink{#1}\@@href}%
\providecommand \@@href[1]{\endgroup#1\@@endlink}%
\providecommand \@sanitize@url [0]{\catcode `\\12\catcode `\$12\catcode `\&12\catcode `\#12\catcode `\^12\catcode `\_12\catcode `\%12\relax}%
\providecommand \@@startlink[1]{}%
\providecommand \@@endlink[0]{}%
\providecommand \url  [0]{\begingroup\@sanitize@url \@url }%
\providecommand \@url [1]{\endgroup\@href {#1}{\urlprefix }}%
\providecommand \urlprefix  [0]{URL }%
\providecommand \Eprint [0]{\href }%
\providecommand \doibase [0]{https://doi.org/}%
\providecommand \selectlanguage [0]{\@gobble}%
\providecommand \bibinfo  [0]{\@secondoftwo}%
\providecommand \bibfield  [0]{\@secondoftwo}%
\providecommand \translation [1]{[#1]}%
\providecommand \BibitemOpen [0]{}%
\providecommand \bibitemStop [0]{}%
\providecommand \bibitemNoStop [0]{.\EOS\space}%
\providecommand \EOS [0]{\spacefactor3000\relax}%
\providecommand \BibitemShut  [1]{\csname bibitem#1\endcsname}%
\let\auto@bib@innerbib\@empty
\bibitem [{\citenamefont {Xie}\ and\ \citenamefont {Grossman}(2018)}]{xie2018crystal}%
  \BibitemOpen
  \bibfield  {author} {\bibinfo {author} {\bibfnamefont {T.}~\bibnamefont {Xie}}\ and\ \bibinfo {author} {\bibfnamefont {J.~C.}\ \bibnamefont {Grossman}},\ }\bibfield  {title} {\bibinfo {title} {Crystal graph convolutional neural networks for an accurate and interpretable prediction of material properties},\ }\href@noop {} {\bibfield  {journal} {\bibinfo  {journal} {Physical review letters}\ }\textbf {\bibinfo {volume} {120}},\ \bibinfo {pages} {145301} (\bibinfo {year} {2018})}\BibitemShut {NoStop}%
\bibitem [{\citenamefont {Schmidt}\ \emph {et~al.}(2019)\citenamefont {Schmidt}, \citenamefont {Marques}, \citenamefont {Botti},\ and\ \citenamefont {Marques}}]{schmidt2019recent}%
  \BibitemOpen
  \bibfield  {author} {\bibinfo {author} {\bibfnamefont {J.}~\bibnamefont {Schmidt}}, \bibinfo {author} {\bibfnamefont {M.~R.}\ \bibnamefont {Marques}}, \bibinfo {author} {\bibfnamefont {S.}~\bibnamefont {Botti}},\ and\ \bibinfo {author} {\bibfnamefont {M.~A.}\ \bibnamefont {Marques}},\ }\bibfield  {title} {\bibinfo {title} {Recent advances and applications of machine learning in solid-state materials science},\ }\href@noop {} {\bibfield  {journal} {\bibinfo  {journal} {npj Computational Materials}\ }\textbf {\bibinfo {volume} {5}},\ \bibinfo {pages} {83} (\bibinfo {year} {2019})}\BibitemShut {NoStop}%
\bibitem [{\citenamefont {Chen}\ \emph {et~al.}(2019)\citenamefont {Chen}, \citenamefont {Ye}, \citenamefont {Zuo}, \citenamefont {Zheng},\ and\ \citenamefont {Ong}}]{chen2019graph}%
  \BibitemOpen
  \bibfield  {author} {\bibinfo {author} {\bibfnamefont {C.}~\bibnamefont {Chen}}, \bibinfo {author} {\bibfnamefont {W.}~\bibnamefont {Ye}}, \bibinfo {author} {\bibfnamefont {Y.}~\bibnamefont {Zuo}}, \bibinfo {author} {\bibfnamefont {C.}~\bibnamefont {Zheng}},\ and\ \bibinfo {author} {\bibfnamefont {S.~P.}\ \bibnamefont {Ong}},\ }\bibfield  {title} {\bibinfo {title} {Graph networks as a universal machine learning framework for molecules and crystals},\ }\href@noop {} {\bibfield  {journal} {\bibinfo  {journal} {Chemistry of Materials}\ }\textbf {\bibinfo {volume} {31}},\ \bibinfo {pages} {3564} (\bibinfo {year} {2019})}\BibitemShut {NoStop}%
\bibitem [{\citenamefont {Tawfik}\ \emph {et~al.}(2020)\citenamefont {Tawfik}, \citenamefont {Isayev}, \citenamefont {Spencer},\ and\ \citenamefont {Winkler}}]{tawfik2020predicting}%
  \BibitemOpen
  \bibfield  {author} {\bibinfo {author} {\bibfnamefont {S.~A.}\ \bibnamefont {Tawfik}}, \bibinfo {author} {\bibfnamefont {O.}~\bibnamefont {Isayev}}, \bibinfo {author} {\bibfnamefont {M.~J.}\ \bibnamefont {Spencer}},\ and\ \bibinfo {author} {\bibfnamefont {D.~A.}\ \bibnamefont {Winkler}},\ }\bibfield  {title} {\bibinfo {title} {Predicting thermal properties of crystals using machine learning},\ }\href@noop {} {\bibfield  {journal} {\bibinfo  {journal} {Advanced Theory and Simulations}\ }\textbf {\bibinfo {volume} {3}},\ \bibinfo {pages} {1900208} (\bibinfo {year} {2020})}\BibitemShut {NoStop}%
\bibitem [{\citenamefont {Andrejevic}\ \emph {et~al.}(2022)\citenamefont {Andrejevic}, \citenamefont {Andrejevic}, \citenamefont {Bernevig}, \citenamefont {Regnault}, \citenamefont {Han}, \citenamefont {Fabbris}, \citenamefont {Nguyen}, \citenamefont {Drucker}, \citenamefont {Rycroft},\ and\ \citenamefont {Li}}]{andrejevic2022machine}%
  \BibitemOpen
  \bibfield  {author} {\bibinfo {author} {\bibfnamefont {N.}~\bibnamefont {Andrejevic}}, \bibinfo {author} {\bibfnamefont {J.}~\bibnamefont {Andrejevic}}, \bibinfo {author} {\bibfnamefont {B.~A.}\ \bibnamefont {Bernevig}}, \bibinfo {author} {\bibfnamefont {N.}~\bibnamefont {Regnault}}, \bibinfo {author} {\bibfnamefont {F.}~\bibnamefont {Han}}, \bibinfo {author} {\bibfnamefont {G.}~\bibnamefont {Fabbris}}, \bibinfo {author} {\bibfnamefont {T.}~\bibnamefont {Nguyen}}, \bibinfo {author} {\bibfnamefont {N.~C.}\ \bibnamefont {Drucker}}, \bibinfo {author} {\bibfnamefont {C.~H.}\ \bibnamefont {Rycroft}},\ and\ \bibinfo {author} {\bibfnamefont {M.}~\bibnamefont {Li}},\ }\bibfield  {title} {\bibinfo {title} {Machine-learning spectral indicators of topology},\ }\href@noop {} {\bibfield  {journal} {\bibinfo  {journal} {Advanced Materials}\ }\textbf {\bibinfo {volume} {34}},\ \bibinfo {pages} {2204113} (\bibinfo {year} {2022})}\BibitemShut {NoStop}%
\bibitem [{\citenamefont {Kong}\ \emph {et~al.}(2022)\citenamefont {Kong}, \citenamefont {Ricci}, \citenamefont {Guevarra}, \citenamefont {Neaton}, \citenamefont {Gomes},\ and\ \citenamefont {Gregoire}}]{kong2022density}%
  \BibitemOpen
  \bibfield  {author} {\bibinfo {author} {\bibfnamefont {S.}~\bibnamefont {Kong}}, \bibinfo {author} {\bibfnamefont {F.}~\bibnamefont {Ricci}}, \bibinfo {author} {\bibfnamefont {D.}~\bibnamefont {Guevarra}}, \bibinfo {author} {\bibfnamefont {J.~B.}\ \bibnamefont {Neaton}}, \bibinfo {author} {\bibfnamefont {C.~P.}\ \bibnamefont {Gomes}},\ and\ \bibinfo {author} {\bibfnamefont {J.~M.}\ \bibnamefont {Gregoire}},\ }\bibfield  {title} {\bibinfo {title} {Density of states prediction for materials discovery via contrastive learning from probabilistic embeddings},\ }\href@noop {} {\bibfield  {journal} {\bibinfo  {journal} {Nature communications}\ }\textbf {\bibinfo {volume} {13}},\ \bibinfo {pages} {949} (\bibinfo {year} {2022})}\BibitemShut {NoStop}%
\bibitem [{\citenamefont {Choudhary}\ \emph {et~al.}(2022)\citenamefont {Choudhary}, \citenamefont {DeCost}, \citenamefont {Chen}, \citenamefont {Jain}, \citenamefont {Tavazza}, \citenamefont {Cohn}, \citenamefont {Park}, \citenamefont {Choudhary}, \citenamefont {Agrawal}, \citenamefont {Billinge} \emph {et~al.}}]{choudhary2022recent}%
  \BibitemOpen
  \bibfield  {author} {\bibinfo {author} {\bibfnamefont {K.}~\bibnamefont {Choudhary}}, \bibinfo {author} {\bibfnamefont {B.}~\bibnamefont {DeCost}}, \bibinfo {author} {\bibfnamefont {C.}~\bibnamefont {Chen}}, \bibinfo {author} {\bibfnamefont {A.}~\bibnamefont {Jain}}, \bibinfo {author} {\bibfnamefont {F.}~\bibnamefont {Tavazza}}, \bibinfo {author} {\bibfnamefont {R.}~\bibnamefont {Cohn}}, \bibinfo {author} {\bibfnamefont {C.~W.}\ \bibnamefont {Park}}, \bibinfo {author} {\bibfnamefont {A.}~\bibnamefont {Choudhary}}, \bibinfo {author} {\bibfnamefont {A.}~\bibnamefont {Agrawal}}, \bibinfo {author} {\bibfnamefont {S.~J.}\ \bibnamefont {Billinge}}, \emph {et~al.},\ }\bibfield  {title} {\bibinfo {title} {Recent advances and applications of deep learning methods in materials science},\ }\href@noop {} {\bibfield  {journal} {\bibinfo  {journal} {npj Computational Materials}\ }\textbf {\bibinfo {volume} {8}},\ \bibinfo {pages} {59} (\bibinfo {year} {2022})}\BibitemShut {NoStop}%
\bibitem [{\citenamefont {Moosavi}\ \emph {et~al.}(2022)\citenamefont {Moosavi}, \citenamefont {Novotny}, \citenamefont {Ongari}, \citenamefont {Moubarak}, \citenamefont {Asgari}, \citenamefont {Kadioglu}, \citenamefont {Charalambous}, \citenamefont {Ortega-Guerrero}, \citenamefont {Farmahini}, \citenamefont {Sarkisov} \emph {et~al.}}]{moosavi2022data}%
  \BibitemOpen
  \bibfield  {author} {\bibinfo {author} {\bibfnamefont {S.~M.}\ \bibnamefont {Moosavi}}, \bibinfo {author} {\bibfnamefont {B.~{\'A}.}\ \bibnamefont {Novotny}}, \bibinfo {author} {\bibfnamefont {D.}~\bibnamefont {Ongari}}, \bibinfo {author} {\bibfnamefont {E.}~\bibnamefont {Moubarak}}, \bibinfo {author} {\bibfnamefont {M.}~\bibnamefont {Asgari}}, \bibinfo {author} {\bibfnamefont {{\"O}.}~\bibnamefont {Kadioglu}}, \bibinfo {author} {\bibfnamefont {C.}~\bibnamefont {Charalambous}}, \bibinfo {author} {\bibfnamefont {A.}~\bibnamefont {Ortega-Guerrero}}, \bibinfo {author} {\bibfnamefont {A.~H.}\ \bibnamefont {Farmahini}}, \bibinfo {author} {\bibfnamefont {L.}~\bibnamefont {Sarkisov}}, \emph {et~al.},\ }\bibfield  {title} {\bibinfo {title} {A data-science approach to predict the heat capacity of nanoporous materials},\ }\href@noop {} {\bibfield  {journal} {\bibinfo  {journal} {Nature materials}\ }\textbf {\bibinfo {volume} {21}},\ \bibinfo {pages} {1419} (\bibinfo {year} {2022})}\BibitemShut {NoStop}%
\bibitem [{\citenamefont {Liu}\ \emph {et~al.}(2023)\citenamefont {Liu}, \citenamefont {Tan}, \citenamefont {Liang}, \citenamefont {Han}, \citenamefont {Xiang},\ and\ \citenamefont {Yan}}]{liu2023machine}%
  \BibitemOpen
  \bibfield  {author} {\bibinfo {author} {\bibfnamefont {Y.}~\bibnamefont {Liu}}, \bibinfo {author} {\bibfnamefont {X.}~\bibnamefont {Tan}}, \bibinfo {author} {\bibfnamefont {J.}~\bibnamefont {Liang}}, \bibinfo {author} {\bibfnamefont {H.}~\bibnamefont {Han}}, \bibinfo {author} {\bibfnamefont {P.}~\bibnamefont {Xiang}},\ and\ \bibinfo {author} {\bibfnamefont {W.}~\bibnamefont {Yan}},\ }\bibfield  {title} {\bibinfo {title} {Machine learning for perovskite solar cells and component materials: key technologies and prospects},\ }\href@noop {} {\bibfield  {journal} {\bibinfo  {journal} {Advanced Functional Materials}\ ,\ \bibinfo {pages} {2214271}} (\bibinfo {year} {2023})}\BibitemShut {NoStop}%
\bibitem [{\citenamefont {Kim}\ \emph {et~al.}(2020)\citenamefont {Kim}, \citenamefont {Noh}, \citenamefont {Gu}, \citenamefont {Aspuru-Guzik},\ and\ \citenamefont {Jung}}]{kim2020generative}%
  \BibitemOpen
  \bibfield  {author} {\bibinfo {author} {\bibfnamefont {S.}~\bibnamefont {Kim}}, \bibinfo {author} {\bibfnamefont {J.}~\bibnamefont {Noh}}, \bibinfo {author} {\bibfnamefont {G.~H.}\ \bibnamefont {Gu}}, \bibinfo {author} {\bibfnamefont {A.}~\bibnamefont {Aspuru-Guzik}},\ and\ \bibinfo {author} {\bibfnamefont {Y.}~\bibnamefont {Jung}},\ }\bibfield  {title} {\bibinfo {title} {Generative adversarial networks for crystal structure prediction},\ }\href {https://doi.org/10.1021/acscentsci.0c00426} {\bibfield  {journal} {\bibinfo  {journal} {ACS central science}\ }\textbf {\bibinfo {volume} {6}},\ \bibinfo {pages} {1412} (\bibinfo {year} {2020})}\BibitemShut {NoStop}%
\bibitem [{\citenamefont {Long}\ \emph {et~al.}(2021)\citenamefont {Long}, \citenamefont {Fortunato}, \citenamefont {Opahle}, \citenamefont {Zhang}, \citenamefont {Samathrakis}, \citenamefont {Shen}, \citenamefont {Gutfleisch},\ and\ \citenamefont {Zhang}}]{long2021constrained}%
  \BibitemOpen
  \bibfield  {author} {\bibinfo {author} {\bibfnamefont {T.}~\bibnamefont {Long}}, \bibinfo {author} {\bibfnamefont {N.~M.}\ \bibnamefont {Fortunato}}, \bibinfo {author} {\bibfnamefont {I.}~\bibnamefont {Opahle}}, \bibinfo {author} {\bibfnamefont {Y.}~\bibnamefont {Zhang}}, \bibinfo {author} {\bibfnamefont {I.}~\bibnamefont {Samathrakis}}, \bibinfo {author} {\bibfnamefont {C.}~\bibnamefont {Shen}}, \bibinfo {author} {\bibfnamefont {O.}~\bibnamefont {Gutfleisch}},\ and\ \bibinfo {author} {\bibfnamefont {H.}~\bibnamefont {Zhang}},\ }\bibfield  {title} {\bibinfo {title} {Constrained crystals deep convolutional generative adversarial network for the inverse design of crystal structures},\ }\href {https://doi.org/10.1038/s41524-021-00526-4} {\bibfield  {journal} {\bibinfo  {journal} {npj Computational Materials}\ }\textbf {\bibinfo {volume} {7}},\ \bibinfo {pages} {66} (\bibinfo {year} {2021})}\BibitemShut {NoStop}%
\bibitem [{\citenamefont {Zhao}\ \emph {et~al.}(2021{\natexlab{a}})\citenamefont {Zhao}, \citenamefont {Al-Fahdi}, \citenamefont {Hu}, \citenamefont {Siriwardane}, \citenamefont {Song}, \citenamefont {Nasiri},\ and\ \citenamefont {Hu}}]{zhao2021high}%
  \BibitemOpen
  \bibfield  {author} {\bibinfo {author} {\bibfnamefont {Y.}~\bibnamefont {Zhao}}, \bibinfo {author} {\bibfnamefont {M.}~\bibnamefont {Al-Fahdi}}, \bibinfo {author} {\bibfnamefont {M.}~\bibnamefont {Hu}}, \bibinfo {author} {\bibfnamefont {E.~M.}\ \bibnamefont {Siriwardane}}, \bibinfo {author} {\bibfnamefont {Y.}~\bibnamefont {Song}}, \bibinfo {author} {\bibfnamefont {A.}~\bibnamefont {Nasiri}},\ and\ \bibinfo {author} {\bibfnamefont {J.}~\bibnamefont {Hu}},\ }\bibfield  {title} {\bibinfo {title} {High-throughput discovery of novel cubic crystal materials using deep generative neural networks},\ }\href {https://onlinelibrary.wiley.com/doi/abs/10.1002/advs.202100566} {\bibfield  {journal} {\bibinfo  {journal} {Advanced Science}\ }\textbf {\bibinfo {volume} {8}},\ \bibinfo {pages} {2100566} (\bibinfo {year} {2021}{\natexlab{a}})}\BibitemShut {NoStop}%
\bibitem [{\citenamefont {Xie}\ \emph {et~al.}(2021)\citenamefont {Xie}, \citenamefont {Fu}, \citenamefont {Ganea}, \citenamefont {Barzilay},\ and\ \citenamefont {Jaakkola}}]{xie2021crystal}%
  \BibitemOpen
  \bibfield  {author} {\bibinfo {author} {\bibfnamefont {T.}~\bibnamefont {Xie}}, \bibinfo {author} {\bibfnamefont {X.}~\bibnamefont {Fu}}, \bibinfo {author} {\bibfnamefont {O.-E.}\ \bibnamefont {Ganea}}, \bibinfo {author} {\bibfnamefont {R.}~\bibnamefont {Barzilay}},\ and\ \bibinfo {author} {\bibfnamefont {T.}~\bibnamefont {Jaakkola}},\ }\bibfield  {title} {\bibinfo {title} {Crystal diffusion variational autoencoder for periodic material generation},\ }\href {https://doi.org/10.48550/arXiv.2110.06197} {\bibfield  {journal} {\bibinfo  {journal} {arXiv preprint arXiv:2110.06197}\ } (\bibinfo {year} {2021})}\BibitemShut {NoStop}%
\bibitem [{\citenamefont {Lyngby}\ and\ \citenamefont {Thygesen}(2022)}]{lyngby2022data}%
  \BibitemOpen
  \bibfield  {author} {\bibinfo {author} {\bibfnamefont {P.}~\bibnamefont {Lyngby}}\ and\ \bibinfo {author} {\bibfnamefont {K.~S.}\ \bibnamefont {Thygesen}},\ }\bibfield  {title} {\bibinfo {title} {Data-driven discovery of 2d materials by deep generative models},\ }\href@noop {} {\bibfield  {journal} {\bibinfo  {journal} {npj Computational Materials}\ }\textbf {\bibinfo {volume} {8}},\ \bibinfo {pages} {232} (\bibinfo {year} {2022})}\BibitemShut {NoStop}%
\bibitem [{\citenamefont {Yang}\ \emph {et~al.}(2023{\natexlab{a}})\citenamefont {Yang}, \citenamefont {Cho}, \citenamefont {Merchant}, \citenamefont {Abbeel}, \citenamefont {Schuurmans}, \citenamefont {Mordatch},\ and\ \citenamefont {Cubuk}}]{yang2023scalable}%
  \BibitemOpen
  \bibfield  {author} {\bibinfo {author} {\bibfnamefont {M.}~\bibnamefont {Yang}}, \bibinfo {author} {\bibfnamefont {K.}~\bibnamefont {Cho}}, \bibinfo {author} {\bibfnamefont {A.}~\bibnamefont {Merchant}}, \bibinfo {author} {\bibfnamefont {P.}~\bibnamefont {Abbeel}}, \bibinfo {author} {\bibfnamefont {D.}~\bibnamefont {Schuurmans}}, \bibinfo {author} {\bibfnamefont {I.}~\bibnamefont {Mordatch}},\ and\ \bibinfo {author} {\bibfnamefont {E.~D.}\ \bibnamefont {Cubuk}},\ }\bibfield  {title} {\bibinfo {title} {Scalable diffusion for materials generation},\ }\href {https://doi.org/10.48550/arXiv.2311.09235} {\bibfield  {journal} {\bibinfo  {journal} {arXiv preprint arXiv:2311.09235}\ } (\bibinfo {year} {2023}{\natexlab{a}})}\BibitemShut {NoStop}%
\bibitem [{\citenamefont {Zhu}\ \emph {et~al.}(2023)\citenamefont {Zhu}, \citenamefont {Nong}, \citenamefont {Yamazaki},\ and\ \citenamefont {Hippalgaonkar}}]{zhu2023wycryst}%
  \BibitemOpen
  \bibfield  {author} {\bibinfo {author} {\bibfnamefont {R.}~\bibnamefont {Zhu}}, \bibinfo {author} {\bibfnamefont {W.}~\bibnamefont {Nong}}, \bibinfo {author} {\bibfnamefont {S.}~\bibnamefont {Yamazaki}},\ and\ \bibinfo {author} {\bibfnamefont {K.}~\bibnamefont {Hippalgaonkar}},\ }\bibfield  {title} {\bibinfo {title} {Wycryst: Wyckoff inorganic crystal generator framework},\ }\href {https://doi.org/10.48550/arXiv.2311.17916} {\bibfield  {journal} {\bibinfo  {journal} {Available at SSRN 4658842}\ } (\bibinfo {year} {2023})}\BibitemShut {NoStop}%
\bibitem [{\citenamefont {Yang}\ \emph {et~al.}(2023{\natexlab{b}})\citenamefont {Yang}, \citenamefont {Zhang}, \citenamefont {Song}, \citenamefont {Hong}, \citenamefont {Xu}, \citenamefont {Zhao}, \citenamefont {Zhang}, \citenamefont {Cui},\ and\ \citenamefont {Yang}}]{yang2023diffusion}%
  \BibitemOpen
  \bibfield  {author} {\bibinfo {author} {\bibfnamefont {L.}~\bibnamefont {Yang}}, \bibinfo {author} {\bibfnamefont {Z.}~\bibnamefont {Zhang}}, \bibinfo {author} {\bibfnamefont {Y.}~\bibnamefont {Song}}, \bibinfo {author} {\bibfnamefont {S.}~\bibnamefont {Hong}}, \bibinfo {author} {\bibfnamefont {R.}~\bibnamefont {Xu}}, \bibinfo {author} {\bibfnamefont {Y.}~\bibnamefont {Zhao}}, \bibinfo {author} {\bibfnamefont {W.}~\bibnamefont {Zhang}}, \bibinfo {author} {\bibfnamefont {B.}~\bibnamefont {Cui}},\ and\ \bibinfo {author} {\bibfnamefont {M.-H.}\ \bibnamefont {Yang}},\ }\bibfield  {title} {\bibinfo {title} {Diffusion models: A comprehensive survey of methods and applications},\ }\href {https://doi.org/10.1145/3626235} {\bibfield  {journal} {\bibinfo  {journal} {ACM Computing Surveys}\ }\textbf {\bibinfo {volume} {56}},\ \bibinfo {pages} {1} (\bibinfo {year} {2023}{\natexlab{b}})}\BibitemShut {NoStop}%
\bibitem [{\citenamefont {Zeni}\ \emph {et~al.}(2023)\citenamefont {Zeni}, \citenamefont {Pinsler}, \citenamefont {Z{\"u}gner}, \citenamefont {Fowler}, \citenamefont {Horton}, \citenamefont {Fu}, \citenamefont {Shysheya}, \citenamefont {Crabb{\'e}}, \citenamefont {Sun}, \citenamefont {Smith} \emph {et~al.}}]{zeni2023mattergen}%
  \BibitemOpen
  \bibfield  {author} {\bibinfo {author} {\bibfnamefont {C.}~\bibnamefont {Zeni}}, \bibinfo {author} {\bibfnamefont {R.}~\bibnamefont {Pinsler}}, \bibinfo {author} {\bibfnamefont {D.}~\bibnamefont {Z{\"u}gner}}, \bibinfo {author} {\bibfnamefont {A.}~\bibnamefont {Fowler}}, \bibinfo {author} {\bibfnamefont {M.}~\bibnamefont {Horton}}, \bibinfo {author} {\bibfnamefont {X.}~\bibnamefont {Fu}}, \bibinfo {author} {\bibfnamefont {S.}~\bibnamefont {Shysheya}}, \bibinfo {author} {\bibfnamefont {J.}~\bibnamefont {Crabb{\'e}}}, \bibinfo {author} {\bibfnamefont {L.}~\bibnamefont {Sun}}, \bibinfo {author} {\bibfnamefont {J.}~\bibnamefont {Smith}}, \emph {et~al.},\ }\bibfield  {title} {\bibinfo {title} {Mattergen: a generative model for inorganic materials design},\ }\href {https://doi.org/10.48550/arXiv.2312.03687} {\bibfield  {journal} {\bibinfo  {journal} {arXiv preprint arXiv:2312.03687}\ } (\bibinfo {year} {2023})}\BibitemShut {NoStop}%
\bibitem [{\citenamefont {Merchant}\ \emph {et~al.}(2023)\citenamefont {Merchant}, \citenamefont {Batzner}, \citenamefont {Schoenholz}, \citenamefont {Aykol}, \citenamefont {Cheon},\ and\ \citenamefont {Cubuk}}]{merchant2023scaling}%
  \BibitemOpen
  \bibfield  {author} {\bibinfo {author} {\bibfnamefont {A.}~\bibnamefont {Merchant}}, \bibinfo {author} {\bibfnamefont {S.}~\bibnamefont {Batzner}}, \bibinfo {author} {\bibfnamefont {S.~S.}\ \bibnamefont {Schoenholz}}, \bibinfo {author} {\bibfnamefont {M.}~\bibnamefont {Aykol}}, \bibinfo {author} {\bibfnamefont {G.}~\bibnamefont {Cheon}},\ and\ \bibinfo {author} {\bibfnamefont {E.~D.}\ \bibnamefont {Cubuk}},\ }\bibfield  {title} {\bibinfo {title} {Scaling deep learning for materials discovery},\ }\href {https://doi.org/10.1038/s41586-023-06735-9} {\bibfield  {journal} {\bibinfo  {journal} {Nature}\ ,\ \bibinfo {pages} {1}} (\bibinfo {year} {2023})}\BibitemShut {NoStop}%
\bibitem [{\citenamefont {Szymanski}\ \emph {et~al.}(2023)\citenamefont {Szymanski}, \citenamefont {Rendy}, \citenamefont {Fei}, \citenamefont {Kumar}, \citenamefont {He}, \citenamefont {Milsted}, \citenamefont {McDermott}, \citenamefont {Gallant}, \citenamefont {Cubuk}, \citenamefont {Merchant} \emph {et~al.}}]{szymanski2023autonomous}%
  \BibitemOpen
  \bibfield  {author} {\bibinfo {author} {\bibfnamefont {N.~J.}\ \bibnamefont {Szymanski}}, \bibinfo {author} {\bibfnamefont {B.}~\bibnamefont {Rendy}}, \bibinfo {author} {\bibfnamefont {Y.}~\bibnamefont {Fei}}, \bibinfo {author} {\bibfnamefont {R.~E.}\ \bibnamefont {Kumar}}, \bibinfo {author} {\bibfnamefont {T.}~\bibnamefont {He}}, \bibinfo {author} {\bibfnamefont {D.}~\bibnamefont {Milsted}}, \bibinfo {author} {\bibfnamefont {M.~J.}\ \bibnamefont {McDermott}}, \bibinfo {author} {\bibfnamefont {M.}~\bibnamefont {Gallant}}, \bibinfo {author} {\bibfnamefont {E.~D.}\ \bibnamefont {Cubuk}}, \bibinfo {author} {\bibfnamefont {A.}~\bibnamefont {Merchant}}, \emph {et~al.},\ }\bibfield  {title} {\bibinfo {title} {An autonomous laboratory for the accelerated synthesis of novel materials},\ }\href {https://doi.org/10.1038/s41586-023-06734-w} {\bibfield  {journal} {\bibinfo  {journal} {Nature}\ ,\ \bibinfo {pages} {1}} (\bibinfo {year} {2023})}\BibitemShut {NoStop}%
\bibitem [{\citenamefont {Piantadosi}(2014)}]{piantadosi2014zipf}%
  \BibitemOpen
  \bibfield  {author} {\bibinfo {author} {\bibfnamefont {S.~T.}\ \bibnamefont {Piantadosi}},\ }\bibfield  {title} {\bibinfo {title} {Zipf’s word frequency law in natural language: A critical review and future directions},\ }\href@noop {} {\bibfield  {journal} {\bibinfo  {journal} {Psychonomic bulletin \& review}\ }\textbf {\bibinfo {volume} {21}},\ \bibinfo {pages} {1112} (\bibinfo {year} {2014})}\BibitemShut {NoStop}%
\bibitem [{\citenamefont {Vaswani}\ \emph {et~al.}(2017)\citenamefont {Vaswani}, \citenamefont {Shazeer}, \citenamefont {Parmar}, \citenamefont {Uszkoreit}, \citenamefont {Jones}, \citenamefont {Gomez}, \citenamefont {Kaiser},\ and\ \citenamefont {Polosukhin}}]{vaswani2017attention}%
  \BibitemOpen
  \bibfield  {author} {\bibinfo {author} {\bibfnamefont {A.}~\bibnamefont {Vaswani}}, \bibinfo {author} {\bibfnamefont {N.}~\bibnamefont {Shazeer}}, \bibinfo {author} {\bibfnamefont {N.}~\bibnamefont {Parmar}}, \bibinfo {author} {\bibfnamefont {J.}~\bibnamefont {Uszkoreit}}, \bibinfo {author} {\bibfnamefont {L.}~\bibnamefont {Jones}}, \bibinfo {author} {\bibfnamefont {A.~N.}\ \bibnamefont {Gomez}}, \bibinfo {author} {\bibfnamefont {{\L}.}~\bibnamefont {Kaiser}},\ and\ \bibinfo {author} {\bibfnamefont {I.}~\bibnamefont {Polosukhin}},\ }\bibfield  {title} {\bibinfo {title} {Attention is all you need},\ }\href {https://doi.org/10.48550/arXiv.1706.03762} {\bibfield  {journal} {\bibinfo  {journal} {Advances in neural information processing systems}\ }\textbf {\bibinfo {volume} {30}} (\bibinfo {year} {2017})}\BibitemShut {NoStop}%
\bibitem [{\citenamefont {Chan}\ \emph {et~al.}(2022)\citenamefont {Chan}, \citenamefont {Santoro}, \citenamefont {Lampinen}, \citenamefont {Wang}, \citenamefont {Singh}, \citenamefont {Richemond}, \citenamefont {McClelland},\ and\ \citenamefont {Hill}}]{chan2022data}%
  \BibitemOpen
  \bibfield  {author} {\bibinfo {author} {\bibfnamefont {S.}~\bibnamefont {Chan}}, \bibinfo {author} {\bibfnamefont {A.}~\bibnamefont {Santoro}}, \bibinfo {author} {\bibfnamefont {A.}~\bibnamefont {Lampinen}}, \bibinfo {author} {\bibfnamefont {J.}~\bibnamefont {Wang}}, \bibinfo {author} {\bibfnamefont {A.}~\bibnamefont {Singh}}, \bibinfo {author} {\bibfnamefont {P.}~\bibnamefont {Richemond}}, \bibinfo {author} {\bibfnamefont {J.}~\bibnamefont {McClelland}},\ and\ \bibinfo {author} {\bibfnamefont {F.}~\bibnamefont {Hill}},\ }\bibfield  {title} {\bibinfo {title} {Data distributional properties drive emergent in-context learning in transformers},\ }\href {https://doi.org/10.48550/arXiv.2205.05055} {\bibfield  {journal} {\bibinfo  {journal} {Advances in Neural Information Processing Systems}\ }\textbf {\bibinfo {volume} {35}},\ \bibinfo {pages} {18878} (\bibinfo {year} {2022})}\BibitemShut {NoStop}%
\bibitem [{\citenamefont {Brown}\ \emph {et~al.}(2020)\citenamefont {Brown}, \citenamefont {Mann}, \citenamefont {Ryder}, \citenamefont {Subbiah}, \citenamefont {Kaplan}, \citenamefont {Dhariwal}, \citenamefont {Neelakantan}, \citenamefont {Shyam}, \citenamefont {Sastry}, \citenamefont {Askell} \emph {et~al.}}]{brown2020language}%
  \BibitemOpen
  \bibfield  {author} {\bibinfo {author} {\bibfnamefont {T.}~\bibnamefont {Brown}}, \bibinfo {author} {\bibfnamefont {B.}~\bibnamefont {Mann}}, \bibinfo {author} {\bibfnamefont {N.}~\bibnamefont {Ryder}}, \bibinfo {author} {\bibfnamefont {M.}~\bibnamefont {Subbiah}}, \bibinfo {author} {\bibfnamefont {J.~D.}\ \bibnamefont {Kaplan}}, \bibinfo {author} {\bibfnamefont {P.}~\bibnamefont {Dhariwal}}, \bibinfo {author} {\bibfnamefont {A.}~\bibnamefont {Neelakantan}}, \bibinfo {author} {\bibfnamefont {P.}~\bibnamefont {Shyam}}, \bibinfo {author} {\bibfnamefont {G.}~\bibnamefont {Sastry}}, \bibinfo {author} {\bibfnamefont {A.}~\bibnamefont {Askell}}, \emph {et~al.},\ }\bibfield  {title} {\bibinfo {title} {Language models are few-shot learners},\ }\href {https://doi.org/10.48550/arXiv.2005.14165} {\bibfield  {journal} {\bibinfo  {journal} {Advances in neural information processing systems}\ }\textbf {\bibinfo {volume} {33}},\ \bibinfo {pages} {1877} (\bibinfo {year} {2020})}\BibitemShut {NoStop}%
\bibitem [{\citenamefont {Devlin}\ \emph {et~al.}(2018)\citenamefont {Devlin}, \citenamefont {Chang}, \citenamefont {Lee},\ and\ \citenamefont {Toutanova}}]{devlin2018bert}%
  \BibitemOpen
  \bibfield  {author} {\bibinfo {author} {\bibfnamefont {J.}~\bibnamefont {Devlin}}, \bibinfo {author} {\bibfnamefont {M.-W.}\ \bibnamefont {Chang}}, \bibinfo {author} {\bibfnamefont {K.}~\bibnamefont {Lee}},\ and\ \bibinfo {author} {\bibfnamefont {K.}~\bibnamefont {Toutanova}},\ }\bibfield  {title} {\bibinfo {title} {Bert: Pre-training of deep bidirectional transformers for language understanding},\ }\href {https://doi.org/10.48550/arXiv.1810.04805} {\bibfield  {journal} {\bibinfo  {journal} {arXiv preprint arXiv:1810.04805}\ } (\bibinfo {year} {2018})}\BibitemShut {NoStop}%
\bibitem [{\citenamefont {Galassi}\ \emph {et~al.}(2021)\citenamefont {Galassi}, \citenamefont {Lippi},\ and\ \citenamefont {Torroni}}]{galassi2020attention}%
  \BibitemOpen
  \bibfield  {author} {\bibinfo {author} {\bibfnamefont {A.}~\bibnamefont {Galassi}}, \bibinfo {author} {\bibfnamefont {M.}~\bibnamefont {Lippi}},\ and\ \bibinfo {author} {\bibfnamefont {P.}~\bibnamefont {Torroni}},\ }\bibfield  {title} {\bibinfo {title} {Attention in natural language processing},\ }\href {https://doi.org/10.1109/TNNLS.2020.3019893} {\bibfield  {journal} {\bibinfo  {journal} {IEEE Transactions on Neural Networks and Learning Systems}\ }\textbf {\bibinfo {volume} {32}},\ \bibinfo {pages} {4291} (\bibinfo {year} {2021})}\BibitemShut {NoStop}%
\bibitem [{\citenamefont {Niu}\ \emph {et~al.}(2021)\citenamefont {Niu}, \citenamefont {Zhong},\ and\ \citenamefont {Yu}}]{niu2021review}%
  \BibitemOpen
  \bibfield  {author} {\bibinfo {author} {\bibfnamefont {Z.}~\bibnamefont {Niu}}, \bibinfo {author} {\bibfnamefont {G.}~\bibnamefont {Zhong}},\ and\ \bibinfo {author} {\bibfnamefont {H.}~\bibnamefont {Yu}},\ }\bibfield  {title} {\bibinfo {title} {A review on the attention mechanism of deep learning},\ }\href@noop {} {\bibfield  {journal} {\bibinfo  {journal} {Neurocomputing}\ }\textbf {\bibinfo {volume} {452}},\ \bibinfo {pages} {48} (\bibinfo {year} {2021})}\BibitemShut {NoStop}%
\bibitem [{\citenamefont {Lewis}\ \emph {et~al.}(2019)\citenamefont {Lewis}, \citenamefont {Liu}, \citenamefont {Goyal}, \citenamefont {Ghazvininejad}, \citenamefont {Mohamed}, \citenamefont {Levy}, \citenamefont {Stoyanov},\ and\ \citenamefont {Zettlemoyer}}]{lewis2019bart}%
  \BibitemOpen
  \bibfield  {author} {\bibinfo {author} {\bibfnamefont {M.}~\bibnamefont {Lewis}}, \bibinfo {author} {\bibfnamefont {Y.}~\bibnamefont {Liu}}, \bibinfo {author} {\bibfnamefont {N.}~\bibnamefont {Goyal}}, \bibinfo {author} {\bibfnamefont {M.}~\bibnamefont {Ghazvininejad}}, \bibinfo {author} {\bibfnamefont {A.}~\bibnamefont {Mohamed}}, \bibinfo {author} {\bibfnamefont {O.}~\bibnamefont {Levy}}, \bibinfo {author} {\bibfnamefont {V.}~\bibnamefont {Stoyanov}},\ and\ \bibinfo {author} {\bibfnamefont {L.}~\bibnamefont {Zettlemoyer}},\ }\bibfield  {title} {\bibinfo {title} {Bart: Denoising sequence-to-sequence pre-training for natural language generation, translation, and comprehension},\ }\href {https://doi.org/10.48550/arXiv.1910.13461} {\bibfield  {journal} {\bibinfo  {journal} {arXiv preprint arXiv:1910.13461}\ } (\bibinfo {year} {2019})}\BibitemShut {NoStop}%
\bibitem [{\citenamefont {Liao}\ \emph {et~al.}(2023)\citenamefont {Liao}, \citenamefont {Wood}, \citenamefont {Das},\ and\ \citenamefont {Smidt}}]{liao2023equiformerv2}%
  \BibitemOpen
  \bibfield  {author} {\bibinfo {author} {\bibfnamefont {Y.-L.}\ \bibnamefont {Liao}}, \bibinfo {author} {\bibfnamefont {B.}~\bibnamefont {Wood}}, \bibinfo {author} {\bibfnamefont {A.}~\bibnamefont {Das}},\ and\ \bibinfo {author} {\bibfnamefont {T.}~\bibnamefont {Smidt}},\ }\bibfield  {title} {\bibinfo {title} {Equiformerv2: Improved equivariant transformer for scaling to higher-degree representations},\ }\href {https://doi.org/10.48550/arXiv.2306.12059} {\bibfield  {journal} {\bibinfo  {journal} {arXiv preprint arXiv:2306.12059}\ } (\bibinfo {year} {2023})}\BibitemShut {NoStop}%
\bibitem [{\citenamefont {Christiano}\ \emph {et~al.}(2017)\citenamefont {Christiano}, \citenamefont {Leike}, \citenamefont {Brown}, \citenamefont {Martic}, \citenamefont {Legg},\ and\ \citenamefont {Amodei}}]{christiano2017deep}%
  \BibitemOpen
  \bibfield  {author} {\bibinfo {author} {\bibfnamefont {P.~F.}\ \bibnamefont {Christiano}}, \bibinfo {author} {\bibfnamefont {J.}~\bibnamefont {Leike}}, \bibinfo {author} {\bibfnamefont {T.}~\bibnamefont {Brown}}, \bibinfo {author} {\bibfnamefont {M.}~\bibnamefont {Martic}}, \bibinfo {author} {\bibfnamefont {S.}~\bibnamefont {Legg}},\ and\ \bibinfo {author} {\bibfnamefont {D.}~\bibnamefont {Amodei}},\ }\bibfield  {title} {\bibinfo {title} {Deep reinforcement learning from human preferences},\ }\href {https://doi.org/10.48550/arXiv.1706.03741} {\bibfield  {journal} {\bibinfo  {journal} {Advances in neural information processing systems}\ }\textbf {\bibinfo {volume} {30}} (\bibinfo {year} {2017})}\BibitemShut {NoStop}%
\bibitem [{\citenamefont {Hautier}\ \emph {et~al.}(2011)\citenamefont {Hautier}, \citenamefont {Fischer}, \citenamefont {Ehrlacher}, \citenamefont {Jain},\ and\ \citenamefont {Ceder}}]{hautier2011data}%
  \BibitemOpen
  \bibfield  {author} {\bibinfo {author} {\bibfnamefont {G.}~\bibnamefont {Hautier}}, \bibinfo {author} {\bibfnamefont {C.}~\bibnamefont {Fischer}}, \bibinfo {author} {\bibfnamefont {V.}~\bibnamefont {Ehrlacher}}, \bibinfo {author} {\bibfnamefont {A.}~\bibnamefont {Jain}},\ and\ \bibinfo {author} {\bibfnamefont {G.}~\bibnamefont {Ceder}},\ }\bibfield  {title} {\bibinfo {title} {Data mined ionic substitutions for the discovery of new compounds},\ }\href@noop {} {\bibfield  {journal} {\bibinfo  {journal} {Inorganic chemistry}\ }\textbf {\bibinfo {volume} {50}},\ \bibinfo {pages} {656} (\bibinfo {year} {2011})}\BibitemShut {NoStop}%
\bibitem [{\citenamefont {Glawe}\ \emph {et~al.}(2016)\citenamefont {Glawe}, \citenamefont {Sanna}, \citenamefont {Gross},\ and\ \citenamefont {Marques}}]{glawe2016optimal}%
  \BibitemOpen
  \bibfield  {author} {\bibinfo {author} {\bibfnamefont {H.}~\bibnamefont {Glawe}}, \bibinfo {author} {\bibfnamefont {A.}~\bibnamefont {Sanna}}, \bibinfo {author} {\bibfnamefont {E.}~\bibnamefont {Gross}},\ and\ \bibinfo {author} {\bibfnamefont {M.~A.}\ \bibnamefont {Marques}},\ }\bibfield  {title} {\bibinfo {title} {The optimal one dimensional periodic table: a modified pettifor chemical scale from data mining},\ }\href@noop {} {\bibfield  {journal} {\bibinfo  {journal} {New Journal of Physics}\ }\textbf {\bibinfo {volume} {18}},\ \bibinfo {pages} {093011} (\bibinfo {year} {2016})}\BibitemShut {NoStop}%
\bibitem [{\citenamefont {Jain}\ \emph {et~al.}(2013)\citenamefont {Jain}, \citenamefont {Ong}, \citenamefont {Hautier}, \citenamefont {Chen}, \citenamefont {Richards}, \citenamefont {Dacek}, \citenamefont {Cholia}, \citenamefont {Gunter}, \citenamefont {Skinner}, \citenamefont {Ceder} \emph {et~al.}}]{jain2013commentary}%
  \BibitemOpen
  \bibfield  {author} {\bibinfo {author} {\bibfnamefont {A.}~\bibnamefont {Jain}}, \bibinfo {author} {\bibfnamefont {S.~P.}\ \bibnamefont {Ong}}, \bibinfo {author} {\bibfnamefont {G.}~\bibnamefont {Hautier}}, \bibinfo {author} {\bibfnamefont {W.}~\bibnamefont {Chen}}, \bibinfo {author} {\bibfnamefont {W.~D.}\ \bibnamefont {Richards}}, \bibinfo {author} {\bibfnamefont {S.}~\bibnamefont {Dacek}}, \bibinfo {author} {\bibfnamefont {S.}~\bibnamefont {Cholia}}, \bibinfo {author} {\bibfnamefont {D.}~\bibnamefont {Gunter}}, \bibinfo {author} {\bibfnamefont {D.}~\bibnamefont {Skinner}}, \bibinfo {author} {\bibfnamefont {G.}~\bibnamefont {Ceder}}, \emph {et~al.},\ }\bibfield  {title} {\bibinfo {title} {Commentary: The materials project: A materials genome approach to accelerating materials innovation},\ }\href@noop {} {\bibfield  {journal} {\bibinfo  {journal} {APL materials}\ }\textbf {\bibinfo {volume} {1}} (\bibinfo {year} {2013})}\BibitemShut {NoStop}%
\bibitem [{\citenamefont {Davies}\ \emph {et~al.}(2019)\citenamefont {Davies}, \citenamefont {Butler}, \citenamefont {Jackson}, \citenamefont {Skelton}, \citenamefont {Morita},\ and\ \citenamefont {Walsh}}]{davies2019smact}%
  \BibitemOpen
  \bibfield  {author} {\bibinfo {author} {\bibfnamefont {D.~W.}\ \bibnamefont {Davies}}, \bibinfo {author} {\bibfnamefont {K.~T.}\ \bibnamefont {Butler}}, \bibinfo {author} {\bibfnamefont {A.~J.}\ \bibnamefont {Jackson}}, \bibinfo {author} {\bibfnamefont {J.~M.}\ \bibnamefont {Skelton}}, \bibinfo {author} {\bibfnamefont {K.}~\bibnamefont {Morita}},\ and\ \bibinfo {author} {\bibfnamefont {A.}~\bibnamefont {Walsh}},\ }\bibfield  {title} {\bibinfo {title} {Smact: Semiconducting materials by analogy and chemical theory},\ }\href@noop {} {\bibfield  {journal} {\bibinfo  {journal} {Journal of Open Source Software}\ }\textbf {\bibinfo {volume} {4}},\ \bibinfo {pages} {1361} (\bibinfo {year} {2019})}\BibitemShut {NoStop}%
\bibitem [{\citenamefont {Ratcliff}\ \emph {et~al.}(2017)\citenamefont {Ratcliff}, \citenamefont {Mohr}, \citenamefont {Huhs}, \citenamefont {Deutsch}, \citenamefont {Masella},\ and\ \citenamefont {Genovese}}]{ratcliff2017challenges}%
  \BibitemOpen
  \bibfield  {author} {\bibinfo {author} {\bibfnamefont {L.~E.}\ \bibnamefont {Ratcliff}}, \bibinfo {author} {\bibfnamefont {S.}~\bibnamefont {Mohr}}, \bibinfo {author} {\bibfnamefont {G.}~\bibnamefont {Huhs}}, \bibinfo {author} {\bibfnamefont {T.}~\bibnamefont {Deutsch}}, \bibinfo {author} {\bibfnamefont {M.}~\bibnamefont {Masella}},\ and\ \bibinfo {author} {\bibfnamefont {L.}~\bibnamefont {Genovese}},\ }\bibfield  {title} {\bibinfo {title} {Challenges in large scale quantum mechanical calculations},\ }\href@noop {} {\bibfield  {journal} {\bibinfo  {journal} {Wiley Interdisciplinary Reviews: Computational Molecular Science}\ }\textbf {\bibinfo {volume} {7}},\ \bibinfo {pages} {e1290} (\bibinfo {year} {2017})}\BibitemShut {NoStop}%
\bibitem [{\citenamefont {Blum}\ \emph {et~al.}(2022)\citenamefont {Blum}, \citenamefont {Rossi}, \citenamefont {Kokott},\ and\ \citenamefont {Scheffler}}]{blum2022fhi}%
  \BibitemOpen
  \bibfield  {author} {\bibinfo {author} {\bibfnamefont {V.}~\bibnamefont {Blum}}, \bibinfo {author} {\bibfnamefont {M.}~\bibnamefont {Rossi}}, \bibinfo {author} {\bibfnamefont {S.}~\bibnamefont {Kokott}},\ and\ \bibinfo {author} {\bibfnamefont {M.}~\bibnamefont {Scheffler}},\ }\bibfield  {title} {\bibinfo {title} {The fhi-aims code: All-electron, ab initio materials simulations towards the exascale},\ }\href@noop {} {\bibfield  {journal} {\bibinfo  {journal} {arXiv preprint arXiv:2208.12335}\ } (\bibinfo {year} {2022})}\BibitemShut {NoStop}%
\bibitem [{\citenamefont {Gao}\ \emph {et~al.}(2019)\citenamefont {Gao}, \citenamefont {Shi}, \citenamefont {Deng}, \citenamefont {Shiring}, \citenamefont {Snaider}, \citenamefont {Liang}, \citenamefont {Yuan}, \citenamefont {Song}, \citenamefont {Janke}, \citenamefont {Liebman-Pel{\'a}ez} \emph {et~al.}}]{gao2019molecular}%
  \BibitemOpen
  \bibfield  {author} {\bibinfo {author} {\bibfnamefont {Y.}~\bibnamefont {Gao}}, \bibinfo {author} {\bibfnamefont {E.}~\bibnamefont {Shi}}, \bibinfo {author} {\bibfnamefont {S.}~\bibnamefont {Deng}}, \bibinfo {author} {\bibfnamefont {S.~B.}\ \bibnamefont {Shiring}}, \bibinfo {author} {\bibfnamefont {J.~M.}\ \bibnamefont {Snaider}}, \bibinfo {author} {\bibfnamefont {C.}~\bibnamefont {Liang}}, \bibinfo {author} {\bibfnamefont {B.}~\bibnamefont {Yuan}}, \bibinfo {author} {\bibfnamefont {R.}~\bibnamefont {Song}}, \bibinfo {author} {\bibfnamefont {S.~M.}\ \bibnamefont {Janke}}, \bibinfo {author} {\bibfnamefont {A.}~\bibnamefont {Liebman-Pel{\'a}ez}}, \emph {et~al.},\ }\bibfield  {title} {\bibinfo {title} {Molecular engineering of organic--inorganic hybrid perovskites quantum wells},\ }\href@noop {} {\bibfield  {journal} {\bibinfo  {journal} {Nature chemistry}\ }\textbf {\bibinfo {volume} {11}},\ \bibinfo {pages} {1151} (\bibinfo {year} {2019})}\BibitemShut {NoStop}%
\bibitem [{\citenamefont {Park}\ \emph {et~al.}(2023)\citenamefont {Park}, \citenamefont {Song}, \citenamefont {Liang}, \citenamefont {Jin}, \citenamefont {Wang}, \citenamefont {Li}, \citenamefont {Shi}, \citenamefont {Gao}, \citenamefont {Zeller}, \citenamefont {Teat} \emph {et~al.}}]{park2023thickness}%
  \BibitemOpen
  \bibfield  {author} {\bibinfo {author} {\bibfnamefont {J.~Y.}\ \bibnamefont {Park}}, \bibinfo {author} {\bibfnamefont {R.}~\bibnamefont {Song}}, \bibinfo {author} {\bibfnamefont {J.}~\bibnamefont {Liang}}, \bibinfo {author} {\bibfnamefont {L.}~\bibnamefont {Jin}}, \bibinfo {author} {\bibfnamefont {K.}~\bibnamefont {Wang}}, \bibinfo {author} {\bibfnamefont {S.}~\bibnamefont {Li}}, \bibinfo {author} {\bibfnamefont {E.}~\bibnamefont {Shi}}, \bibinfo {author} {\bibfnamefont {Y.}~\bibnamefont {Gao}}, \bibinfo {author} {\bibfnamefont {M.}~\bibnamefont {Zeller}}, \bibinfo {author} {\bibfnamefont {S.~J.}\ \bibnamefont {Teat}}, \emph {et~al.},\ }\bibfield  {title} {\bibinfo {title} {Thickness control of organic semiconductor-incorporated perovskites},\ }\href@noop {} {\bibfield  {journal} {\bibinfo  {journal} {Nature Chemistry}\ }\textbf {\bibinfo {volume} {15}},\ \bibinfo {pages} {1745} (\bibinfo {year} {2023})}\BibitemShut {NoStop}%
\bibitem [{\citenamefont {Song}\ \emph {et~al.}(2023)\citenamefont {Song}, \citenamefont {Liu}, \citenamefont {Kanai}, \citenamefont {Mitzi},\ and\ \citenamefont {Blum}}]{song2023structure}%
  \BibitemOpen
  \bibfield  {author} {\bibinfo {author} {\bibfnamefont {R.}~\bibnamefont {Song}}, \bibinfo {author} {\bibfnamefont {C.}~\bibnamefont {Liu}}, \bibinfo {author} {\bibfnamefont {Y.}~\bibnamefont {Kanai}}, \bibinfo {author} {\bibfnamefont {D.~B.}\ \bibnamefont {Mitzi}},\ and\ \bibinfo {author} {\bibfnamefont {V.}~\bibnamefont {Blum}},\ }\bibfield  {title} {\bibinfo {title} {Structure and electronic tunability of acene alkylamine based layered hybrid organic-inorganic perovskites from first principles},\ }\href@noop {} {\bibfield  {journal} {\bibinfo  {journal} {Physical Review Materials}\ }\textbf {\bibinfo {volume} {7}},\ \bibinfo {pages} {084601} (\bibinfo {year} {2023})}\BibitemShut {NoStop}%
\bibitem [{\citenamefont {Pederson}\ \emph {et~al.}(2022)\citenamefont {Pederson}, \citenamefont {Kozlowski}, \citenamefont {Song}, \citenamefont {Beall}, \citenamefont {Ganahl}, \citenamefont {Hauru}, \citenamefont {Lewis}, \citenamefont {Yao}, \citenamefont {Mallick}, \citenamefont {Blum} \emph {et~al.}}]{pederson2022large}%
  \BibitemOpen
  \bibfield  {author} {\bibinfo {author} {\bibfnamefont {R.}~\bibnamefont {Pederson}}, \bibinfo {author} {\bibfnamefont {J.}~\bibnamefont {Kozlowski}}, \bibinfo {author} {\bibfnamefont {R.}~\bibnamefont {Song}}, \bibinfo {author} {\bibfnamefont {J.}~\bibnamefont {Beall}}, \bibinfo {author} {\bibfnamefont {M.}~\bibnamefont {Ganahl}}, \bibinfo {author} {\bibfnamefont {M.}~\bibnamefont {Hauru}}, \bibinfo {author} {\bibfnamefont {A.~G.}\ \bibnamefont {Lewis}}, \bibinfo {author} {\bibfnamefont {Y.}~\bibnamefont {Yao}}, \bibinfo {author} {\bibfnamefont {S.~B.}\ \bibnamefont {Mallick}}, \bibinfo {author} {\bibfnamefont {V.}~\bibnamefont {Blum}}, \emph {et~al.},\ }\bibfield  {title} {\bibinfo {title} {Large scale quantum chemistry with tensor processing units},\ }\href@noop {} {\bibfield  {journal} {\bibinfo  {journal} {Journal of Chemical Theory and Computation}\ }\textbf {\bibinfo {volume} {19}},\ \bibinfo {pages} {25} (\bibinfo {year} {2022})}\BibitemShut {NoStop}%
\bibitem [{\citenamefont {Cramer}\ and\ \citenamefont {Truhlar}(2009)}]{cramer2009density}%
  \BibitemOpen
  \bibfield  {author} {\bibinfo {author} {\bibfnamefont {C.~J.}\ \bibnamefont {Cramer}}\ and\ \bibinfo {author} {\bibfnamefont {D.~G.}\ \bibnamefont {Truhlar}},\ }\bibfield  {title} {\bibinfo {title} {Density functional theory for transition metals and transition metal chemistry},\ }\href@noop {} {\bibfield  {journal} {\bibinfo  {journal} {Physical Chemistry Chemical Physics}\ }\textbf {\bibinfo {volume} {11}},\ \bibinfo {pages} {10757} (\bibinfo {year} {2009})}\BibitemShut {NoStop}%
\bibitem [{\citenamefont {Xue}\ \emph {et~al.}(2023)\citenamefont {Xue}, \citenamefont {Song}, \citenamefont {Guo}, \citenamefont {Sung}, \citenamefont {Lortz}, \citenamefont {Williams},\ and\ \citenamefont {Lu}}]{xue2023extended}%
  \BibitemOpen
  \bibfield  {author} {\bibinfo {author} {\bibfnamefont {J.}~\bibnamefont {Xue}}, \bibinfo {author} {\bibfnamefont {R.}~\bibnamefont {Song}}, \bibinfo {author} {\bibfnamefont {Z.}~\bibnamefont {Guo}}, \bibinfo {author} {\bibfnamefont {H.~H.}\ \bibnamefont {Sung}}, \bibinfo {author} {\bibfnamefont {R.}~\bibnamefont {Lortz}}, \bibinfo {author} {\bibfnamefont {I.~D.}\ \bibnamefont {Williams}},\ and\ \bibinfo {author} {\bibfnamefont {H.}~\bibnamefont {Lu}},\ }\bibfield  {title} {\bibinfo {title} {Extended honeycomb metal chloride with tunable antiferromagnetic correlations},\ }\href@noop {} {\bibfield  {journal} {\bibinfo  {journal} {Chemistry of Materials}\ }\textbf {\bibinfo {volume} {36}},\ \bibinfo {pages} {551} (\bibinfo {year} {2023})}\BibitemShut {NoStop}%
\bibitem [{\citenamefont {Wang}\ \emph {et~al.}(2023)\citenamefont {Wang}, \citenamefont {McWhorter}, \citenamefont {McKeown~Wessler}, \citenamefont {Yao}, \citenamefont {Song}, \citenamefont {Mitzi},\ and\ \citenamefont {Blum}}]{wang2023exploration}%
  \BibitemOpen
  \bibfield  {author} {\bibinfo {author} {\bibfnamefont {T.}~\bibnamefont {Wang}}, \bibinfo {author} {\bibfnamefont {T.~M.}\ \bibnamefont {McWhorter}}, \bibinfo {author} {\bibfnamefont {G.~C.}\ \bibnamefont {McKeown~Wessler}}, \bibinfo {author} {\bibfnamefont {Y.}~\bibnamefont {Yao}}, \bibinfo {author} {\bibfnamefont {R.}~\bibnamefont {Song}}, \bibinfo {author} {\bibfnamefont {D.~B.}\ \bibnamefont {Mitzi}},\ and\ \bibinfo {author} {\bibfnamefont {V.}~\bibnamefont {Blum}},\ }\bibfield  {title} {\bibinfo {title} {Exploration, prediction, and experimental verification of structure and optoelectronic properties in i2-eu-iv-x4 (i= li, cu, ag; iv= si, ge, sn; x= s, se) chalcogenide semiconductors},\ }\href@noop {} {\bibfield  {journal} {\bibinfo  {journal} {Chemistry of Materials}\ }\textbf {\bibinfo {volume} {36}},\ \bibinfo {pages} {340} (\bibinfo {year} {2023})}\BibitemShut {NoStop}%
\bibitem [{\citenamefont {Blum}\ \emph {et~al.}(2009)\citenamefont {Blum}, \citenamefont {Gehrke}, \citenamefont {Hanke}, \citenamefont {Havu}, \citenamefont {Havu}, \citenamefont {Ren}, \citenamefont {Reuter},\ and\ \citenamefont {Scheffler}}]{blum2009ab}%
  \BibitemOpen
  \bibfield  {author} {\bibinfo {author} {\bibfnamefont {V.}~\bibnamefont {Blum}}, \bibinfo {author} {\bibfnamefont {R.}~\bibnamefont {Gehrke}}, \bibinfo {author} {\bibfnamefont {F.}~\bibnamefont {Hanke}}, \bibinfo {author} {\bibfnamefont {P.}~\bibnamefont {Havu}}, \bibinfo {author} {\bibfnamefont {V.}~\bibnamefont {Havu}}, \bibinfo {author} {\bibfnamefont {X.}~\bibnamefont {Ren}}, \bibinfo {author} {\bibfnamefont {K.}~\bibnamefont {Reuter}},\ and\ \bibinfo {author} {\bibfnamefont {M.}~\bibnamefont {Scheffler}},\ }\bibfield  {title} {\bibinfo {title} {Ab initio molecular simulations with numeric atom-centered orbitals},\ }\href@noop {} {\bibfield  {journal} {\bibinfo  {journal} {Computer Physics Communications}\ }\textbf {\bibinfo {volume} {180}},\ \bibinfo {pages} {2175} (\bibinfo {year} {2009})}\BibitemShut {NoStop}%
\bibitem [{\citenamefont {Perdew}\ \emph {et~al.}(1996)\citenamefont {Perdew}, \citenamefont {Burke},\ and\ \citenamefont {Ernzerhof}}]{perdew1996generalized}%
  \BibitemOpen
  \bibfield  {author} {\bibinfo {author} {\bibfnamefont {J.~P.}\ \bibnamefont {Perdew}}, \bibinfo {author} {\bibfnamefont {K.}~\bibnamefont {Burke}},\ and\ \bibinfo {author} {\bibfnamefont {M.}~\bibnamefont {Ernzerhof}},\ }\bibfield  {title} {\bibinfo {title} {Generalized gradient approximation made simple},\ }\href@noop {} {\bibfield  {journal} {\bibinfo  {journal} {Physical review letters}\ }\textbf {\bibinfo {volume} {77}},\ \bibinfo {pages} {3865} (\bibinfo {year} {1996})}\BibitemShut {NoStop}%
\bibitem [{\citenamefont {Huhn}\ and\ \citenamefont {Blum}(2017)}]{huhn2017one}%
  \BibitemOpen
  \bibfield  {author} {\bibinfo {author} {\bibfnamefont {W.~P.}\ \bibnamefont {Huhn}}\ and\ \bibinfo {author} {\bibfnamefont {V.}~\bibnamefont {Blum}},\ }\bibfield  {title} {\bibinfo {title} {One-hundred-three compound band-structure benchmark of post-self-consistent spin-orbit coupling treatments in density functional theory},\ }\href@noop {} {\bibfield  {journal} {\bibinfo  {journal} {Physical Review Materials}\ }\textbf {\bibinfo {volume} {1}},\ \bibinfo {pages} {033803} (\bibinfo {year} {2017})}\BibitemShut {NoStop}%
\bibitem [{\citenamefont {Zhao}\ \emph {et~al.}(2021{\natexlab{b}})\citenamefont {Zhao}, \citenamefont {Yu}, \citenamefont {Zhang}, \citenamefont {Xiao}, \citenamefont {Zhang},\ and\ \citenamefont {Blum}}]{zhao2021quasi}%
  \BibitemOpen
  \bibfield  {author} {\bibinfo {author} {\bibfnamefont {R.}~\bibnamefont {Zhao}}, \bibinfo {author} {\bibfnamefont {V.~W.-z.}\ \bibnamefont {Yu}}, \bibinfo {author} {\bibfnamefont {K.}~\bibnamefont {Zhang}}, \bibinfo {author} {\bibfnamefont {Y.}~\bibnamefont {Xiao}}, \bibinfo {author} {\bibfnamefont {Y.}~\bibnamefont {Zhang}},\ and\ \bibinfo {author} {\bibfnamefont {V.}~\bibnamefont {Blum}},\ }\bibfield  {title} {\bibinfo {title} {Quasi-four-component method with numeric atom-centered orbitals for relativistic density functional simulations of molecules and solids},\ }\href@noop {} {\bibfield  {journal} {\bibinfo  {journal} {Physical Review B}\ }\textbf {\bibinfo {volume} {103}},\ \bibinfo {pages} {245144} (\bibinfo {year} {2021}{\natexlab{b}})}\BibitemShut {NoStop}%
\bibitem [{\citenamefont {Havu}\ \emph {et~al.}(2009)\citenamefont {Havu}, \citenamefont {Blum}, \citenamefont {Havu},\ and\ \citenamefont {Scheffler}}]{havu2009efficient}%
  \BibitemOpen
  \bibfield  {author} {\bibinfo {author} {\bibfnamefont {V.}~\bibnamefont {Havu}}, \bibinfo {author} {\bibfnamefont {V.}~\bibnamefont {Blum}}, \bibinfo {author} {\bibfnamefont {P.}~\bibnamefont {Havu}},\ and\ \bibinfo {author} {\bibfnamefont {M.}~\bibnamefont {Scheffler}},\ }\bibfield  {title} {\bibinfo {title} {Efficient o (n) integration for all-electron electronic structure calculation using numeric basis functions},\ }\href@noop {} {\bibfield  {journal} {\bibinfo  {journal} {Journal of Computational Physics}\ }\textbf {\bibinfo {volume} {228}},\ \bibinfo {pages} {8367} (\bibinfo {year} {2009})}\BibitemShut {NoStop}%
\end{thebibliography}%


\begin{thebibliography}{13}%
\makeatletter
\providecommand \@ifxundefined [1]{%
 \@ifx{#1\undefined}
}%
\providecommand \@ifnum [1]{%
 \ifnum #1\expandafter \@firstoftwo
 \else \expandafter \@secondoftwo
 \fi
}%
\providecommand \@ifx [1]{%
 \ifx #1\expandafter \@firstoftwo
 \else \expandafter \@secondoftwo
 \fi
}%
\providecommand \natexlab [1]{#1}%
\providecommand \enquote  [1]{``#1''}%
\providecommand \bibnamefont  [1]{#1}%
\providecommand \bibfnamefont [1]{#1}%
\providecommand \citenamefont [1]{#1}%
\providecommand \href@noop [0]{\@secondoftwo}%
\providecommand \href [0]{\begingroup \@sanitize@url \@href}%
\providecommand \@href[1]{\@@startlink{#1}\@@href}%
\providecommand \@@href[1]{\endgroup#1\@@endlink}%
\providecommand \@sanitize@url [0]{\catcode `\\12\catcode `\$12\catcode `\&12\catcode `\#12\catcode `\^12\catcode `\_12\catcode `\%12\relax}%
\providecommand \@@startlink[1]{}%
\providecommand \@@endlink[0]{}%
\providecommand \url  [0]{\begingroup\@sanitize@url \@url }%
\providecommand \@url [1]{\endgroup\@href {#1}{\urlprefix }}%
\providecommand \urlprefix  [0]{URL }%
\providecommand \Eprint [0]{\href }%
\providecommand \doibase [0]{https://doi.org/}%
\providecommand \selectlanguage [0]{\@gobble}%
\providecommand \bibinfo  [0]{\@secondoftwo}%
\providecommand \bibfield  [0]{\@secondoftwo}%
\providecommand \translation [1]{[#1]}%
\providecommand \BibitemOpen [0]{}%
\providecommand \bibitemStop [0]{}%
\providecommand \bibitemNoStop [0]{.\EOS\space}%
\providecommand \EOS [0]{\spacefactor3000\relax}%
\providecommand \BibitemShut  [1]{\csname bibitem#1\endcsname}%
\let\auto@bib@innerbib\@empty
\bibitem [{\citenamefont {Chan}\ \emph {et~al.}(2022)\citenamefont {Chan}, \citenamefont {Santoro}, \citenamefont {Lampinen}, \citenamefont {Wang}, \citenamefont {Singh}, \citenamefont {Richemond}, \citenamefont {McClelland},\ and\ \citenamefont {Hill}}]{chan2022data}%
  \BibitemOpen
  \bibfield  {author} {\bibinfo {author} {\bibfnamefont {S.}~\bibnamefont {Chan}}, \bibinfo {author} {\bibfnamefont {A.}~\bibnamefont {Santoro}}, \bibinfo {author} {\bibfnamefont {A.}~\bibnamefont {Lampinen}}, \bibinfo {author} {\bibfnamefont {J.}~\bibnamefont {Wang}}, \bibinfo {author} {\bibfnamefont {A.}~\bibnamefont {Singh}}, \bibinfo {author} {\bibfnamefont {P.}~\bibnamefont {Richemond}}, \bibinfo {author} {\bibfnamefont {J.}~\bibnamefont {McClelland}},\ and\ \bibinfo {author} {\bibfnamefont {F.}~\bibnamefont {Hill}},\ }\bibfield  {title} {\bibinfo {title} {Data distributional properties drive emergent in-context learning in transformers},\ }\href {https://doi.org/10.48550/arXiv.2205.05055} {\bibfield  {journal} {\bibinfo  {journal} {Advances in Neural Information Processing Systems}\ }\textbf {\bibinfo {volume} {35}},\ \bibinfo {pages} {18878} (\bibinfo {year} {2022})}\BibitemShut {NoStop}%
\bibitem [{\citenamefont {Devlin}\ \emph {et~al.}(2018)\citenamefont {Devlin}, \citenamefont {Chang}, \citenamefont {Lee},\ and\ \citenamefont {Toutanova}}]{devlin2018bert}%
  \BibitemOpen
  \bibfield  {author} {\bibinfo {author} {\bibfnamefont {J.}~\bibnamefont {Devlin}}, \bibinfo {author} {\bibfnamefont {M.-W.}\ \bibnamefont {Chang}}, \bibinfo {author} {\bibfnamefont {K.}~\bibnamefont {Lee}},\ and\ \bibinfo {author} {\bibfnamefont {K.}~\bibnamefont {Toutanova}},\ }\bibfield  {title} {\bibinfo {title} {Bert: Pre-training of deep bidirectional transformers for language understanding},\ }\href {https://doi.org/10.48550/arXiv.1810.04805} {\bibfield  {journal} {\bibinfo  {journal} {arXiv preprint arXiv:1810.04805}\ } (\bibinfo {year} {2018})}\BibitemShut {NoStop}%
\bibitem [{\citenamefont {Brown}\ \emph {et~al.}(2020)\citenamefont {Brown}, \citenamefont {Mann}, \citenamefont {Ryder}, \citenamefont {Subbiah}, \citenamefont {Kaplan}, \citenamefont {Dhariwal}, \citenamefont {Neelakantan}, \citenamefont {Shyam}, \citenamefont {Sastry}, \citenamefont {Askell} \emph {et~al.}}]{brown2020language}%
  \BibitemOpen
  \bibfield  {author} {\bibinfo {author} {\bibfnamefont {T.}~\bibnamefont {Brown}}, \bibinfo {author} {\bibfnamefont {B.}~\bibnamefont {Mann}}, \bibinfo {author} {\bibfnamefont {N.}~\bibnamefont {Ryder}}, \bibinfo {author} {\bibfnamefont {M.}~\bibnamefont {Subbiah}}, \bibinfo {author} {\bibfnamefont {J.~D.}\ \bibnamefont {Kaplan}}, \bibinfo {author} {\bibfnamefont {P.}~\bibnamefont {Dhariwal}}, \bibinfo {author} {\bibfnamefont {A.}~\bibnamefont {Neelakantan}}, \bibinfo {author} {\bibfnamefont {P.}~\bibnamefont {Shyam}}, \bibinfo {author} {\bibfnamefont {G.}~\bibnamefont {Sastry}}, \bibinfo {author} {\bibfnamefont {A.}~\bibnamefont {Askell}}, \emph {et~al.},\ }\bibfield  {title} {\bibinfo {title} {Language models are few-shot learners},\ }\href {https://doi.org/10.48550/arXiv.2005.14165} {\bibfield  {journal} {\bibinfo  {journal} {Advances in neural information processing systems}\ }\textbf {\bibinfo {volume} {33}},\ \bibinfo {pages} {1877} (\bibinfo {year} {2020})}\BibitemShut {NoStop}%
\bibitem [{\citenamefont {Vaswani}\ \emph {et~al.}(2017)\citenamefont {Vaswani}, \citenamefont {Shazeer}, \citenamefont {Parmar}, \citenamefont {Uszkoreit}, \citenamefont {Jones}, \citenamefont {Gomez}, \citenamefont {Kaiser},\ and\ \citenamefont {Polosukhin}}]{vaswani2017attention}%
  \BibitemOpen
  \bibfield  {author} {\bibinfo {author} {\bibfnamefont {A.}~\bibnamefont {Vaswani}}, \bibinfo {author} {\bibfnamefont {N.}~\bibnamefont {Shazeer}}, \bibinfo {author} {\bibfnamefont {N.}~\bibnamefont {Parmar}}, \bibinfo {author} {\bibfnamefont {J.}~\bibnamefont {Uszkoreit}}, \bibinfo {author} {\bibfnamefont {L.}~\bibnamefont {Jones}}, \bibinfo {author} {\bibfnamefont {A.~N.}\ \bibnamefont {Gomez}}, \bibinfo {author} {\bibfnamefont {{\L}.}~\bibnamefont {Kaiser}},\ and\ \bibinfo {author} {\bibfnamefont {I.}~\bibnamefont {Polosukhin}},\ }\bibfield  {title} {\bibinfo {title} {Attention is all you need},\ }\href {https://doi.org/10.48550/arXiv.1706.03762} {\bibfield  {journal} {\bibinfo  {journal} {Advances in neural information processing systems}\ }\textbf {\bibinfo {volume} {30}} (\bibinfo {year} {2017})}\BibitemShut {NoStop}%
\bibitem [{\citenamefont {Galassi}\ \emph {et~al.}(2021)\citenamefont {Galassi}, \citenamefont {Lippi},\ and\ \citenamefont {Torroni}}]{galassi2020attention}%
  \BibitemOpen
  \bibfield  {author} {\bibinfo {author} {\bibfnamefont {A.}~\bibnamefont {Galassi}}, \bibinfo {author} {\bibfnamefont {M.}~\bibnamefont {Lippi}},\ and\ \bibinfo {author} {\bibfnamefont {P.}~\bibnamefont {Torroni}},\ }\bibfield  {title} {\bibinfo {title} {Attention in natural language processing},\ }\href {https://doi.org/10.1109/TNNLS.2020.3019893} {\bibfield  {journal} {\bibinfo  {journal} {IEEE Transactions on Neural Networks and Learning Systems}\ }\textbf {\bibinfo {volume} {32}},\ \bibinfo {pages} {4291} (\bibinfo {year} {2021})}\BibitemShut {NoStop}%
\bibitem [{\citenamefont {Niu}\ \emph {et~al.}(2021)\citenamefont {Niu}, \citenamefont {Zhong},\ and\ \citenamefont {Yu}}]{niu2021review}%
  \BibitemOpen
  \bibfield  {author} {\bibinfo {author} {\bibfnamefont {Z.}~\bibnamefont {Niu}}, \bibinfo {author} {\bibfnamefont {G.}~\bibnamefont {Zhong}},\ and\ \bibinfo {author} {\bibfnamefont {H.}~\bibnamefont {Yu}},\ }\bibfield  {title} {\bibinfo {title} {A review on the attention mechanism of deep learning},\ }\href@noop {} {\bibfield  {journal} {\bibinfo  {journal} {Neurocomputing}\ }\textbf {\bibinfo {volume} {452}},\ \bibinfo {pages} {48} (\bibinfo {year} {2021})}\BibitemShut {NoStop}%
\bibitem [{\citenamefont {Hautier}\ \emph {et~al.}(2011)\citenamefont {Hautier}, \citenamefont {Fischer}, \citenamefont {Ehrlacher}, \citenamefont {Jain},\ and\ \citenamefont {Ceder}}]{hautier2011data}%
  \BibitemOpen
  \bibfield  {author} {\bibinfo {author} {\bibfnamefont {G.}~\bibnamefont {Hautier}}, \bibinfo {author} {\bibfnamefont {C.}~\bibnamefont {Fischer}}, \bibinfo {author} {\bibfnamefont {V.}~\bibnamefont {Ehrlacher}}, \bibinfo {author} {\bibfnamefont {A.}~\bibnamefont {Jain}},\ and\ \bibinfo {author} {\bibfnamefont {G.}~\bibnamefont {Ceder}},\ }\bibfield  {title} {\bibinfo {title} {Data mined ionic substitutions for the discovery of new compounds},\ }\href@noop {} {\bibfield  {journal} {\bibinfo  {journal} {Inorganic chemistry}\ }\textbf {\bibinfo {volume} {50}},\ \bibinfo {pages} {656} (\bibinfo {year} {2011})}\BibitemShut {NoStop}%
\bibitem [{\citenamefont {Glawe}\ \emph {et~al.}(2016)\citenamefont {Glawe}, \citenamefont {Sanna}, \citenamefont {Gross},\ and\ \citenamefont {Marques}}]{glawe2016optimal}%
  \BibitemOpen
  \bibfield  {author} {\bibinfo {author} {\bibfnamefont {H.}~\bibnamefont {Glawe}}, \bibinfo {author} {\bibfnamefont {A.}~\bibnamefont {Sanna}}, \bibinfo {author} {\bibfnamefont {E.}~\bibnamefont {Gross}},\ and\ \bibinfo {author} {\bibfnamefont {M.~A.}\ \bibnamefont {Marques}},\ }\bibfield  {title} {\bibinfo {title} {The optimal one dimensional periodic table: a modified pettifor chemical scale from data mining},\ }\href@noop {} {\bibfield  {journal} {\bibinfo  {journal} {New Journal of Physics}\ }\textbf {\bibinfo {volume} {18}},\ \bibinfo {pages} {093011} (\bibinfo {year} {2016})}\BibitemShut {NoStop}%
\bibitem [{\citenamefont {Liao}\ \emph {et~al.}(2023)\citenamefont {Liao}, \citenamefont {Wood}, \citenamefont {Das},\ and\ \citenamefont {Smidt}}]{liao2023equiformerv2}%
  \BibitemOpen
  \bibfield  {author} {\bibinfo {author} {\bibfnamefont {Y.-L.}\ \bibnamefont {Liao}}, \bibinfo {author} {\bibfnamefont {B.}~\bibnamefont {Wood}}, \bibinfo {author} {\bibfnamefont {A.}~\bibnamefont {Das}},\ and\ \bibinfo {author} {\bibfnamefont {T.}~\bibnamefont {Smidt}},\ }\bibfield  {title} {\bibinfo {title} {Equiformerv2: Improved equivariant transformer for scaling to higher-degree representations},\ }\href {https://doi.org/10.48550/arXiv.2306.12059} {\bibfield  {journal} {\bibinfo  {journal} {arXiv preprint arXiv:2306.12059}\ } (\bibinfo {year} {2023})}\BibitemShut {NoStop}%
\bibitem [{\citenamefont {Jain}\ \emph {et~al.}(2013)\citenamefont {Jain}, \citenamefont {Ong}, \citenamefont {Hautier}, \citenamefont {Chen}, \citenamefont {Richards}, \citenamefont {Dacek}, \citenamefont {Cholia}, \citenamefont {Gunter}, \citenamefont {Skinner}, \citenamefont {Ceder} \emph {et~al.}}]{jain2013commentary}%
  \BibitemOpen
  \bibfield  {author} {\bibinfo {author} {\bibfnamefont {A.}~\bibnamefont {Jain}}, \bibinfo {author} {\bibfnamefont {S.~P.}\ \bibnamefont {Ong}}, \bibinfo {author} {\bibfnamefont {G.}~\bibnamefont {Hautier}}, \bibinfo {author} {\bibfnamefont {W.}~\bibnamefont {Chen}}, \bibinfo {author} {\bibfnamefont {W.~D.}\ \bibnamefont {Richards}}, \bibinfo {author} {\bibfnamefont {S.}~\bibnamefont {Dacek}}, \bibinfo {author} {\bibfnamefont {S.}~\bibnamefont {Cholia}}, \bibinfo {author} {\bibfnamefont {D.}~\bibnamefont {Gunter}}, \bibinfo {author} {\bibfnamefont {D.}~\bibnamefont {Skinner}}, \bibinfo {author} {\bibfnamefont {G.}~\bibnamefont {Ceder}}, \emph {et~al.},\ }\bibfield  {title} {\bibinfo {title} {Commentary: The materials project: A materials genome approach to accelerating materials innovation},\ }\href@noop {} {\bibfield  {journal} {\bibinfo  {journal} {APL materials}\ }\textbf {\bibinfo {volume} {1}} (\bibinfo {year} {2013})}\BibitemShut {NoStop}%
\bibitem [{\citenamefont {Li}\ \emph {et~al.}(2021)\citenamefont {Li}, \citenamefont {Liu}, \citenamefont {Baronett}, \citenamefont {Liu}, \citenamefont {Wang}, \citenamefont {Li}, \citenamefont {Chen}, \citenamefont {Li}, \citenamefont {Zhu},\ and\ \citenamefont {Chen}}]{li2021computation}%
  \BibitemOpen
  \bibfield  {author} {\bibinfo {author} {\bibfnamefont {J.}~\bibnamefont {Li}}, \bibinfo {author} {\bibfnamefont {J.}~\bibnamefont {Liu}}, \bibinfo {author} {\bibfnamefont {S.~A.}\ \bibnamefont {Baronett}}, \bibinfo {author} {\bibfnamefont {M.}~\bibnamefont {Liu}}, \bibinfo {author} {\bibfnamefont {L.}~\bibnamefont {Wang}}, \bibinfo {author} {\bibfnamefont {R.}~\bibnamefont {Li}}, \bibinfo {author} {\bibfnamefont {Y.}~\bibnamefont {Chen}}, \bibinfo {author} {\bibfnamefont {D.}~\bibnamefont {Li}}, \bibinfo {author} {\bibfnamefont {Q.}~\bibnamefont {Zhu}},\ and\ \bibinfo {author} {\bibfnamefont {X.-Q.}\ \bibnamefont {Chen}},\ }\bibfield  {title} {\bibinfo {title} {Computation and data driven discovery of topological phononic materials},\ }\href@noop {} {\bibfield  {journal} {\bibinfo  {journal} {Nature communications}\ }\textbf {\bibinfo {volume} {12}},\ \bibinfo {pages} {1204} (\bibinfo {year} {2021})}\BibitemShut {NoStop}%
\bibitem [{\citenamefont {Merchant}\ \emph {et~al.}(2023)\citenamefont {Merchant}, \citenamefont {Batzner}, \citenamefont {Schoenholz}, \citenamefont {Aykol}, \citenamefont {Cheon},\ and\ \citenamefont {Cubuk}}]{merchant2023scaling}%
  \BibitemOpen
  \bibfield  {author} {\bibinfo {author} {\bibfnamefont {A.}~\bibnamefont {Merchant}}, \bibinfo {author} {\bibfnamefont {S.}~\bibnamefont {Batzner}}, \bibinfo {author} {\bibfnamefont {S.~S.}\ \bibnamefont {Schoenholz}}, \bibinfo {author} {\bibfnamefont {M.}~\bibnamefont {Aykol}}, \bibinfo {author} {\bibfnamefont {G.}~\bibnamefont {Cheon}},\ and\ \bibinfo {author} {\bibfnamefont {E.~D.}\ \bibnamefont {Cubuk}},\ }\bibfield  {title} {\bibinfo {title} {Scaling deep learning for materials discovery},\ }\href {https://doi.org/10.1038/s41586-023-06735-9} {\bibfield  {journal} {\bibinfo  {journal} {Nature}\ ,\ \bibinfo {pages} {1}} (\bibinfo {year} {2023})}\BibitemShut {NoStop}%
\bibitem [{\citenamefont {Xie}\ \emph {et~al.}(2021)\citenamefont {Xie}, \citenamefont {Fu}, \citenamefont {Ganea}, \citenamefont {Barzilay},\ and\ \citenamefont {Jaakkola}}]{xie2021crystal}%
  \BibitemOpen
  \bibfield  {author} {\bibinfo {author} {\bibfnamefont {T.}~\bibnamefont {Xie}}, \bibinfo {author} {\bibfnamefont {X.}~\bibnamefont {Fu}}, \bibinfo {author} {\bibfnamefont {O.-E.}\ \bibnamefont {Ganea}}, \bibinfo {author} {\bibfnamefont {R.}~\bibnamefont {Barzilay}},\ and\ \bibinfo {author} {\bibfnamefont {T.}~\bibnamefont {Jaakkola}},\ }\bibfield  {title} {\bibinfo {title} {Crystal diffusion variational autoencoder for periodic material generation},\ }\href {https://doi.org/10.48550/arXiv.2110.06197} {\bibfield  {journal} {\bibinfo  {journal} {arXiv preprint arXiv:2110.06197}\ } (\bibinfo {year} {2021})}\BibitemShut {NoStop}%
\end{thebibliography}%

\end{document}


\title{Supplementary Materials:\\ Self-Supervised Generative Models for Crystal Structures}

\author{Fangze~Liu}
\affiliation{Department of Physics, Stanford University, Stanford, CA 94305, USA}
\affiliation{Stanford Institute for Materials and Energy Sciences, SLAC National Accelerator Laboratory, Menlo Park, CA 94025, USA}

\author{Zhantao~Chen}
\affiliation{Stanford Institute for Materials and Energy Sciences, SLAC National Accelerator Laboratory, Menlo Park, CA 94025, USA}
\affiliation{Linac Coherent Light Source, SLAC National Accelerator Laboratory, Menlo Park, CA 94025, USA}

\author{Tianyi~Liu}
\affiliation{Department of Chemistry, Stanford University, Stanford, CA 94305, USA}
\affiliation{Stanford Institute for Materials and Energy Sciences, SLAC National Accelerator Laboratory, Menlo Park, CA 94025, USA}

\author{Ruyi~Song}
\affiliation{Stanford Institute for Materials and Energy Sciences, SLAC National Accelerator Laboratory, Menlo Park, CA 94025, USA}

\author{Yu~Lin}
\affiliation{Stanford Institute for Materials and Energy Sciences, SLAC National Accelerator Laboratory, Menlo Park, CA 94025, USA}

\author{Joshua~J.~Turner}
\affiliation{Stanford Institute for Materials and Energy Sciences, SLAC National Accelerator Laboratory, Menlo Park, CA 94025, USA}
\affiliation{Linac Coherent Light Source, SLAC National Accelerator Laboratory, Menlo Park, CA 94025, USA}

\author{Chunjing~Jia}
\affiliation{Stanford Institute for Materials and Energy Sciences, SLAC National Accelerator Laboratory, Menlo Park, CA 94025, USA}
\affiliation{Department of Physics, University of Florida, Gainesville, FL 32611, USA}

\maketitle

\tableofcontents

\section{Introduction}

\begin{figure*}[htb!]
\includegraphics[width=1\linewidth]{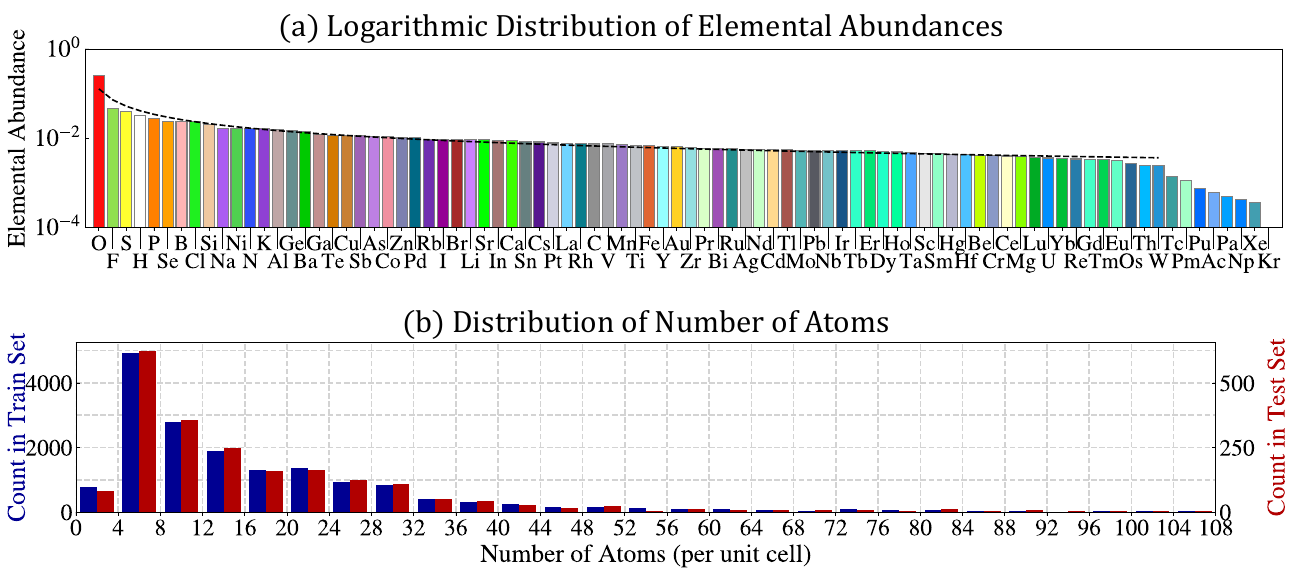}
\caption{\label{data_distribution}
\textbf{Training Data Distributions.} Panel (a) shows the distribution of elemental abundance on a logarithmic scale within our training samples. The distribution follows a power law (skewed Zipfian) distribution $p(X=x) \propto 1/x^\alpha$, where $X$ is the rank of an element class and $\alpha=0.8163$. Note that the fitting of the distribution excluded the eight least frequently occurring elements. This resulting distribution, depicted as a dashed line in the plot, closely resembles the distribution of tokens in a natural language corpora, which typically exhibit an exponent close to $1$~\cite{chan2022data}. Panel (b) presents the distribution of the number of atoms per structural unit cell within our training set (red, left axis) and test set (blue, right axis).}
\end{figure*}

In this study, we propose a new framework for autonomous material generation. The distinctive feature of our approach is the minimal reliance on pre-existing human pre-knowledge during the training process: the generative model is trained exclusively with the atomic numbers of species and their positions for each material structure. This approach not only significantly reduces computational costs and mitigates biases arising from prior assumptions, but also facilitates the study of mappings from material structures to complex physical properties. \\

\textbf{Motivation.} Our work draws inspiration from the field of natural language processing (NLP), which has seen remarkable advancements in recent years. These improvements have been largely driven by the adoption of self-supervised learning strategies~\cite{devlin2018bert} and, though more resource-intensive, in-context learning~\cite{brown2020language}, as well as the self-attention mechanism in large language models (LLMs)~\cite{vaswani2017attention, galassi2020attention, niu2021review}. Besides these aspects, the data distribution has been identified as a crucial factor in driving emergent behaviors in transformer-based LLMs~\cite{chan2022data}. As shown in Fig.~\ref{data_distribution}a, we observe that the distribution of elemental abundance in crystal structures is close to that of vocabulary abundance in natural languages, following a power law distribution---a common pattern in natural datasets like languages. This observation suggests that the success of LLMs could be beneficial in material prediction research. Therefore, we trained a generative model for crystal structures in a similar manner, as detailed in Section~\ref{sec:pre-training}.\\

\textbf{Obstacles and detours.} The challenge, however, lies in the absence of a critic to evaluate the generated structures. We propose utilizing the actor-critic learning framework to guide generative training, similar to applying reinforcement learning from human feedback (RLHF) in NLP. Ideally, numerical methods like DFT calculations or even human evaluations could serve as the critic, while employing GANs offers a computationally efficient solution for early-stage training. More importantly, our objectives are not limited to merely aiding in material generation; we are also keen on exploring the insights that can be extracted from crystal structures. To ensure our training process remain unbiased and uninfluenced by pre-existing knowledge, we deliberately avoided using DFT and similar methods during the training phase. As a result, we choose to integrate the GAN architecture into our training pipeline to fully exploit existing data without relying on external sources like DFT-calculated stability labels (refer to Section~\ref{sec:gan}). \\

\textbf{Solution.} To autonomously generate physically meaningful crystals, the model is expected to self-learn the principles of crystal formation and potentially gaining inferential capabilities. The training of material prediction can be mainly divided into two stages: 
(1) \textit{Pre-training stage}: we guide the model through fill-in-the-blank and sentence-arranging exercises, using known crystal structures as reference. (2) \textit{Generative adversarial stage}: two sufficient trained models play opposing roles. One model endeavors to generate structures indistinguishable from real ones, while the other one attempts to distinguish between the generated and real ones. 
Beyond generative learning, our model also demonstrate the adaptability for down-stream regression and classification tasks through supervised fine-tuning (see Section~\ref{sec:fine-tuning}). \\

\textbf{Generation or reconstruction?} Our current model focuses on generating the optimal crystal structure under given crystallographic design---it may not yield the globally optimal structure if the specified conditions do not fall within the convex of global optima. It worth emphasizing that our models has never seen and been trained by the test samples. Therefore, if it performs well in ``reconstructing" a never-before-seen corrupted inputs, we would also expect it to perform well with any input—without the need to resemble any existing material structure. We demonstrate the models' reconstruction ability in Section~\ref{sec:analysis}.\\

\textbf{Comparison.} The prevailing popular models for crystal structure generation utilize diffusion process and variational autoencoders (VAEs). Diffusion models are typically applied to generate crystals from given composition, unit cell information, and null position embeddings. Our training method asymptotically approaches to a diffusion process as the noise introduced to the inputs goes to a large limit (requiring an additional modification to the loss function). Currently, our generative model is presently confined to scenarios where the basic design of the desired crystal structure is predefined. In other words, while the diffusion-based studies target the global optimal structures for certain compositions, our work focuses on identifying local optima within a defined framework. Nonetheless, our training strategy can be adapted for global optimization by introducing higher noise levels. VEAs follow a distinct approach: projecting crystal structures into a latent space through an encoder and then decoding it back to crystal structures, enabling material generation through sampling over the latent space.\\

\textbf{Probabilistic model.} Our generative model functions as a conditional probabilistic model, as it outputs the probability of specific atomic species at each masked site. Therefore, this model offers a probabilistic understanding into the nature of crystals in an interpretable manner, delivering richer information compared to other unconditional probabilistic models~\cite{hautier2011data, glawe2016optimal}. Details are provided in Section~\ref{subsec:novelty}. \\

In this Supplemental Material, we provide a comprehensive explanation of our simulations, covering (\ref{sec:pre-training}) self-supervised pre-training, (\ref{sec:fine-tuning}) supervised fine-tuning on regression and classification tasks, (\ref{sec:gan}) the application of a generative adversarial network to further improve the performance of our generative model, and (\ref{sec:analysis}) a comparative analysis of generative models focusing on metrics including validity, similarity and novelty.

\section{Self-supervised Pre-training}\label{sec:pre-training}

\inlinesubsection{Model setup}
Our study utilizes the most recent equivariant graph neural network with Transformer architecture, EquiformerV2~\cite{liao2023equiformerv2}, for our pre-trained model. 
Equivariance is a property of an operator $\Phi: X\rightarrow Y$ mapping between vector spaces $X$ and $Y$ by which it commutes with the group action of a group $(G,\circ)$, \textit{i.e.}, $\Phi \circ \rho^X(g) = \rho^Y(g) \circ \Phi$, where $\rho^X(g)$ is the group representation $\rho$ of $g \in G$ action on $X$. 
For our purposes, the equivariant graph neural network acts as the operator $\Phi$, being equivariant to E(3) group actions---3D translations, rotations and inversions---in processing 3D atomistic structures $X$ and $Y$. This allows the model to inherently capture symmetries in the data, eliminating the need for data augmentation. 
During the pre-training, we trained EquiformerV2 with $12$ Transformer blocks, $8$ attention heads, $118$ output channels, a maximum degree of $6$ and a maximum order of $2$. We used the Adam optimizer with a learning rate of $1\times 10^{-4}$. Batch sizes were adjusted through the training process: 1 for early epochs (below 25), 3 for mid-stage epochs (25-45), and 5 for later epochs (beyond 45). \\

\inlinesubsection{Dataset} 
We collect a dataset comprising $21{,}250$ diverse, stable crystal structures from the Materials Project \cite{jain2013commentary}, allocating $80\%$ for training and $10\%$ each for validation and testing. Our selection criteria prioritize samples with available electronic band structures and density of states to enhance the quality and reliability of the training data. 
The pre-train model receives input crystal structures undergoing augmentation through two operations: (1) with a $50\%$ probability, randomly selecting a species and masking all its atoms, and with a $50\%$  probability, masking $15\%$ of the atoms within a unit cell; (2) introducing random perturbations to the positions of all atoms. The perturbations follow a Gaussian distribution with a standard deviation of $\tilde{\sigma}_{\text{noise}}\times\min(\text{edges})$, where $\tilde{\sigma}_{\text{noise}}=0.1$ is set for the pre-training phase.
Our model is designed to reconstruct these corrupted crystal structure by revealing the masked atoms and accurately adjusting the positions of all atoms. \\

In our representation, the atomic species of $N$ atoms within a unit cell of a real crystal structure are represented as $Z \in \mathbb{R}^{N\times C}$. Here, $Z$ constitutes a set of one-hot vectors, each of length $C=118$, corresponding to the $N$ atoms in a unit cell. The perturbations applied to the real atomic positions in $D=3$ dimensional space are denoted as $P \in \mathbb{R}^{N\times D}$. 
Consequently, the output of our model for predicting atomic species is denoted as $\tilde{\mathcal{Z}}\in \mathbb{R}^{N\times C}$, which is the type-$0$ output of the equivariant neural network with $C$ output channels, and the output for positional perturbations is $\tilde{P} \in \mathbb{R}^{N\times D}$, which is the type-$1$ output of the equivariant graph neural network.\\

\begin{figure}[htb!]
\includegraphics[width=0.75\linewidth]{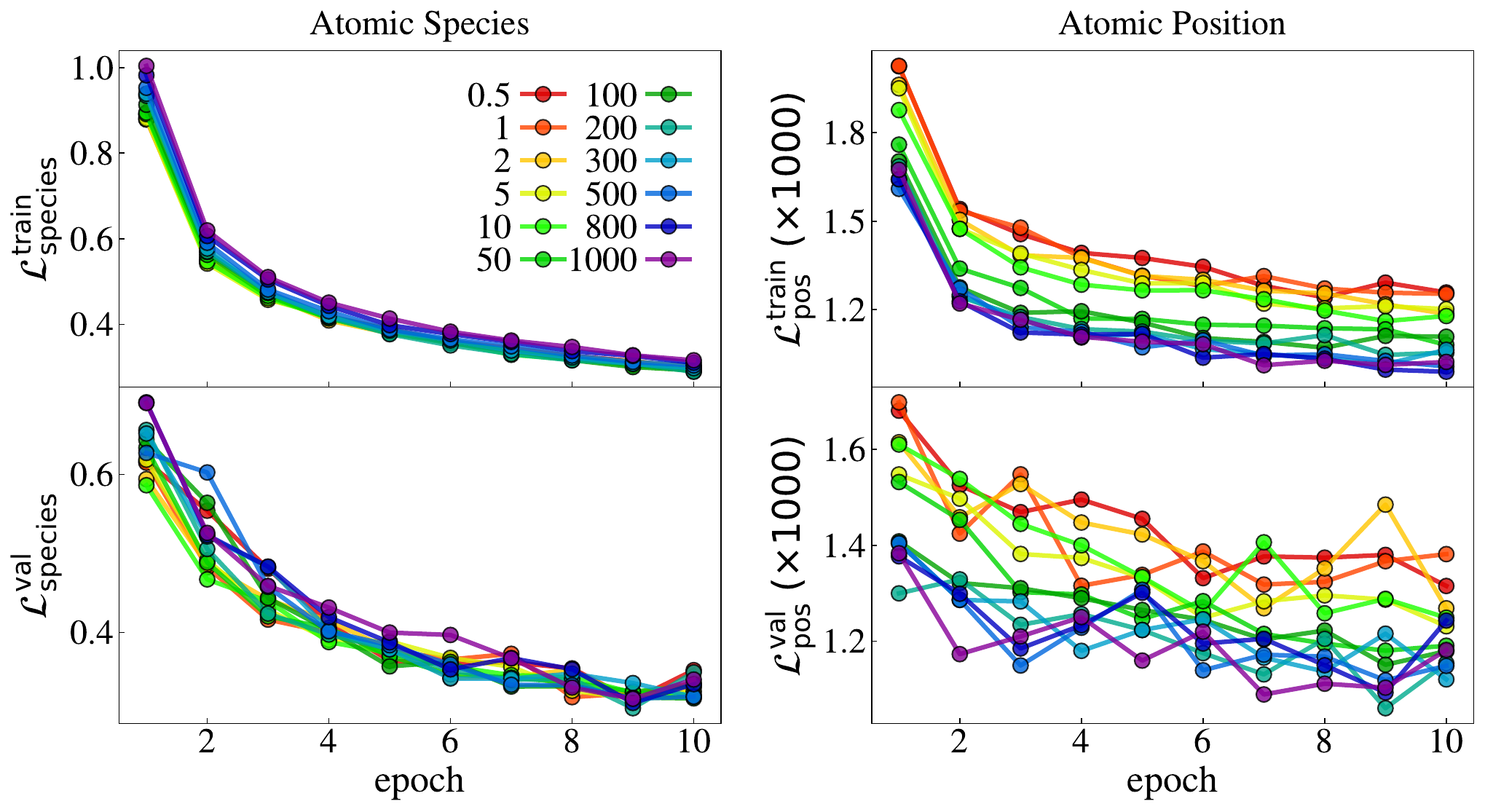}
\caption{\label{loss_pretrain_weights}
\textbf{Training and Validation Losses of Pre-trained Model in Relation to $w_{\text{pos}}$.} The loss is defined as $L_{\text{pretrain}} = L_{\text{species}} + w_{\text{pos}} \times L_{\text{pos}}$ in Eqn.~\ref{eq:loss_pretrain}.
}
\end{figure}

\begin{figure}[htb!]
\includegraphics[width=0.7\linewidth]{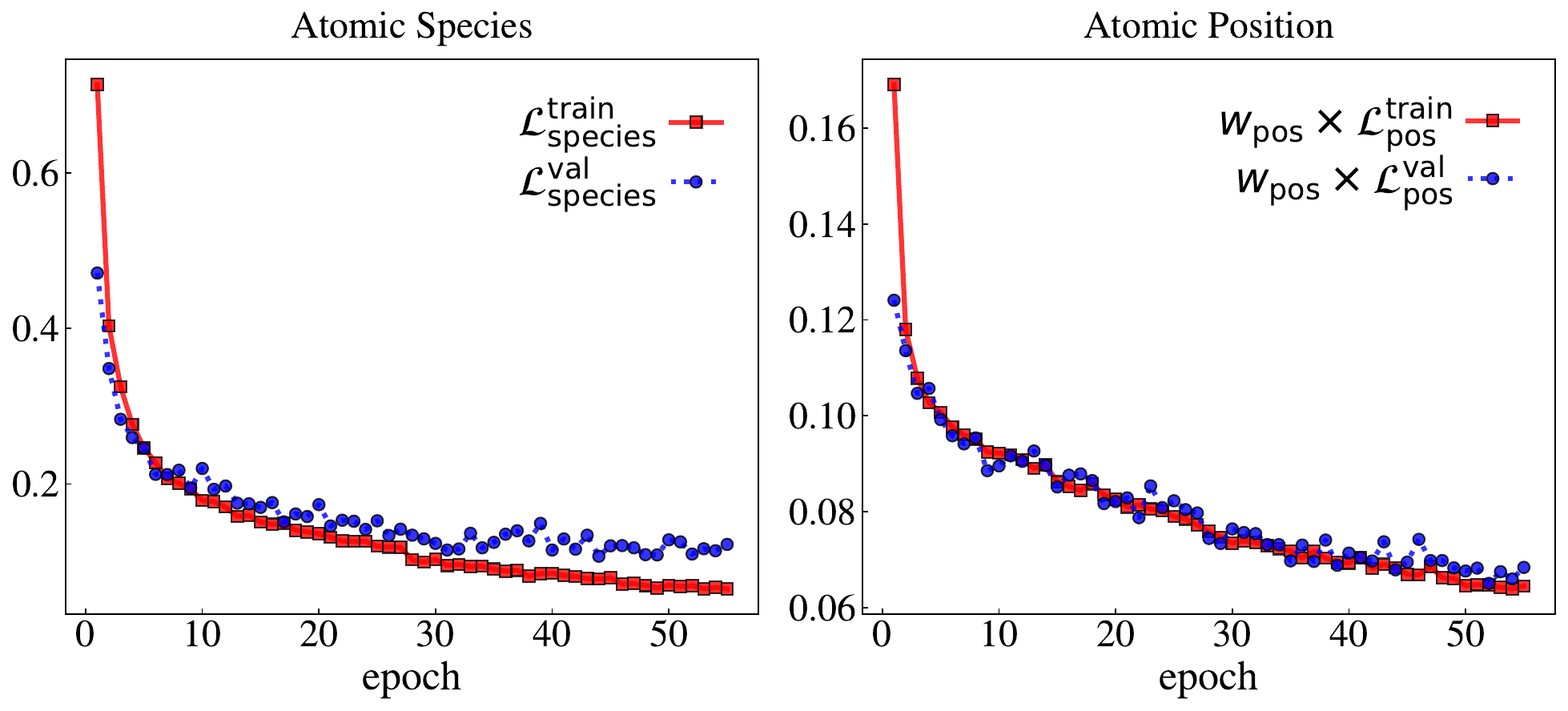}
\caption{\label{loss_pretrain}
\textbf{Pre-trained Model Losses.} The weighted average of $\mathcal{L}_{\text{species}}$ and $\mathcal{L}_{\text{pos}}$ for both training and validation stages, with each batch's loss scaled by the number of atoms in each batch. Training was done with the weight parameter $w_{\text{pos}}=200$.
}
\end{figure}

\inlinesubsection{Pre-training losses} 
To optimize the model weights for this task, the training loss for each sample is defined as 
\begin{equation} \label{eq:loss_pretrain}
\begin{aligned}
    & \mathcal{L}_{\text{pretrain}} = \mathcal{L}_{\text{species}} + w_{\text{pos}}  \times \mathcal{L}_{\text{pos}}, \\
    & \mathcal{L}_{\text{species}} = -\sum_{i=1}^{N} \sum_{j=1}^{C} Z_{ij}  \text{LogSoftmax}\big( \tilde{\mathcal{Z}}_{i} \big)_{j} = -\sum_{i=1}^{N} \sum_{j=1}^{C} Z_{ij} \log{\frac{\exp \tilde{\mathcal{Z}}_{ij}}{\sum_{k=1}^C \exp\tilde{\mathcal{Z}}_{ik}}}, \\
    & \mathcal{L}_{\text{pos}} = \frac{1}{N} \sum_{i=1}^{N} \sum_{j=1}^{D} (P_{ij} - \tilde{P}_{ij})^2. \\
\end{aligned}
\end{equation}

In Fig.~\ref{loss_pretrain_weights}, we investigated the impact of the weight $w_{\text{pos}}$ during pre-training. $\mathcal{L}_{\text{species}}$ remains unaffected by changes in $w_{\text{pos}}$, whereas $\mathcal{L}_{\text{pos}}$ decreases with an increase in $w_{\text{pos}}$ and reaches a plateau when $w_{\text{pos}}\approx 200$. Based on this observations, we selected $w_{\text{pos}}=200$ for our pre-training.  The pre-training losses, normalized by the number of atoms in each batch, is shown in Fig.~\ref{loss_pretrain}. \\

\section{Supervised Fine-tuning}\label{sec:fine-tuning}

\begin{figure}[htb!]
\includegraphics[width=1\linewidth]{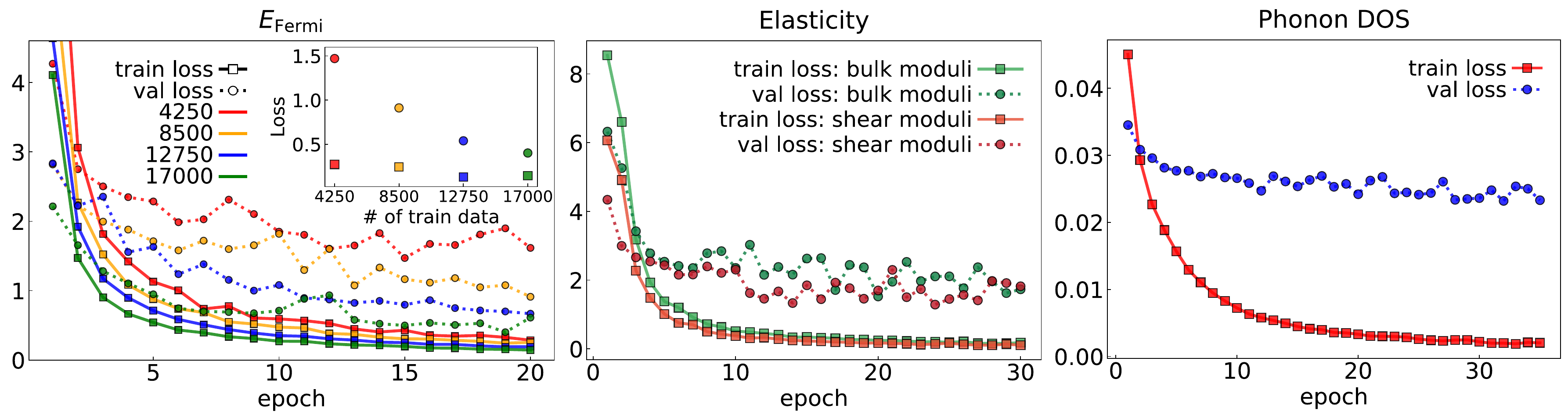}
\caption{\label{loss_regression}
\textbf{Regression Tasks Losses.}  Showcasing the adaptability and versatility of the pre-trained generative model in regression tasks through fine-tuning, including the prediction of Fermi energy, bulk and shear moduli, and phonon density-of-states.
}
\end{figure}

\begin{figure}[htb!]
\includegraphics[width=0.9\linewidth]{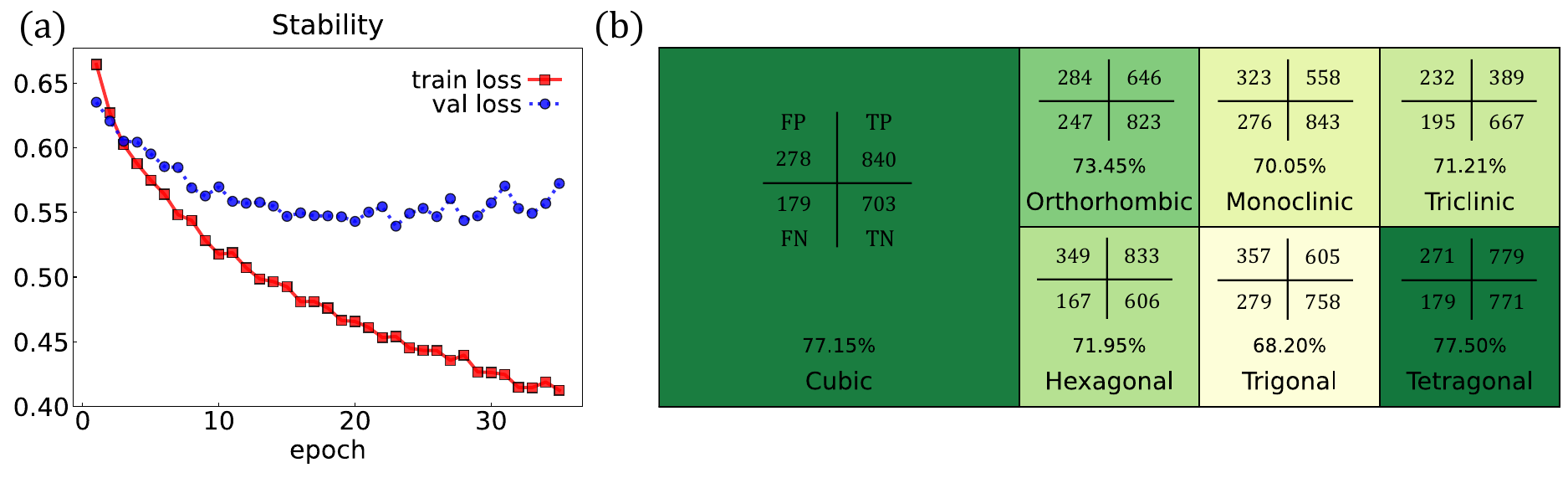}
\caption{\label{stability}
\textbf{Stability Classification Prediction Performance}.
(a) Training and validation losses and (b) testing accuracy of the fine-tuned model specialized in stability classification. The model was trained sorely on cubic crystal structures, and tested on various crystal structures. (b) displays classification results (True Positives, True Negatives, False Negatives, and False Positives) in a clockwise order, with accuracy indicated by percentage numbers and reflected by colors. 
}
\end{figure}

Through pre-training on a broad and diverse dataset, the model develops a fundamental understanding of the principles of materials composition and encodes this information in the model parameters. The pre-trained model can be more quickly and efficiently tailored for a variety of downstream tasks than a model trained from scratch. 
For a specific supervised task, the pre-trained model is merged with a feed-forward layer, which is randomly initialized and is designed to adapt the output of the pre-trained model to the requirement of the new task. During fine-tuning, both the pre-trained models and the feed-forward layer are trained on a task-specific dataset.\\ 

\inlinesubsection{Fine-tuning on regression tasks} 
For regression tasks, we evaluate our model on predicting (1) the Fermi energy, (2) bulk moduli and shear moduli, and (3) phonon density-of-states (DOS) based solely on the crystal structure, using Mean Squared Error (MSE) as the loss function. 
(1) For the prediction of Fermi energy, we utilize the same training set as used in the pre-training phase to prevent any information from the validation set being inadvertently exposed to the model. We further explore the influence of the training set size, depicted in Fig.~\ref{loss_regression}, reveals that the validation loss is significantly affected by the size of train data. 
(2) To investigate the behavior of fine-tuning on smaller datasets, we fine-tune the model for elasticity properties like bulk moduli and shear moduli with units of $\log{\text{GPa}}$. This fine-tuning is performed using a dataset comprising $1915$ training samples and $638$ validation samples from Materials Project. 
(3) The pre-trained model is originally trained by predicting atomic displacement in crystal structures, which may capture how local structural perturbations influence the vibration properties of materials, and establishing a direct correlation between structural changes and macroscopic phononic behavior. Therefore, we leverage our pre-trained model to predict the phonon DOS, as shown in Fig.~\ref{loss_regression}, with $1{,}524$ training samples and $304$ validation samples from \cite{li2021computation}. \\

\inlinesubsection{Fine-tuning on classification tasks}
For classification tasks, the loss function is Negative Log-Likelihood Loss (NLLLoss) applied after LogSoftmax, as $\mathcal{L}_{\text{species}}$ in Eqn.~\ref{eq:loss_pretrain}. 
To assess our pre-trained model's capability in classifying crystal structures, we naturally focus on identifying the stability of crystal structures. We fine-tune the model by using $4{,}017$ stable and $4{,}065$ unstable cubic crystal structures from Materials Project. It is worth noting this model was initially pre-trained exclusively on stable crystal structures. 
For evaluation, we utilize $2{,}000$ samples from each crystal types, except for triclinic crystal structures, which have only $1{,}483$ samples. These samples are nearly evenly split between stable and unstable structures. The accuracy of stability prediction is shown in Fig.~\ref{stability}. Remarkably, although the model's fine-tuning was confined to cubic crystals, it demonstrates extrapolative capabilities to other types of crystal structures.\\

\section{Generative Adversarial Network}\label{sec:gan}

\begin{figure}[htb!]
\includegraphics[width=0.8\linewidth]{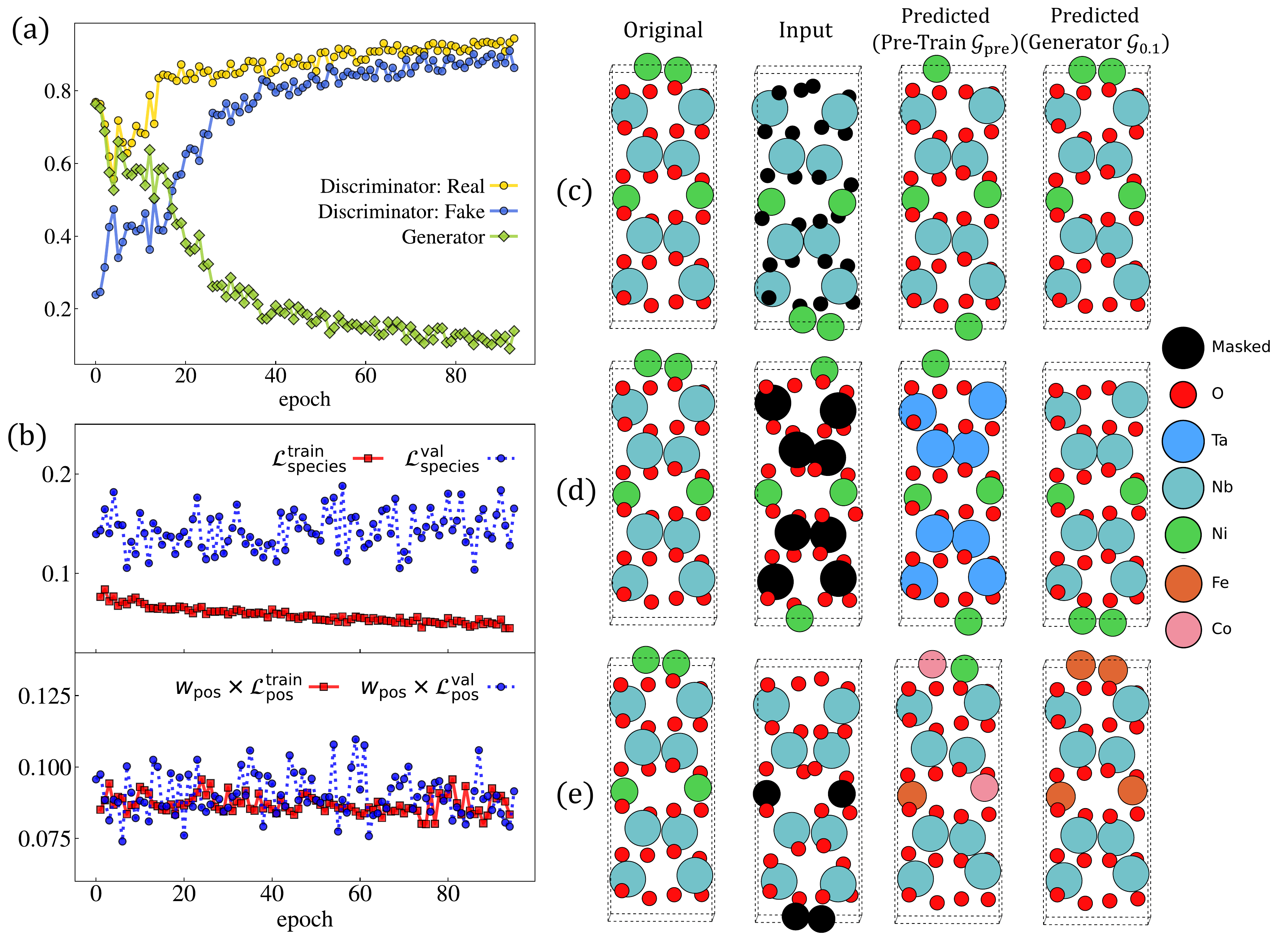}
\caption{\label{gan_sigma0.1}
\textbf{GAN Performance with $\tilde{\sigma}_{\text{noise}}=0.1$}. 
(a) shows the validation accuracy of discriminator and generator trained with input structures perturbed by noise normally distributed with standard deviation of $0.1\times\min(\text{edges})$. (b) illustrates losses of the generator, defined as $L^{\text{G}}_{\text{total}} = 0.1 \times L^{\text{G}}_{\text{label}} + L^{\text{G}}_{\text{atom}} + 200 \times L^{\text{G}}_{\text{pos}}$. (c-e) demonstrate the ability of this generator, comparing with the pre-trained model, to reconstruct the original crystal structure from inputs with masked atoms and perturbed positions, and potentially discover novel crystal structures.
}
\end{figure}

\begin{figure}[htb!]
\includegraphics[width=0.8\linewidth]{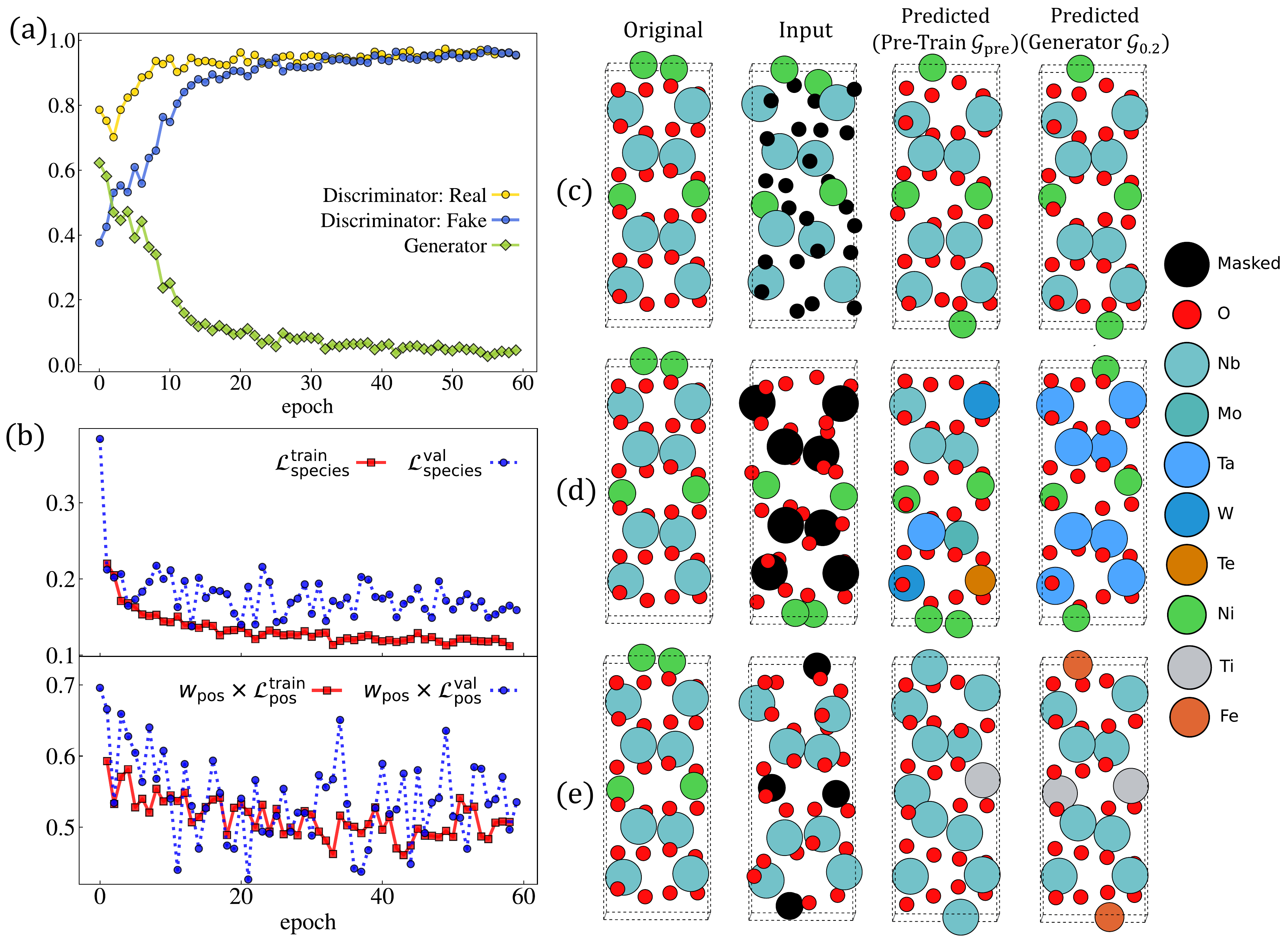}
\caption{\label{gan_sigma0.2}
\textbf{GAN Performance with $\tilde{\sigma}_{\text{noise}}=0.2$}. This figure extends the analysis of Fig.~\ref{gan_sigma0.1}, focusing on the impact of increased positional noise in the training, validation, and testing inputs for GAN. For context, the pre-trained generative model for comparison was trained with a lower positional noise level $\tilde{\sigma}_{\text{noise}}=0.1$.
}
\end{figure}

The objective of our generative model extends beyond merely reconstruct the original crystal structure from its contaminated counterpart, while also exploring novel crystal structures that have not been discovered before. 
The challenge, however, lies in the limitation of the loss function, which can only optimize the reconstruction goals, and the absence of a tool for discriminating generated structures.\\ 

We consider the framework of actor-critic learning to guide the generative training. Based on this framework, we have two options: self-supervised generative adversarial networks (GANs) or incorporating human pre-knowledge of crystal structures as a critic. Compared to GANs, the later one, such as utilizing density functional theory (DFT), are computationally expensive, with duration ranging from a few minutes to several hours for each crystal graph. Furthermore, the critic neural network model supervised by substantial DFT data has been sufficiently trained in previous studies~\cite{merchant2023scaling}. Such well-trained critic model could serve as a valuable tool for the future enhancement of our generative model. \\

Another concern involves utilizing generative training to explore how much information is embedded in the crystal structures. This topic is analogous to examining whether a child, without any formal teaching on recognizing sentiments, can discern the sentimental tones in a sentence after extensive reading. To ensure rigorous control over variables, we tend to isolate the teaching from humans. \\

\inlinesubsection{Model setup} For the above reasons, we finally select self-supervised GANs. In this setup, the discriminative neural network is trained to distinguish the crystal structures produced by the generative model from the real crystal structures. Concurrently, the generative neural network is optimized to deceive the discriminator. We initiated the generator with pre-trained model parameters and used the fine-tuning model parameters for stability classification in downstream tasks as the discriminator's starting point, which is chosen because this task shares some commonalities with the discriminator's role, despite their tasks not being entirely aligned. \\

\inlinesubsection{GAN training losses} Having found that the discriminator tends to outperform generator rapidly while using cross entropy as the generator loss, and leads to gradient vanishing in our tasks, we use Wasserstein distance as an alternative:
\begin{equation}
\begin{aligned}
\mathcal{L}_{\text{Wasserstein}} = \frac{1}{M}\sum_{m=1}^{M}\sum_{k=1}^{K} \abs{y_{k}^{(m)} - \tilde{y}_{k}^{(m)}},
\end{aligned}
\end{equation}
where $M$ is the number of samples in each batch and $K=2$ for binary classification. The Wasserstein loss for the generator $\hat{G}: \mathbb{R}^{N\times (C+D)} \rightarrow \mathbb{R}^{N\times (C+D)}$ is formulated as
\begin{equation}
\begin{aligned}
\mathcal{L}^G_{\text{Wasserstein}} = \frac{1}{M}\sum_{m=1}^{M}\sum_{k=1}^{2} \Bigl|y_{k}^{(m)}- \text{LogSoftmax}\Big(\hat{D}\big(\hat{G}(X^{(m)})\big)\Big)_{k}\Bigr|,
\end{aligned}
\end{equation}
with $y^{(m)} = (0,1)$ indicating the label vector for each sample $m$. The input of the generator, $X^{(m)}$, consists of masked atomic species $\tilde{\mathcal{Z}}^{(m)} \in \mathbb{R}^{N\times C}$ and perturbed positions $\tilde{P}^{(m)} \in \mathbb{R}^{N\times D}$ augmented based on the $m$-th crystal structure. 

On the other hand, for the Wasserstein loss in the discriminator $\hat{D}: \mathbb{R}^{N\times (C+D)} \rightarrow \mathbb{R}^{2}$, $y^{(m)}$ is set to $(1,0)$ for real structure inputs, \textit{i.e.}, $\tilde{y}^{(m)}=\text{LogSoftmax}\left(\hat{D}\left(X^{(m)}\right)\right)$; $y^{(m)}=(0,1)$ for inputs from the generated structures, \textit{i.e.}, $\tilde{y}^{(m)}=\text{LogSoftmax}\left(\hat{D}\left(\hat{G}(X^{(m)})\right)\right)$. When training the generator, we keep the parameters of the discriminator frozen, and vice versa.\\

To prevent the generator from producing features specifically tailored to the weakness of the discriminator, rather than learning to produce the realistic crystal structures, we incorporate $\mathcal{L}_{\text{species}}$ and $\mathcal{L}_{\text{pos}}$, as specified Eqn.~\ref{eq:loss_pretrain}, into the generator loss:
\begin{equation}
\begin{aligned}
\mathcal{L}^G = w_{\text{W}} \times \mathcal{L}^G_{\text{Wasserstein}} + \mathcal{L}_{\text{species}} + w_{\text{pos}} \times \mathcal{L}_{\text{pos}}.
\end{aligned}
\end{equation}
Note that we use $w_{\text{pos}}=200$ as for pre-training and GAN. The training of GAN is done with $w_{\text{W}}=0.1$, though both $w_{\text{W}}=1$ and $0.1$ have been tested with no significant difference observed.\\

Modifications to the training procedure are also made to alleviate the issue of the discriminator outperforming the generator, such as reducing the learning rate of the discriminator (lr$_{G}=10^{-4}$ and lr$_{D}=10^{-5}$), positional noise adding to the real crystal structures (the noise follows a normal distribution with a standard deviation of $0.001\times \min(\text{edges})$), and adjusting the training schedule to update the generator more frequently than the discriminator.\\

\inlinesubsection{Results} In the generative adversarial stage, both the training and validation datasets employed are identical to those used in the pre-training, and the same level of positional noise is introduced to the input structures, \textit{i.e.}, $\sigma_{\text{noise}}=0.1\times\min(\text{edges})$, as shown in Fig.~\ref{gan_sigma0.1}. 
Despite observing minimal reduction in the loss $\mathcal{L}_{\text{species}}$ and $\mathcal{L}_{\text{pos}}$ across increasing training epochs, the GAN achieves a better performance in generating physically reasonable crystal structures, at least visually, compared to the pre-train model. \\

To make our model compatible with larger displacements, we experimented positional noise with $\sigma_{\text{noise}}=0.2\times\min(\text{edges})$ during GAN training in Fig.~\ref{gan_sigma0.2}. This adjustment led to notable decreases in $\mathcal{L}_{\text{species}}$ and $\mathcal{L}_{\text{pos}}$. We also evaluated the performance of the pre-train model (trained with noise level $\sigma_{\text{noise}}=0.1\times\min(\text{edges})$) and the generator (trained with noise level $\sigma_{\text{noise}}=0.2\times\min(\text{edges})$) on reconstructing structures perturbed with noise at $\sigma_{\text{noise}}=0.2\times\min(\text{edges})$. As illustrated in Fig.~\ref{gan_sigma0.2}(c), the pre-train model shows a decline in predicting atomic species, while the GAN generator exhibited an improved performance in handling the structures with a higher noise level. \\

\begin{figure}[htb!]
\includegraphics[width=1\linewidth]{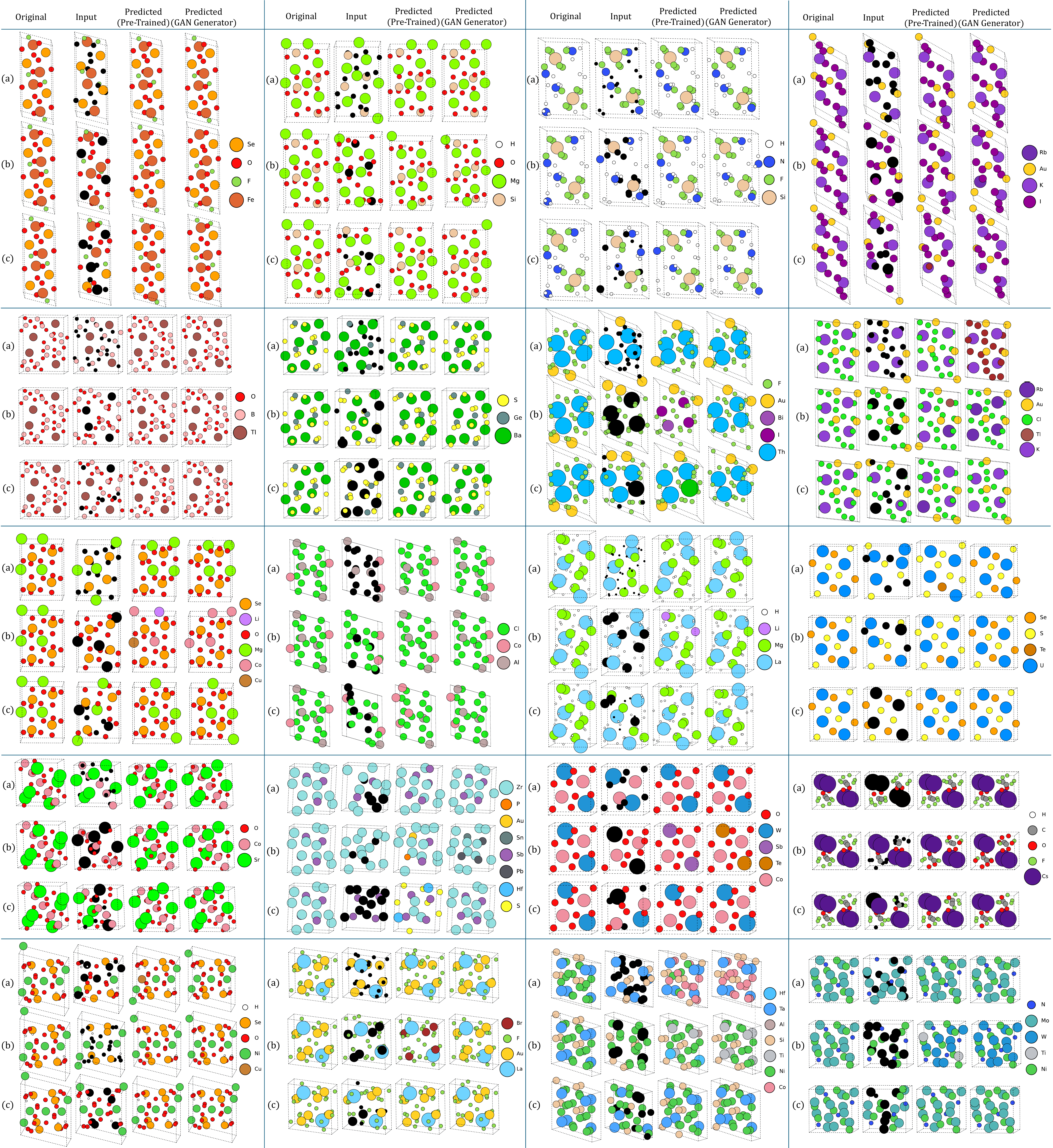}
\caption{\label{more_generative_results}
\textbf{Generative Capability of the Pre-Trained Model and GAN Generator.} 
More examples to demonstrate the capability of the pre-train model and the generator from GAN to reconstruct or generate crystal structures. The inputs involve masked atomic species, with masks applied to (a) frequent occurring species, (b) infrequent occurring species, and (c) a random section of $30\%$ of atoms, along with positions perturbed by noise following a normal distribution with a standard deviation of $\tilde{\sigma}_{\text{noise}}\times\min(\text{edges})$, where $\tilde{\sigma}_{\text{noise}}=0.2$. Note that the pre-trained generative model is trained with a positional noise level of $\tilde{\sigma}_{\text{noise}}=0.1$, while the GAN generator with $\tilde{\sigma}_{\text{noise}}=0.2$.
}
\end{figure}

In addition to the main text, we provide additional visual illustrations to demonstrate the generative capabilities of our models, as shown in Fig.~\ref{more_generative_results}.

\section{Comparative Analysis of Generative Capability} \label{sec:analysis}

\begin{figure}[htb!]
\includegraphics[width=0.8\linewidth]{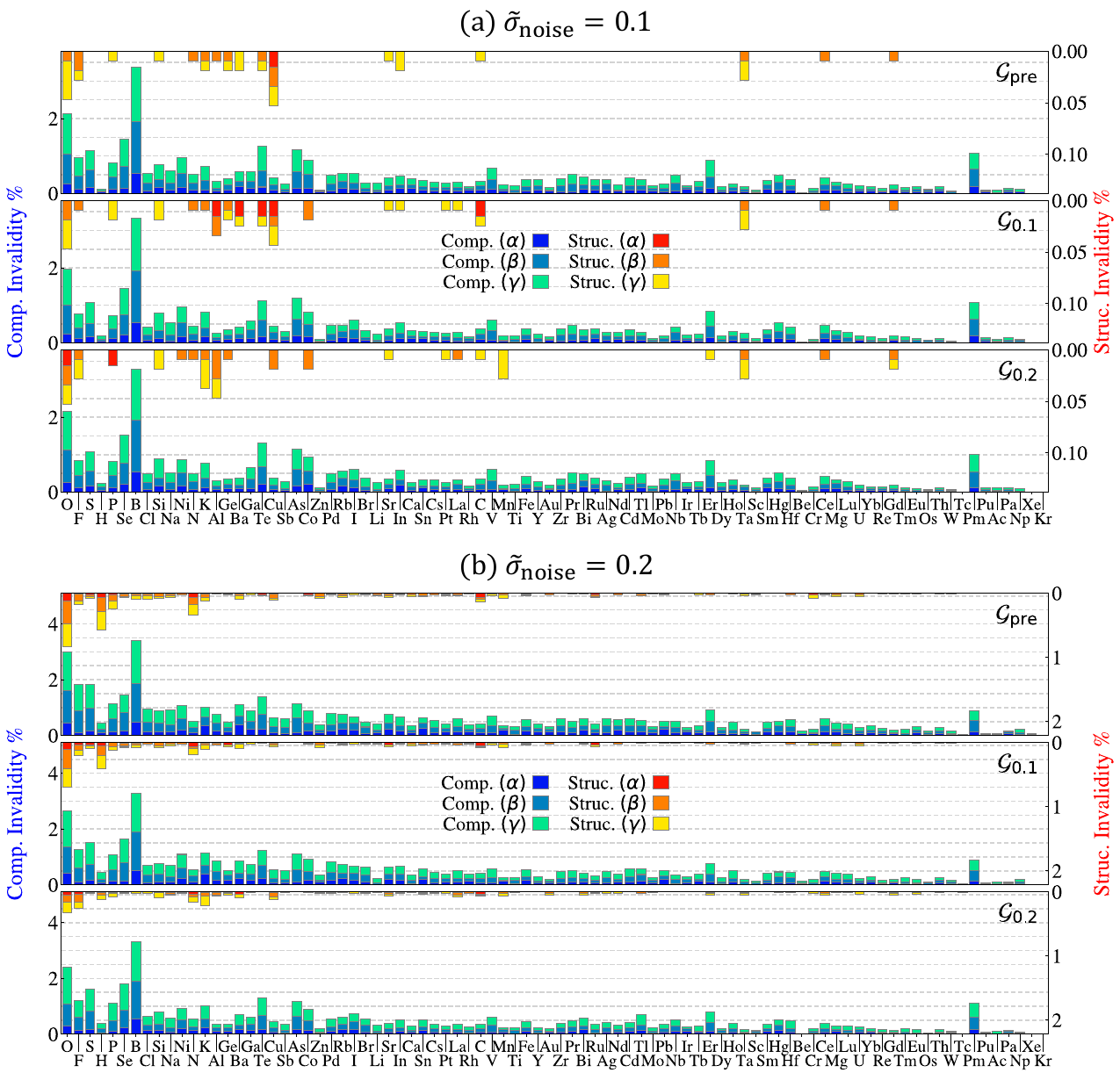}
\caption{\label{invalidity}
\textbf{Species Occurrence in Invalid Structures Generated by Generative Models.}
Here, we display the number of times each species occurs in invalid structures, assessed using Composition and Structure Validity calculations, generated by the pre-train generative model ($\mathcal{G}_{\text{pre}}$) and GAN generators ($\mathcal{G}_{0.1}$ and $\mathcal{G}_{0.2}$). The vertical axis on the left represents Composition Invalidity in percentage. On the right, the vertical axis indicates Structure Invalidity in percentage, increasing downwards. Species are arranged based on their abundance in the training set, correlating with the data distribution shown in Fig.~\ref{data_distribution}. 
Three mask types are applied to the inputs: ($\alpha$) masking all atoms of a specific species within each crystal structure, repeated for each species; ($\beta$) randomly selecting and masking $15\%$ of atoms, repeated five times per structure; ($\gamma$) a similar approach with $30\%$ of atoms. Perturbations to the positions of input crystal structures follows a normal distribution with standard deviation of $\tilde{\sigma}_{\text{noise}}\times\min(\text{edges})$. 
Evaluation is performed on a test set augmented from $2$k samples to approximately $6.5$k for mask type ($\alpha$) and $10$k for mask type ($\beta$) and ($\gamma$).
}
\end{figure}

\begin{figure}[htb!]
\includegraphics[width=0.8\linewidth]{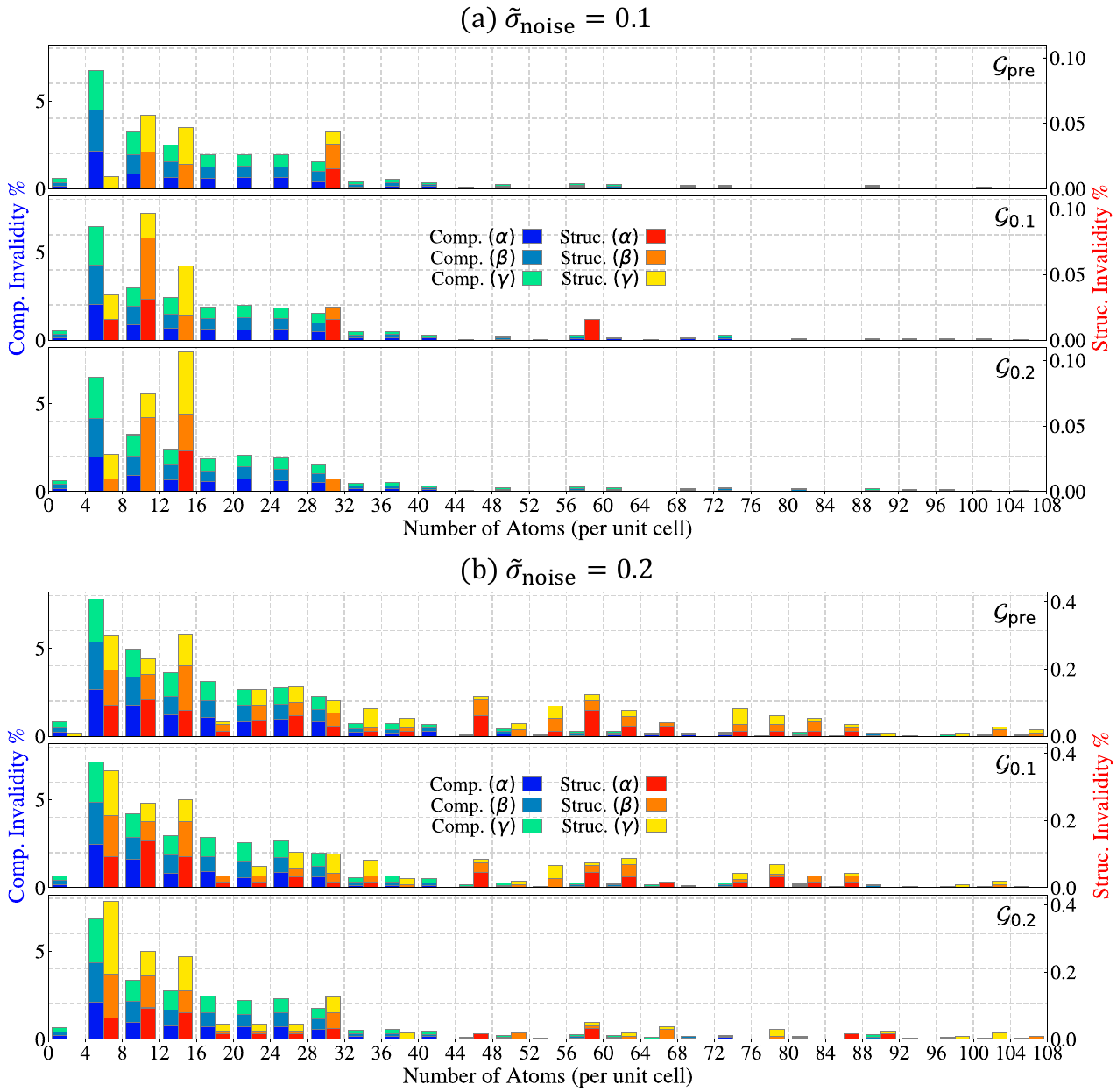}
\caption{\label{invalidity_vs_natoms}
\textbf{Invalidity Versus Number of Atoms (per unit cell).}
Panels (a-b) illustrate the occurrence of the number of atoms of the invalid crystal structures, assessed using Composition and Structure Validity calculations, generated by the pre-train generative model ($\mathcal{G}_{\text{pre}}$) and GAN generators ($\mathcal{G}_{0.1}$ and $\mathcal{G}_{0.2}$). 
The left vertical axis denotes Composition Invalidity in percentage, and the right axis denotes Structure Invalidity in percentage. The same augmented test set as Fig.~\ref{invalidity} is used. 
}
\end{figure}

\inlinesubsection{Validity} To quantitatively evaluate the crystal structures generated by our generative models, we present analysis of \textit{structural validity}, which depends on distance between atoms, and \textit{composition validity}, based on the overall charge~\cite{xie2021crystal}. The results are shown in the main text. Here, we further analyze the generated structures that do not pass the validity tests by analyzing the number of atoms per unit cell and the elements in the invalid structures in Fig.~\ref{invalidity} and Fig.~\ref{invalidity_vs_natoms}. These figures infer the following arguments:

\begin{enumerate}
    \item When the noise level in the testing inputs matches that of the training inputs, the GAN generator, though not outperforming the pre-trained model in constructing structurally valid crystals, demonstrates enhanced robustness in generating compositionally valid structures. Despite all generative models being trained on inputs with only $15\%$ of atoms masked, the GAN approach improves extrapolation capability, especially in scenarios where $30\%$ atoms are randomly masked (mask type $\gamma$).
    \item When the models are tested with a higher noise level, as depicted in Fig.~\ref{invalidity}b and \ref{invalidity_vs_natoms}b, the GAN generators surpasses the pre-trained in creating compositionally and structurally valid crystals, especially obtaining a better performance when the number of atoms is large.
    \item Interestingly, the GAN generator trained with a higher noise level ($\mathcal{G}_{0.2}$) fails to achieve the highest validity on the the test set with a lower noise level ($\tilde{\sigma}_{\text{noise}}=0.1$). This observation suggests that diversifying the test set by including variations in $\tilde{\sigma}_{\text{noise}}$ may enhance the model's versatility. 
    \item Figure~\ref{invalidity_vs_natoms} depicts the relationship between the crystal structures that models fail to successfully generate and the number of atoms (per unit cell) within these structures. The similarity in the distribution of the number of atoms in compositionally invalid structures to that in original structures (Fig.~\ref{data_distribution}b) suggests a weak correlation between composition validity and the number of atoms. In contrast, the distribution for structure invalidity deviates from that observed in realistic structures. Although it is not perfectly reliable to draw statistically significant conclusions for rigorous analysis from the structure invalidity distribution, due to the relatively small number (approximately $10$ per set) of invalid structures generated by our models in each augmented dataset, the trends observed still demonstrate the GAN generator's enhanced capability in structure validity with increased $\tilde{\sigma}_{\text{noise}}$.
    \item The pattern of invalidity distribution provides insights for future model improvements, for example, by training with crystal structures that are similar to those fall within the lower validity ranges.
\end{enumerate}

\begin{figure}[htb!]
\includegraphics[width=0.9\linewidth]{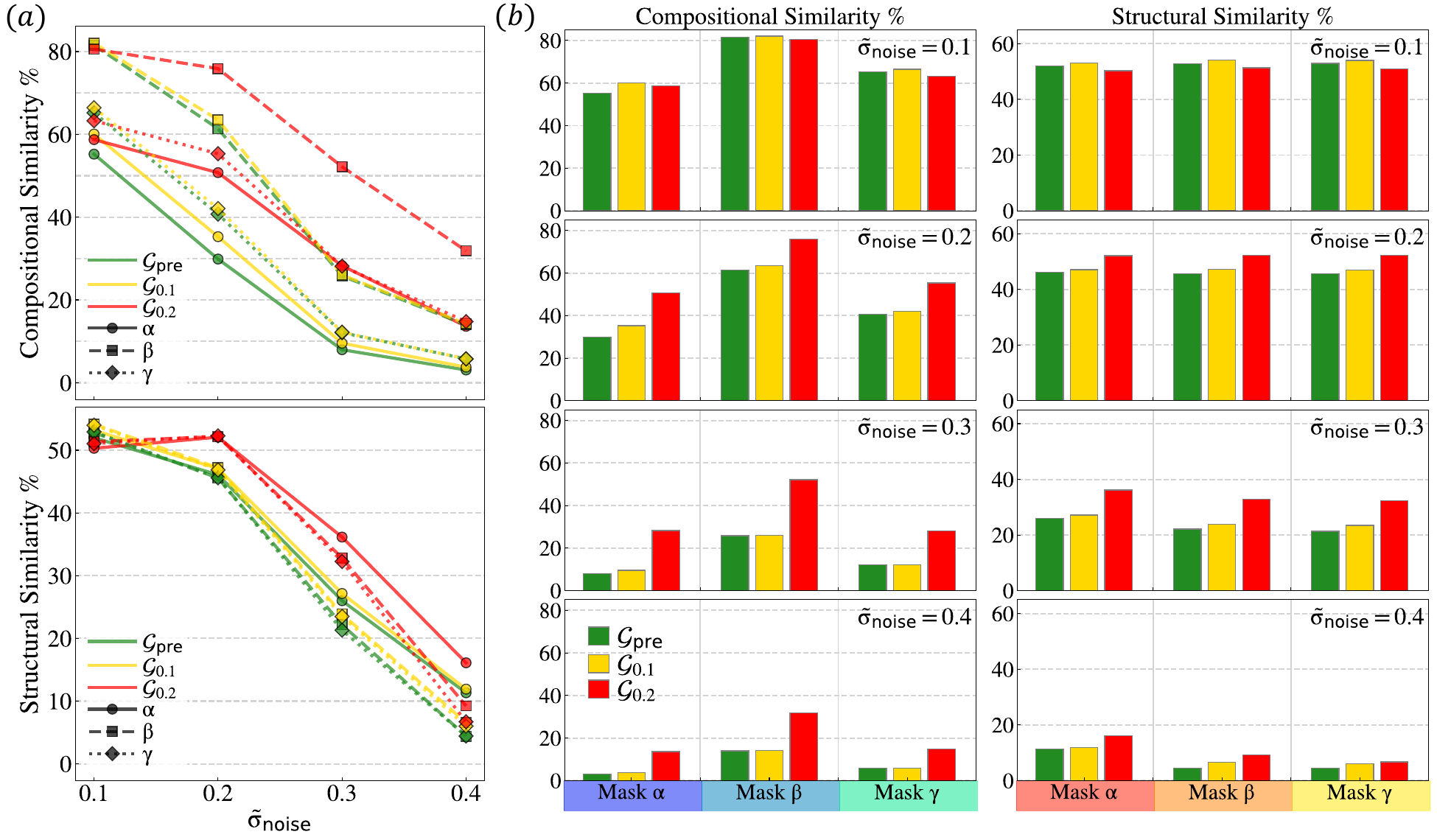}
\caption{\label{similarity}
\textbf{Compositional and Structural Similarity.} We show the compositional and structural similarity, expressed as percentages, for generative models ($\mathcal{G_{\text{pre}}}, \mathcal{G}_{0.1}, \mathcal{G}_{0.2}$). These models are evaluated using test samples corrupted with various mask types ($\alpha, \beta, \gamma$) and positional noise levels ($\tilde{\sigma}_{\text{noise}}=0.1, 0.2, 0.3, 0.4$) introduced to the test samples. Panels (a) and (b) present complementary perspectives on the data: (a) displays the distributions of similarities across noise levels, whereas (b) focuses on direct comparisons of similarity metrics within identical noise level. Measurements are done with in total $\sim 28$k corrupted samples.
}
\end{figure}

\begin{figure}[htb!]
\includegraphics[width=1\linewidth]{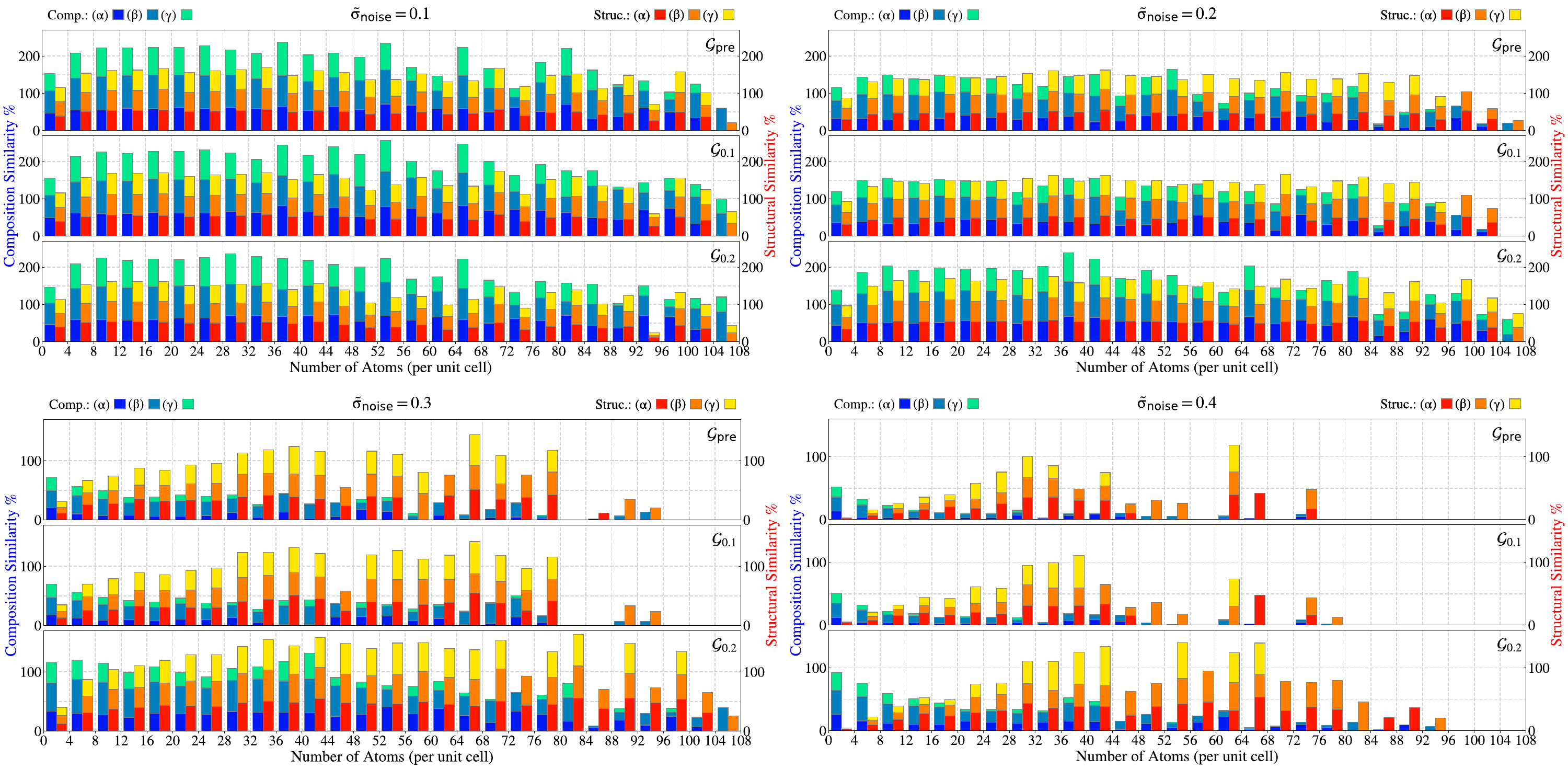}
\caption{\label{similarity_vs_natoms}
\textbf{Similarities Versus Number of Atoms (per unit cell).} This figure examines how the number of atoms per unit cell within test crystal structures affects the compositional similarity (shown as a blue series on the left axis) and structural similarity (shown as a red series on the right axis). The effects of different mask type ($\alpha, \beta, \gamma$) and positional noise levels ($\tilde{\sigma}_{\text{noise}} = 0.1, 0.2, 0.3, 0.4$) on the test samples are depicted. Each histogram bar represents the average similarity observed across test samples with a specific number of atoms.
}
\end{figure}

\begin{figure}[htb!]
\includegraphics[width=1\linewidth]{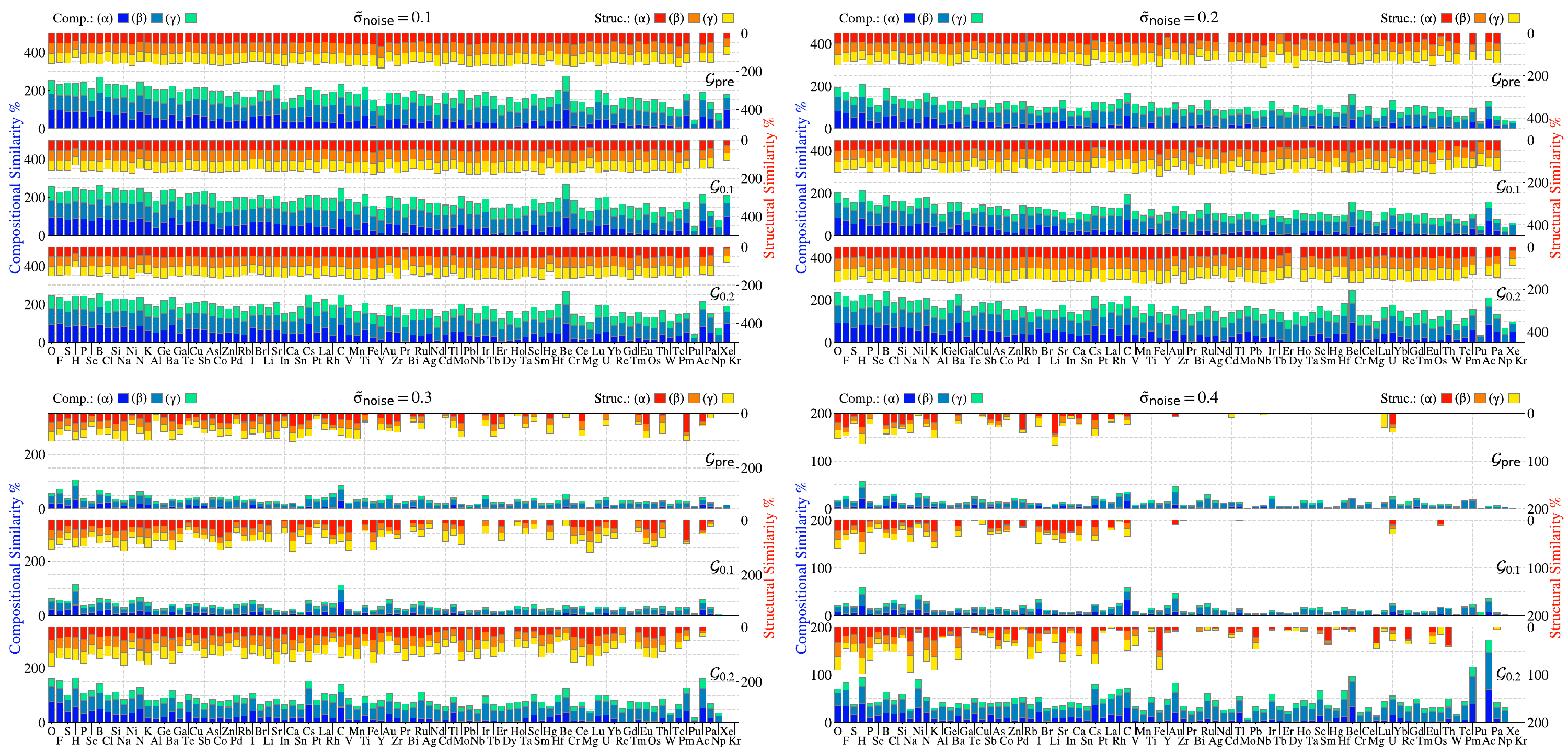}
\caption{\label{similarity_vs_mspecie}
\textbf{Similarities Versus Masked Atomic Species.} This figure explores the impact of masking specific atomic species on compositional similarity (blue histograms with bottom-to-top axis on the left) and structural similarity (red histograms with top-to-bottom axis on the right). Each histogram bar is calculated with evaluations conducted on samples that had certain atomic species masked. 
}
\end{figure}

\inlinesubsection{Similarity} 
We introduce two metrics designed to evaluate the effectiveness of generative models in accurately reproducing the properties of ground truth materials. These metrics, namely compositional similarity and structural similarity, quantify the accuracy with which the models replicate material structures in alignment with their original counterparts.\\

\textit{Compositional similarity} is defined as the average percentage match between the generated and original material compositions across the test set:
\begin{equation}
    \text{Compositional Similarity} = \frac{1}{|\text{test set}|} \sum_{m=1}^{|\text{test set}|} \delta(Z^{(m)}-\tilde{Z}^{(m)}),
\end{equation}
Here $\tilde{Z}^{(m)}_{ij}=\delta \left(j-\arg \max_k \left(\text{LogSoftmax} \left(\tilde{\mathcal{Z}}^{(m)}_{i}\right)_k \right) \right)$, where $\tilde{\mathcal{Z}}^{(m)}\in \mathbb{R}^{N\times C}$ denotes the compositional output from the generative model, as shown in Eqn.~\ref{eq:loss_pretrain}. This metric provides insight into the model's ability to accurately restore the composition of materials. However, it is important to note that while high compositional similarity indicates model accuracy, it is contradictory to assessing its capability to innovate novel structures. 

It is important to clarify that, for computational simplicity, a generated structure is considered compositionally identical to the original structure only if $Z_{i}=\tilde{Z}_{i}$ for all $i$. This definition, however, overlooks the scenario where two atoms of different species ($Z_{i}\neq Z_{j}$) may have their locations exchanged due to the large positional noise. Such an exchange would lead the generative models to swap their species and positions, resulting in $\tilde{Z}_{i} = Z_{j}$, $\tilde{Z}_{j} = Z_{i}$, $\tilde{P}_{i} = P_{j}$, and $\tilde{P}_{j} = P_{i}$. In this case, the generated structure should still be considered equivalent to the original counterpart. When noise level is sufficiently low, the likelihood of this scenario occurring is minimal, but as our analysis in Fig.~\ref{similarity} includes large noise levels, it is conceivable that these conditions may induce inaccuracies in assessing compositional similarity.\\

\textit{Structural similarity}, on the other hand, aims to evaluate the model's ability in replicating the structural integrity when the model decides to generate compositions identical to the originals. It is defined as:
\begin{equation}
    \text{Structural Similarity} = 1- \frac{1}{|\text{test set}|} \sum_{m=1}^{|\text{test set}|} \delta(Z^{(m)}-\tilde{Z}^{(m)}) \left( \frac{1}{N_m}\sum_{i=1}^{N_m} \sqrt{\sum_{j=1}^{D} \left(P^{(m)}_{ij}-\tilde{P}^{(m)}_{ij} \right)^2} \right)/S^{(m)}.
\end{equation}
Here, the equation considers the Euclidean distance between the positions in the original structure ($P^{(m)}\in \mathbb{R}^{N\times D}$) and generated structure ($\tilde{P}^{(m)}\in \mathbb{R}^{N\times D}$) for the $m$-th test sample, normalized by the average positional deviation introduced to the input ($S^{(m)}\in \mathbb{R}$). A higher structural similarity score, approaching 1, indicates a closer match to the original material's structural positioning, offering a direct measure of the model performance in restoring material structure. \\

To facilitate an evaluation of model performance on various corrupted test samples, we illustrate the distribution of the compositional and structural similarity across noise levels in Fig.~\ref{similarity}a. We further conduct a comparative analysis of the performances of three generative models within the same test dataset in Fig.~\ref{similarity}b. From these analyses, we can infer the following key insights:
\begin{enumerate}
    \item The distribution of compositional and structural similarities approximates a Gaussian distribution, with the mean values locating at the standard deviation of the noise distribution introduced to the training samples, $\sigma_{\text{noise}}=\tilde{\sigma}_{\text{noise}}\times \min{(\text{edges})}$.
    \item GAN generators perform better across all test sets when compared to the pre-trained model $\mathcal{G}_{\text{pre}}$.
    \item When handling the test samples with mask type $\beta$ (randomly selecting and masking $15\%$ of atoms), generative models are more inclined to replicate the original crystals. When the model has decided to restore the originals, their performances on restoring the atoms back to the original positions are almost unaffected by the specific mask types used in the test samples.
    \item As the noise level of the test set increases to $\tilde{\sigma}_{\text{noise}}=0.4$, the models quickly fail to revert the compositions and structures back to their originals. Our model serves as a mechanism to find the local minima in incomplete crystal structure, rather than finding a global minimum. At sufficiently high noise levels, the input structure is pushed into the other local minima, thus making the model unlikely to reconstruct a structure that matches the ground truth exactly.
    \item $\mathcal{G}_{0.2}$ uses a training set with $\tilde{\sigma}_{\text{noise}}=0.2$ during GAN training, but since it utilizes the model parameters of the pre-trained model $\mathcal{G}_{\text{pre}}$, hence it has been effectively exposed to training set with different noise levels. This exposure helps the model perform better on training sets with higher noise levels without losing performance on the training set at the lower noise levels. It suggests that sampling the standard deviation of noise distribution of the training set can enable the model to handle test samples with different noise levels, potentially improving its accuracy in finding both local and global minima in the crystal structures.
\end{enumerate}

Figure ~\ref{similarity_vs_natoms} and~\ref{similarity_vs_mspecie} visualize the effects of the size of crystals and the species that the models should unmask on models' performance. The insights drawn from these figures are summarized as follows:
\begin{enumerate}
    \item The generative models demonstrate increased difficulty in replicating the compositions of the crystals that have a large number of atoms and contain species that are less frequent in the test set. 
    \item Despite the non-uniform distribution of the size of crystals and species abundance within the datasets (as shown in Fig.~\ref{data_distribution}), the structural similarity scores for all models are marginally affected by the number of atoms and the types of species masked at low noise levels. 
    \item At high noise levels, the improvements of $\mathcal{G}_{0.2}$ are evident in its enhanced capability in reconstructing crystal structures compared to other generative models, assuming our original crystal structures represent the global optima. This improvement is particularly noticeable when tested with a crystal having a larger number of atoms and less common species being masked.
\end{enumerate}

In the table below, we present a summary of the \textbf{validity} and \textbf{similarity} scores, highlighting the highest score achieved in each test set, in which samples have been masked and perturbed in different ways:

\begin{center}
\begin{tabular}{|c|c|c|c|c|c|c|c|c|c|c|c|c|c|}
    \hline
    \multicolumn{2}{|c|}{\text{Test Input}} & \multicolumn{3}{c|}{\text{Comp. Validity \%}} & \multicolumn{3}{|c|}{\text{Struc. Validity \%}} & \multicolumn{3}{c|}{\text{Comp. Similarity \%}} & \multicolumn{3}{|c|}{\text{Struc. Similarity \%}} \\
    \hline
    \text{Mask} & $\tilde{\sigma}_{\text{noise}}$ & $\mathcal{G_{\text{pre}}}$ & $\mathcal{G}_{0.1}$ & $\mathcal{G}_{0.2}$ & $\mathcal{G_{\text{pre}}}$ & $\mathcal{G}_{0.1}$ & $\mathcal{G}_{0.2}$ & $\mathcal{G_{\text{pre}}}$ & $\mathcal{G}_{0.1}$ & $\mathcal{G}_{0.2}$ & $\mathcal{G_{\text{pre}}}$ & $\mathcal{G}_{0.1}$ & $\mathcal{G}_{0.2}$ \\
    \hlineB{4}
    \multirow{4}{*}{$\alpha$} 
    & 0.1 & 92.94 & \colorbox{lime}{93.15} & 93.05 & \colorbox{yellow}{99.98} & 99.92 & 99.97 
          & 55.22 & \colorbox{lime}{60.06} & 58.72 & 51.99 & \colorbox{yellow}{53.18} & 50.31 \\
    \cline{2-14}
    & 0.2 & 88.97 & 90.93 & \colorbox{lime}{92.25} & 99.19 & 99.33 & \colorbox{yellow}{99.60} 
          & 29.88 & 35.28 & \colorbox{lime}{50.75} & 46.08 &  47.08 &  \colorbox{yellow}{52.09} \\
    \cline{2-14}
    & 0.3 & 83.25 & 84.99 & \colorbox{lime}{89.37} & 96.18 & 96.90 & \colorbox{yellow}{98.89} 
          & 7.98 & 9.56 & \colorbox{lime}{28.26} & 26.00 & 27.16 & \colorbox{yellow}{36.16} \\
    \cline{2-14}
    & 0.4 & 79.35 & 81.50 & \colorbox{lime}{86.75} & 94.03 & 95.11 & \colorbox{yellow}{98.36} 
          & 3.08 & 3.77 & \colorbox{lime}{13.63} & 11.32 & 11.90 & \colorbox{yellow}{16.10}  \\
    \hlineB{4}
    \multirow{4}{*}{$\beta$} 
    & 0.1 & 91.98 & \colorbox{lime}{92.33} & 92.08 & \colorbox{yellow}{99.93} & 99.92 & 99.90 
          & 81.54 & \colorbox{lime}{82.03} & 80.58 & 52.80 & \colorbox{yellow}{54.16} & 51.38 \\
    \cline{2-14}
    & 0.2 & 89.37 & 90.55 & \colorbox{lime}{91.50} & 99.24 & 99.38 & \colorbox{yellow}{99.52} 
          &  61.37 & 63.50 & \colorbox{lime}{75.87} & 45.57 & 47.21 & \colorbox{yellow}{52.19} \\
    \cline{2-14}
    & 0.3 & 84.27 & 84.94 & \colorbox{lime}{87.98} & 96.76 & 97.55 & \colorbox{yellow}{98.70}  
          &  25.79 & 26.04 & \colorbox{lime}{52.18} & 22.16 & 23.87 & \colorbox{yellow}{32.82} \\
    \cline{2-14}
    & 0.4 & 83.44 & 83.33 & \colorbox{lime}{85.77} & 94.80 & 97.81 & \colorbox{yellow}{98.44}  
          & 13.96 & 14.24 & \colorbox{lime}{31.86} & 4.41 & 6.55 & \colorbox{yellow}{9.23} \\
    \hlineB{4}
    \multirow{4}{*}{$\gamma$} 
    & 0.1 & 91.47 & \colorbox{lime}{91.94} & 91.56 & \colorbox{yellow}{99.92} & \colorbox{yellow}{99.92} & \colorbox{yellow}{99.92} 
          & 65.20 & \colorbox{lime}{66.43} & 63.28 & 52.94 & \colorbox{yellow}{54.00} & 51.06 \\
    \cline{2-14}
    & 0.2 & 88.95 & 89.60 & \colorbox{lime}{90.32} & 99.27 & \colorbox{yellow}{99.38} & \colorbox{yellow}{99.38}
          &  40.72 & 42.07 & \colorbox{lime}{55.32} & 45.61 & 46.87 & \colorbox{yellow}{52.30} \\
    \cline{2-14}
    & 0.3 & 84.81 & 85.77 & \colorbox{lime}{87.60} & 96.64 & 97.42 & \colorbox{yellow}{98.76} 
          &  12.11 & 12.22 & \colorbox{lime}{28.17} & 21.31 & 23.49 & \colorbox{yellow}{32.26}\\
    \cline{2-14}
    & 0.4 & 84.09 & 85.78 & \colorbox{lime}{86.59} & 94.27 & 95.31 & \colorbox{yellow}{98.54} 
          & 5.77 & 5.77 & \colorbox{lime}{14.77} & 4.42 & 6.00 & \colorbox{yellow}{6.70} \\
    \hline
\end{tabular}
\end{center}

\begin{figure}[htb!]
\includegraphics[width=1\linewidth]{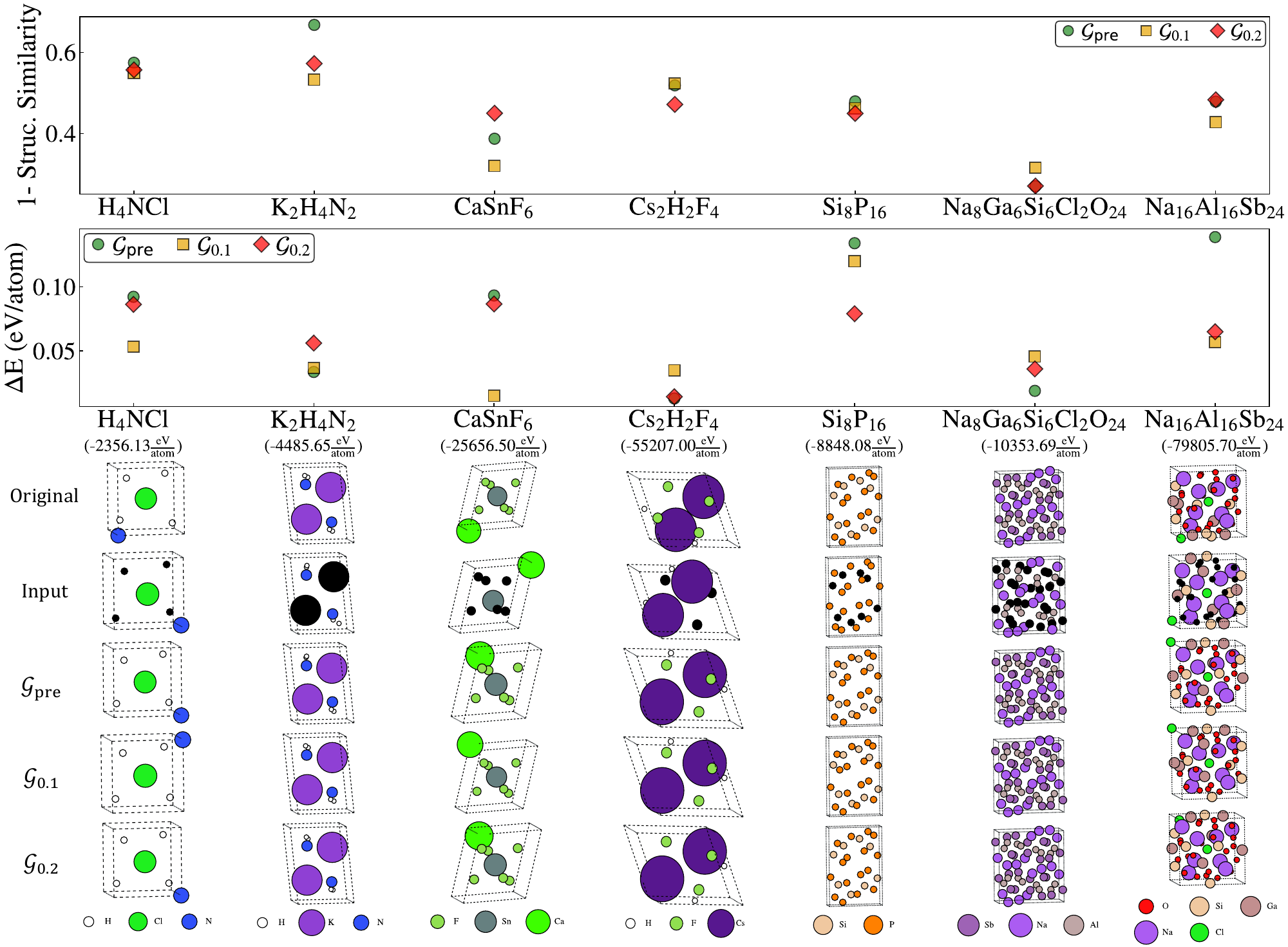}
\caption{\label{dft}
\textbf{DFT calculations.} A comparison of the structural similarity, relative total energy, calculated via DFT and denoted as $\Delta E$ (relative to the total energy of the original crystal structure, with values shown in brackets) and visualizations for crystal structures generated from randomly sampled crystals in the test set, with positions perturbed by $\tilde{\sigma}_{\textbf{noise}}=0.1$. 
}
\end{figure}

\inlinesubsection{DFT calculations} In Fig.~\ref{dft}, we present a comparison of the structural similarity and relative total energy of generated crystal structures. These structures are generated by various generative models with inputting contaminated structures randomly sampled from the test set. We draw the following conclusions:
\begin{enumerate}
    \item The generated structure exhibiting the lowest relative total energy $\Delta E$ shows the highest similarity for each crystal. This observation suggests that structural similarity can serve as an indicator of part of the information provided by DFT calculations. 
    \item Structures produced by GAN generators generally demonstrate a lower $\Delta E$ and a higher structural similarity, indicating the effectiveness of GANs in reconstructing structures with lower total energy compared to those generated by the pre-trained model.
    \item Although the generated structures are nearly visually indistinguishable from their original counterparts, their structural similarities are around $50\%$. This is attributed to the normalization of this metric by the average noise.
\end{enumerate}

\begin{figure}[htb!]
\includegraphics[width=0.85\linewidth]{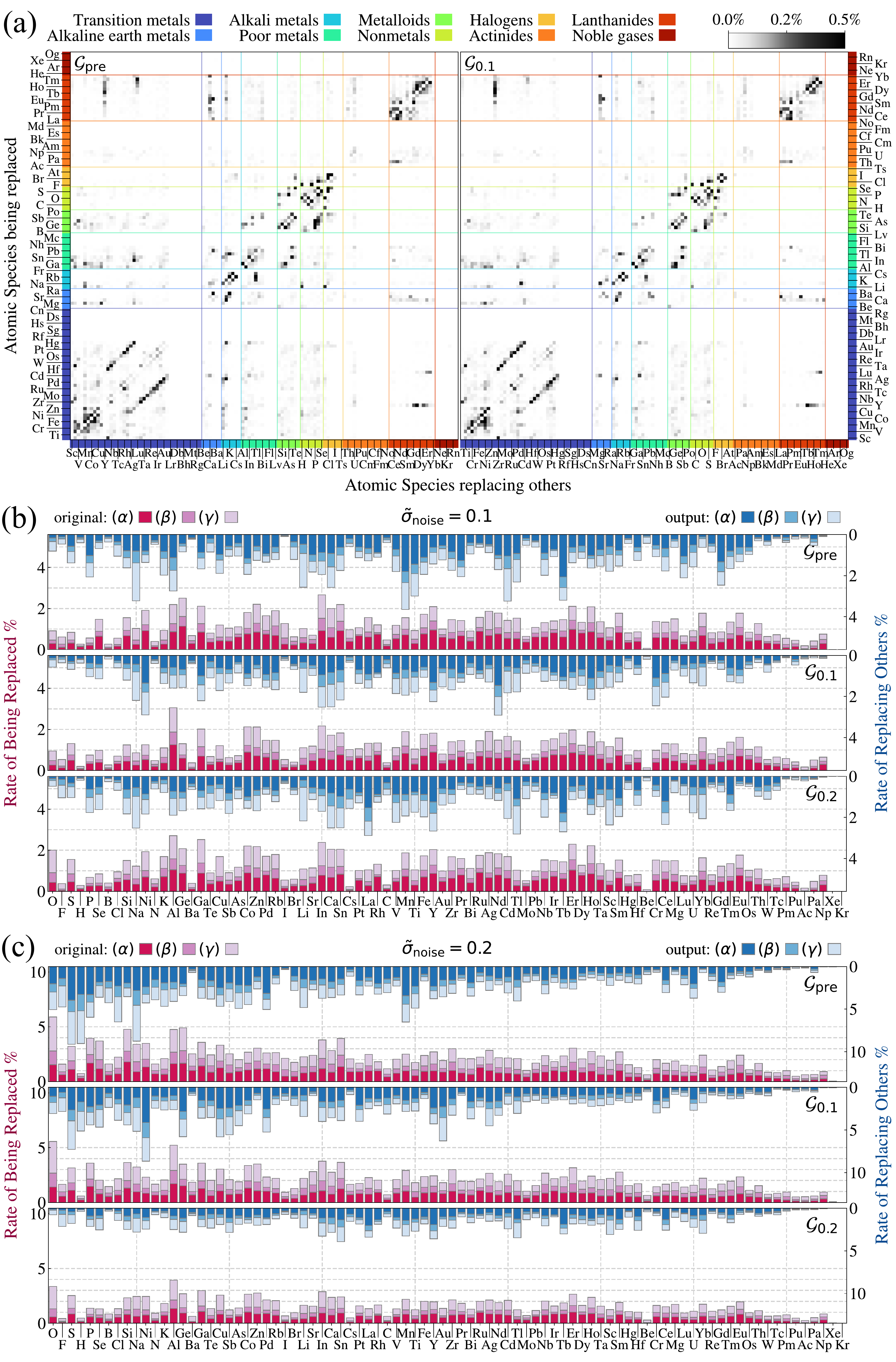}
\caption{\label{novelty_supp}
\textbf{Replacement Rates in Novel Generated Structures.} (a) shows the correlation between the species being replaced and those replacing them in the valid, novel generated crystal structure. The color bar displays the average correlations across augmented test data, encompassing all mask types and noise levels. Here, species are arranged based on their group classification. The crystal generation employed $\mathcal{G}_{\text{pre}}$ and $\mathcal{G}_{\text{0.1}}$. 
(b-c) displays the distribution of $\text{Rate}_{\text{replace(d)}}^{(\mathcal{G},\delta)}(X)$ in Eqn.~\ref{eq:replacement}, arranged by the occurrence probability, across different mask types ($\alpha,\beta,\gamma$) and positional noise level $\tilde{\sigma}_{\text{noise}}=0.1$ and $0.2$, for models ($\mathcal{G}_{\text{pre}}$, $\mathcal{G}_{\text{0.1}}$, and $\mathcal{G}_{\text{0.2}}$). 
Test data is identical to that utilized in our validity calculation.
}
\end{figure}

\inlinesubsection{Novelty}\label{subsec:novelty} In this study so far, we have primarily demonstrated the model's capability for reconstruction, but have not yet characterized its potential for generation. Therefore, we define a new quantity termed \textit{novelty}. A generated structure is considered novel if it is both compositionally and structurally valid, and it exhibits a distinct composition from the original. In Fig.~\ref{novelty_supp}, we present the rate of atomic species being replaced or replacing other species under three different masking types. 
Here, we denote $n_{\text{replaced}}^{(\mathcal{G},\delta)}(X)$ the number of novel crystals generated by our generative model $\mathcal{G}$, under mask type $(\delta)$, where species $s$ is replaced, and $n_{\text{replace}}^{(\mathcal{G},\delta)}(X)$ as the counterpart for the species $X$ substituting original species, and $N^{(\mathcal{G},\delta)}_{\text{novel}}$ as the total number of novel crystals generated under mask type $(\delta)$. The rate of species $X$ replacing other species (or being replaced) can be defined as 
\begin{equation} \label{eq:replacement}
\begin{aligned}
\text{Rate}_{\text{replace(d)}}^{(\mathcal{G},\delta)}(X) = n_{\text{replace(d)}}^{(\mathcal{G},\delta)}(X)/ N^{(\mathcal{G},\delta)}_{\text{novel}}.
\end{aligned}
\end{equation}\\

\begin{figure}[htb!]
\includegraphics[width=0.9\linewidth]{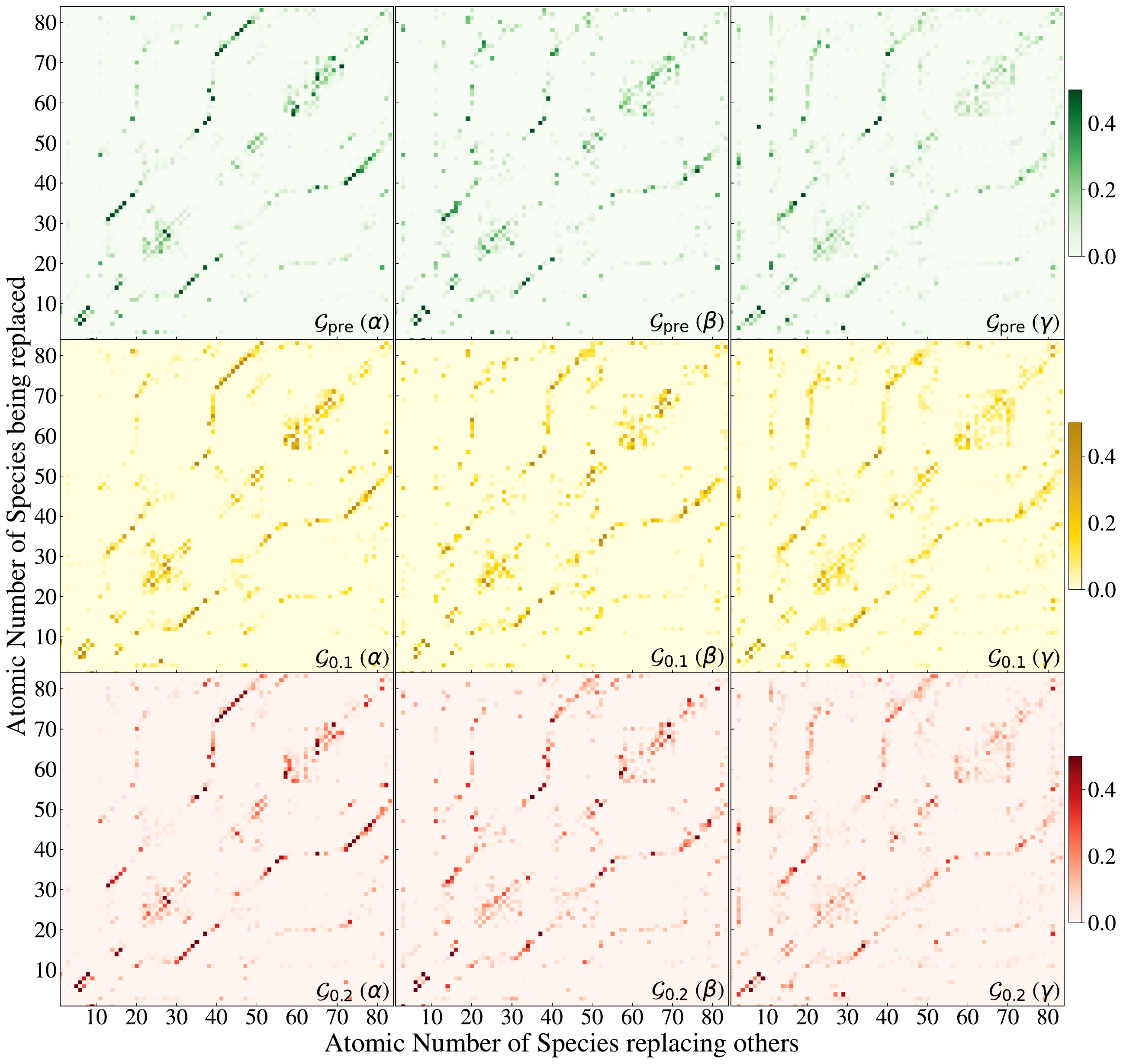}
\caption{\label{novelty_details}
\textbf{Replacement Rates in Novel Generated Structures.} Different from Fig.~\ref{novelty_supp}, species are organized by their atomic numbers, and the color bar indicates the average correlations across all novel samples with each corresponding species being replaced. Each column corresponds to various mask types employed to augment the test set, and each row is for different generative models.
}
\end{figure}

We notice that simple substitution methods~\cite{glawe2016optimal} compute the likelihood of substituting species $A$ with $B$, denoted as $P(A,B)$, through data mining. This method is particularly useful when the probability of one species being replaced by another remains constant, irrespective of different atomic and positional information within various crystals---although such scenarios that rarely applies. On the other hand, our generative model is essentially a more sophisticated probabilistic model, capable of offering conditional probabilities $P(A,B|\text{condition})$ under diverse conditions, such as varying structures and compositions. When conditions are overlooked, as shown in Fig.~\ref{novelty_details}, the averaged probabilities $\frac{1}{|\text{condition}|}\sum_{\text{condition}} P(A,B|\text{condition})$ given by our generative models bears a resemblance to those produced by simple substitution. \\

In Fig.~\ref{novelty_supp}a, we expand our analysis by exploring the correlation between the species being replaced and those replacing them in the valid, novel crystal structure, generated by the pre-trained model $\mathcal{G}_{\text{pre}}$ and the GAN generator  $\mathcal{G}_{\text{0.1}}$ (refer to the main text for $\mathcal{G}_{\text{0.2}}$), computed over augmented test data for all mask types and positional noise levels. This results, as depicted in the figure, indicate that within each group block, the correlations are predominant. This suggests that our generative models are capable of categorizing elements into groups in a manner analogous to the classification system used in the periodic table.\\

Figure ~\ref{novelty_supp}b-c presents the detailed distribution of $\text{Rate}_{\text{replace(d)}}^{(\mathcal{G},\delta)}(X)$ across different mask types, noise levels, and models. In this figure, the elements are ordered based on their occurrence probability in our dataset. The relatively uniform distributions implies that the model is not affected by the uneven distribution of species occurrences in the training set, thereby reducing potential bias. 
When the positional noise in the test input crystals exceeds that of the training inputs, we observe a tendency of the models to generate novel crystals. However, it is worth noting that while these crystals have passed the composition and structural validity tests, rigorous testing is required to prove the stability of their chemical structures.\\

\bibliography{ref}